# Spinel Ferrite-Based Materials for Electrochemical Applications: Synthesis, Applications, and Future Perspectives

**Sakhi Tiwari[1], Ashish Dubey[1], Pragyanand Prajapati[2], Akhilesh Kumar Singh[3], Ashutosh Kumar[4,*], Jai Singh[1,$]**

[1]Department of Pure and Applied Physics, Guru Ghasidas Vishwavidyalaya (A Central University), Bilaspur, C.G. 495006, India

[2]Government Girls Degree College Samadi, Ahiraula, Azamgarh, Uttar Pradesh-276001, India

[3]School of Materials Science and Technology, Indian Institute of Technology (BHU), Varanasi 221005, India

[4]Functional Materials Laboratory, Department of Materials Science and Metallurgical Engineering, Indian Institute of Technology Bhilai, Chhattisgarh 491002, India

*Email: ashutosh@iitbhilai.ac.in
#Email: jai.bhu@gmail.com

## Table of Content

## Abstract

Spinel ferrites have shown wide range of applications in various fields, including supercapacitors, Li-ion batteries, water splitting, chemical sensors, catalytic activity, high-frequency magnetic devices, and biomedical, hyperthermia, drug delivery applications., etc. This review focuses on the latest progress and trends on design of spinel ferrites ($MFe_2O_4$; M = Divalent transition metal ion) based materials for electrochemical applications. Spinel ferrites exhibit good chemical stability, high surface area and excellent electrochemical behaviour with their multiple oxidation states, making them suitable for the detection of a wide range of gaseous analytes including volatile organic compounds, heavy metal ions, biomolecules, and environmental pollutants. It also makes spinel ferrites a great choice for efficient energy storage and utility in supercapacitors. The electrochemical performance of spinel ferrite-based electrode materials can be effectively tuned via morphology control, incorporation of carbon-based materials, compositional substitution, doping and forming composition systems, especially in nanostructured form. This review will serve as a comprehensive resource for researchers interested in the synthesis and various enhancement techniques for electrode material composition of electrochemical device applications specifically, gas sensors and supercapacitors, which are two of the highest emerging functional applications for the new sustainability directed world, utilizing these advanced materials.

**Keywords:**

Spinel Ferrite, Nanostructures, Electrochemical Sensors, Electrochemical capacitors

## 1. Introduction

Spinel ferrites currently have attracted considerable popularity due to its multifunctional roles in energy technologies, catalysis, sensing, electronics, environmental remediation, and biomedical fields. Spinels were initially employed as precious stones, such as red spinels containing $Cr^{3+}$ and blue spinels containing $Fe^{2+}$ and $Zn^{2+}$. Little known until the year 1915, Bragg and Nishikawa authenticated that spinels have a general chemical formula $A_xB_{3-x}O_4$, in which A and B are metal cations that typically have $2^+$ and $3^+$ charges and occupy tetrahedral and octahedral sites, respectively.[1] The compounds with Fe as B cation are known as spinel ferrites with chemical formula $AFe_2O_4$ where A can be Zn, Ni, Mn, Fe, Cu, Co etc. Spinel structures, in general, feature a closely packed cubic arrangement, where half of the octahedral sites and 1/8 of the tetrahedral sites are occupied by cations A and B, respectively, while the remaining sites are filled with oxygen anions. This configuration results in spinels keeping four octahedral and eight tetrahedral voids in effect and can include a variety of cation types, even in higher oxidation states.[2]

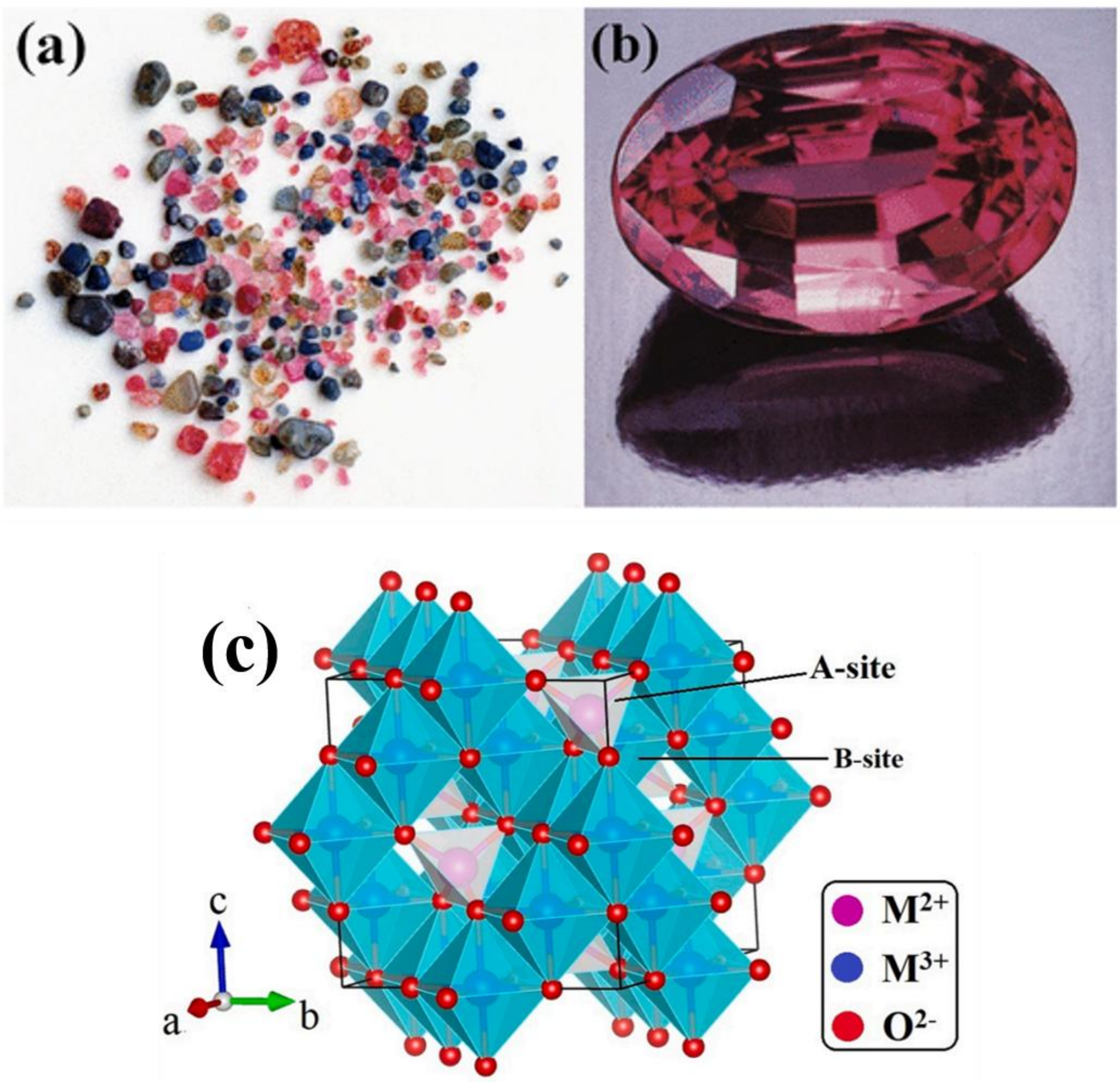

Fig. 1 (a) Natural Sapphire-Ruby[3] (b) Pink jewellery stone in spinel structure[4] (c) Illustration of spinel structure with tetragonal A site with $M^{2+}$ divalent metal cations, orthogonal B site $M^{3+}$ trivalent metal cations and remaining cubic sites filled with oxygen anions[5]

Spinel ferrite materials have been extensively reviewed in several detailed review studies.[6] Several reviews discussed its use in photovoltaic-electrolysis and photo-electrochemical water splitting systems[7], lithium-ion batteries, supercapacitors[8], and microwave absorption.[9] In recent years, Bimetallic spinel ferrite-based materials have been extensively reviewed for their use in electrochemical devices due to due to their low toxicity, the ability to easily control their morphology using straightforward synthesis methods, stable chemical composition, remarkable electrochemical behaviour, and commendable mechanical stability. The unique intrinsic properties of spinel ferrite can be utilized depending on its application. In electrochemical sensors, spinel ferrites provide increase in active sites for catalytic effects. Their rich redox chemistry and enhanced electron transfer kinetics leads to improved sensitivity. These enhancements arise from more effective interfacial charge transfer processes and stronger interactions between the analyte and the electrode surface, allowing for faster reaction kinetics and improved detection.[10] For example, Zhang et al. provided a comprehensive overview of the advancements in gas sensors employing zinc ferrite ($ZnFe_2O_4$).[11] Similarly, in case of energy-storage devices, especially for the supercapacitor applications, spinel ferrites mainly contribute multiple oxidation states and stable structures providing for a high energy density, specific capacitance and fast-rate capability. These improvements are governed by enhanced charge storage mechanisms involving both electrical double-layer formation and fast surface redox reactions.[12] Verma and Das fabricated $CuFe_2O_4$ based electrode for supercapacitor applications.[13] There has been a noticeable rise in papers in recent years that provide fascinating insights into these materials' electrochemical performance as shown in Fig. 2.

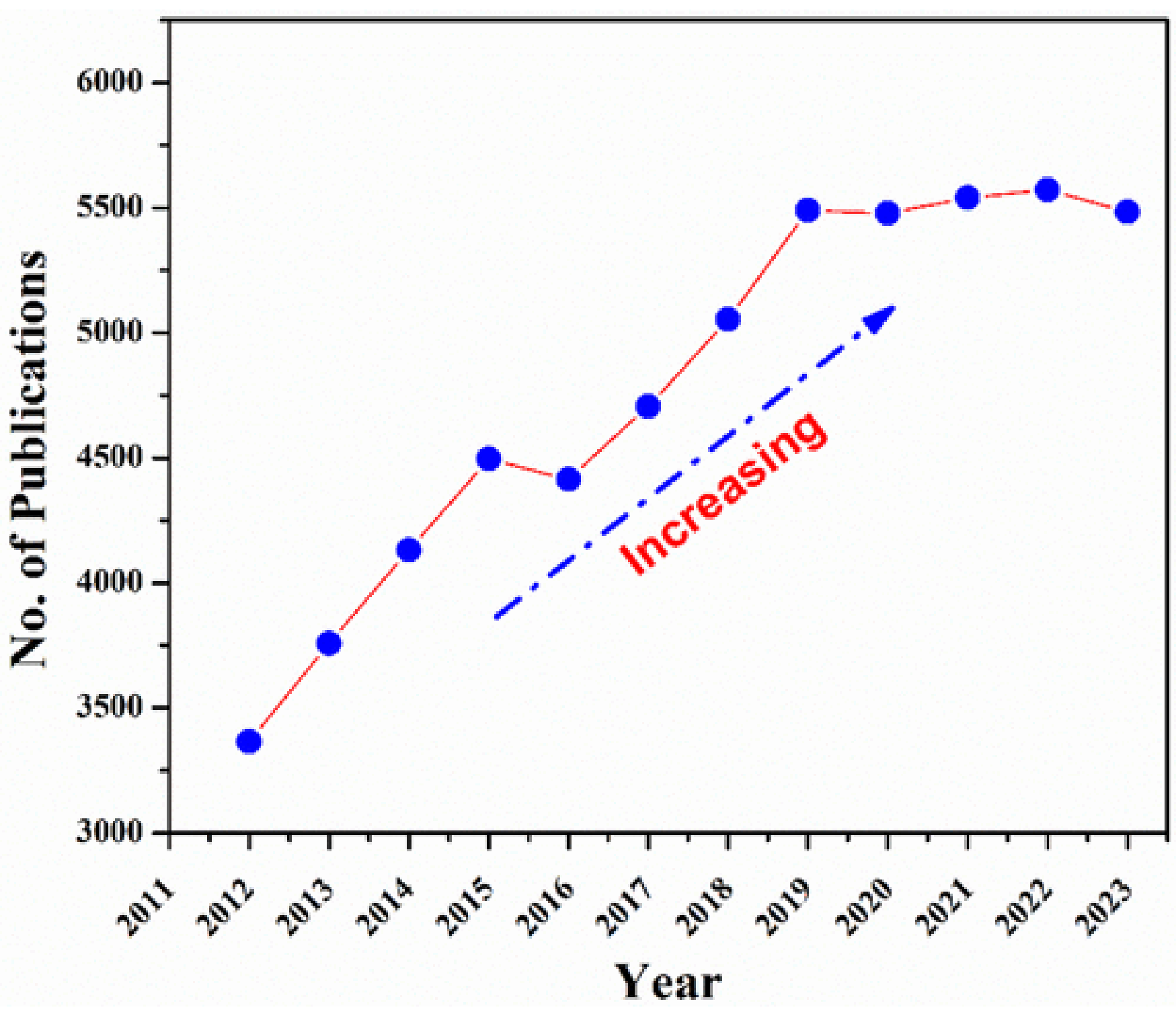

Fig. 2 Number of publications with 'spinel ferrites'[14]

Although numerous review articles have discussed spinel ferrites in the context of magnetic materials, photocatalysis, batteries, electrocatalysis, and biomedical applications, a comprehensive understanding of the relationship between synthesis strategy, defect engineering, structural evolution, and electrochemical performance for gas sensing and supercapacitor applications remains limited. Most existing reviews primarily summarize individual applications without critically correlating how synthesis-induced modifications in crystallinity, morphology, cation distribution, oxygen vacancies, and interfacial engineering collectively govern charge-transfer kinetics and electrochemical activity. Moreover, recent advances involving graphene-based hybridization, hierarchical nanostructures, heterointerface engineering, and data-driven materials design have not been systematically integrated within a unified framework.

In this review, we comprehensively discuss the synthesis–structure–property–performance relationship of spinel ferrites, emphasizing how different synthesis strategies influence crystallinity, morphology, defect chemistry, cation distribution, and electronic structure, which collectively govern their electrochemical performance. Particular focus is placed on gas-sensing and supercapacitor applications, highlighting the underlying electrochemical mechanisms and the role of nanostructuring, carbonaceous hybridization, and transition metal and rare-earth doping in enhancing device performance. Furthermore, current challenges, emerging strategies, and future perspectives, including flexible and wearable electronics, biomedical and environmental monitoring, scalable fabrication, commercialization, and AI-assisted materials design, are critically discussed to provide guidelines for the rational development of next-generation spinel ferrite-based electrochemical devices.

## 2. Synthesis approaches of spinel ferrite nanomaterials

Traditionally, spinel oxides have been synthesized using a conventional solid-state technique, which frequently needs high temperatures for the thermal breakdown of precursors. However, materials made by this method tend to show larger crystallite size and grains resulting in lower specific surface area, making them less suitable for electrochemical energy storage and gas sensing applications where high surface electrochemistry is required.[15] To overcome this limitation, various low-temperature synthesis procedures have been developed. These procedures involve wet chemical routes like hydrothermal, sol-gel, co-precipitation methods which leads to formation of numerous innovative morphologies. These morphologies include from dimensional structures like nanoneedles, nanospheres, and nanocubes to complex and striking flower-like, urchin-like morphologies. The hierarchical nano/microarchitectures like core-shell structure or hollow tube, porous sphere or yolk-shell structure, all of which contribute to improved electrochemical performance.[16]

To synthesis nanomaterials, the two main categories of methods are top-down and bottom-up approaches. In the top-down approach, bulk precursors are broken down into nanoscale materials, whereas in the bottom-up method, smaller chemical species assemble to form larger functional structures.[17] Among bottom-up approaches, several techniques such as hydrothermal, sol-gel, template-assisted, electrospinning and microwave-assisted methods have gained considerable attention for the synthesis of ferrite ($AFe_2O_4$) nanomaterials due to their ability to precisely control material properties. Importantly, the choice of synthesis method plays a crucial role in tailoring $AFe_2O_4$ materials to meet the specific requirements of electrochemical applications, as it governs key characteristics such as particle size, surface area, porosity, and morphology, which directly influence electrochemical performance.[18,19]

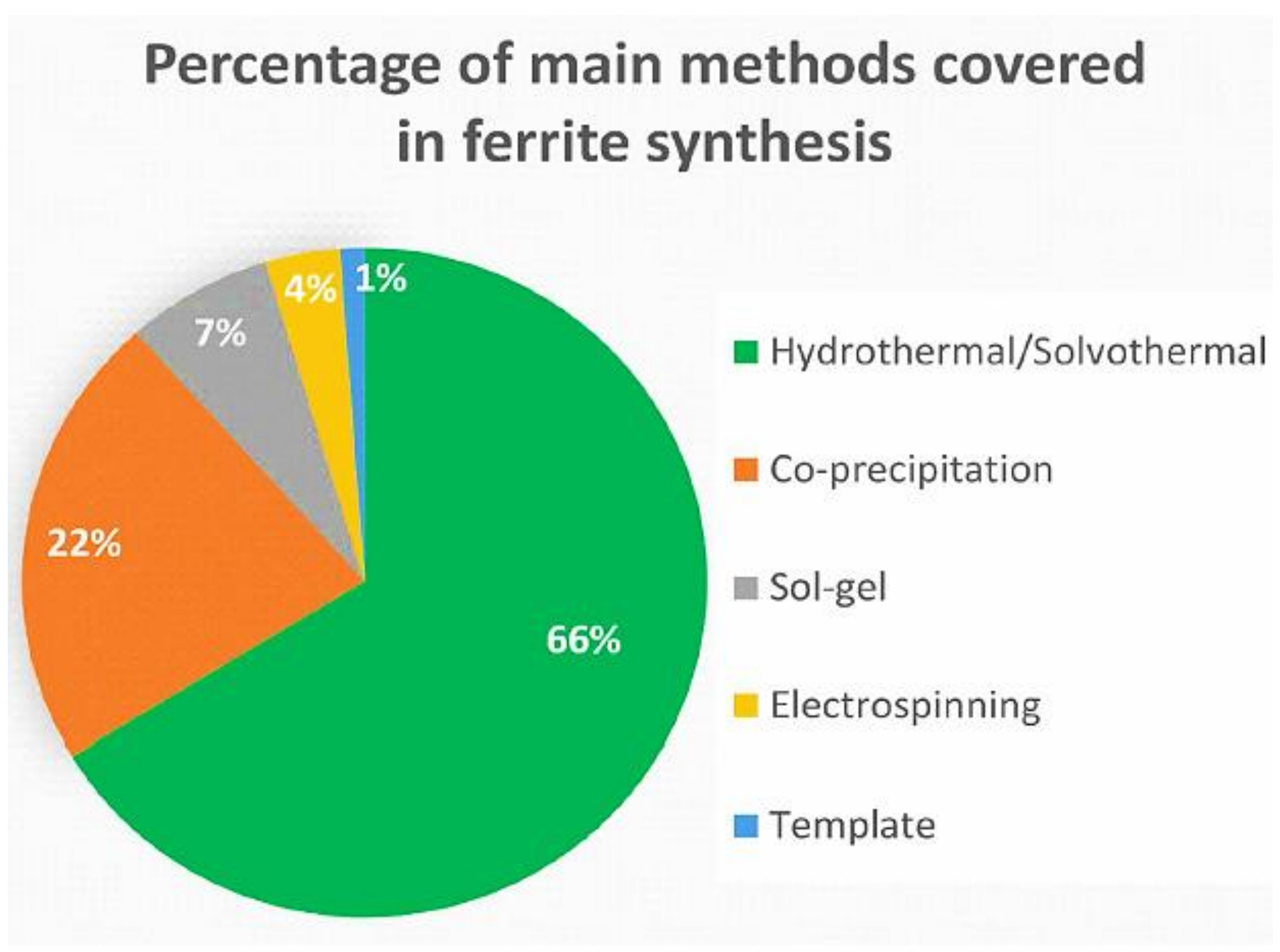

Fig. 3 Percentage of preparatory techniques applied to the synthesis of spinel ferrites[20]

Various systematic comparisons have shown how the synthesis method affects the structure and morphology of $MFe_2O_4$ (M = Fe, Zn, Mn, Co, Ni, Cu, etc.) nanopowders. Coprecipitation produces homogenous nanoparticles but agglomerated and irregular sizes with little or no morphology control, while sol-gel produces highly homogenous and crystalline nanoparticles usually spherical shapes with moderate agglomeration. Finally, the hydrothermal route produces a more complex and morphology-controlled nanomaterials. So, for example, the agglomeration reduces conductivity, thus, various synthesis routes highlight how tailoring them improves material properties.[21] While conventional techniques such as sol-gel and hydrothermal synthesis remain widely employed for the preparation of spinel ferrite nanoparticles, more emerging strategies including electrospinning, template and microwave-assisted methods have gained significant attention in recent years.

### 2.1 Co-precipitation

The co-precipitation method is one of the simplest and most widely adopted approaches for synthesizing metallic nanoparticles and has been extensively used for the preparation of spinel ferrites ($AFe_2O_4$). In this method, soluble metal salts such as nitrates, acetates, or chlorides are dissolved in an aqueous medium, followed by the addition of a suitable precipitating agent (e.g., NaOH or $NH_4OH$) to induce the formation of hydroxide precursors, which then transforms into the desired ferrite phase upon drying and calcination. The synthesis conditions have an immense impact on the structural and morphological properties of the resultant nanoparticles, particularly the precursor chemistry and precipitation kinetics. The pH and temperature regulate nucleation and crystal growth behaviour, whereas precursor concentration and stirring conditions influence particle homogeneity and phase formation. Thus, precise control over stoichiometry and reaction conditions is essential to yield phase-pure ferrite nanostructures with controlled morphology. Due to its simple processing, cost-effectiveness, and ability to produce chemically homogeneous nanostructures under relatively mild conditions, the co-precipitation method has emerged as a widely preferred route for synthesizing nanoparticles on a large scale. Owing to these merits, the method has been widely utilized for synthesizing ferrite nanomaterials for supercapacitor, sensing, catalytic, and magnetic applications. However, rapid nucleation and uncontrolled crystal growth may often result in particle agglomeration and broad particle-size distribution, which can adversely affect electrochemical performance.[22–24]

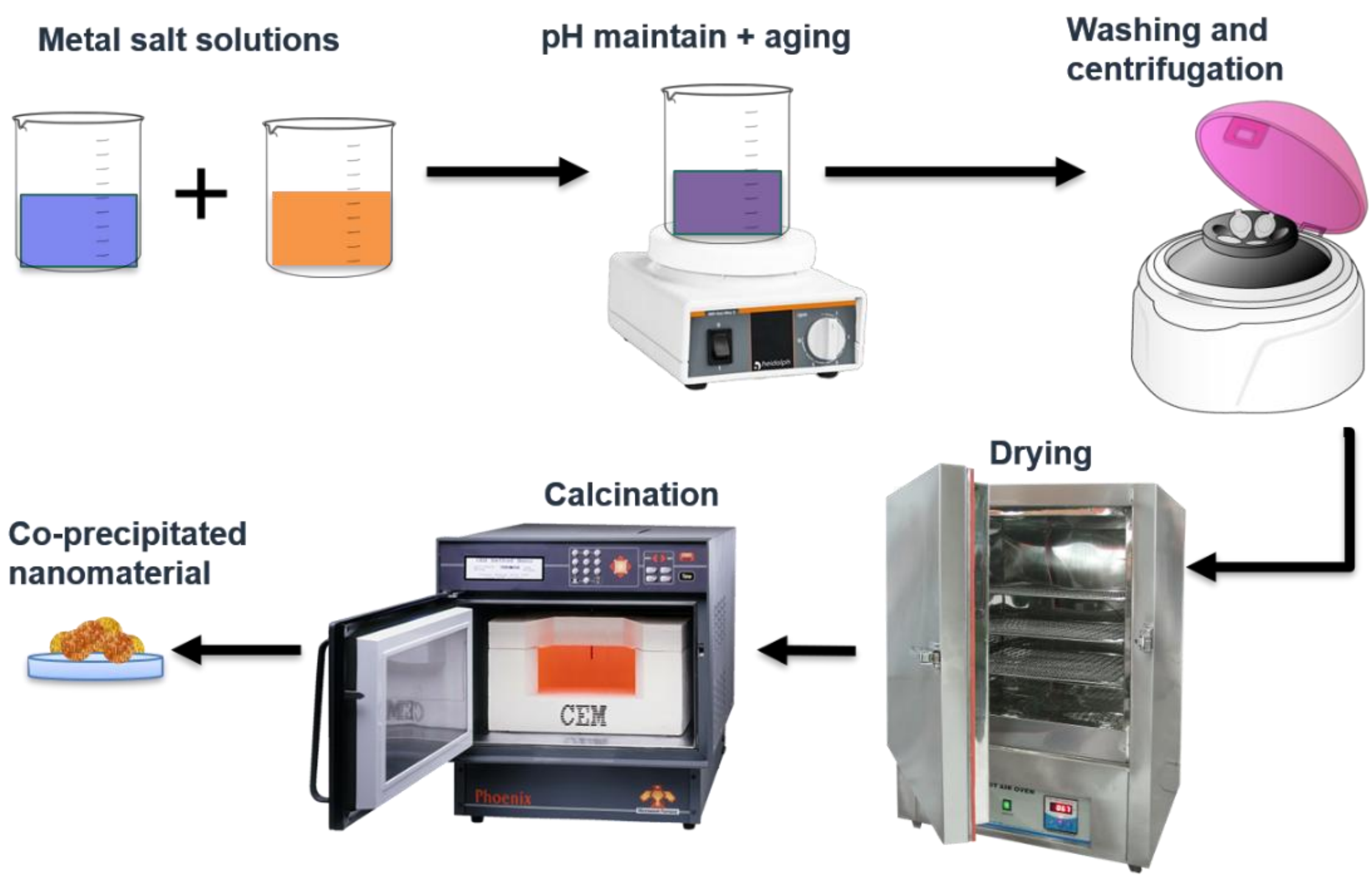

Fig. 4 Synthesis schematic of spinel ferrite nanomaterials via co-precipitation method

Zhang et al. synthesized $NiFe_2O_4$ nanoparticles using nickel chloride and ferric chloride precursors in the presence of oleic acid as a surfactant under alkaline conditions (pH=10). Their investigation revealed that increasing calcination temperature promoted particle agglomeration and grain growth, thereby reducing the effective surface area of the material. Such morphological evolution significantly influences ion diffusion and electrochemical activity in ferrite-based electrodes.[25] Similar co-precipitation approaches have also been employed for the synthesis of $ZnFe_2O_4$, $MnFe_2O_4$, $CoFe_2O_4$, $MgFe_2O_4$.[26–29] The technique is utilized for numerous additional compositions based on spinel ferrites as well. For instance, $NiFe_2O_4$ with Sn substitution, copper oxide@$CuFe_2O_4$, and MWCNT supported $CuFe_2O_4$ nanocomposites. etc., are employed in various electrochemical and sensing applications.[30–32]

In recent years, co-precipitation strategies have further been integrated with external stimuli such as ultrasonic irradiation to improve precursor mixing, nucleation rate, and particle dispersion. For instance, ultrasound-assisted synthesis of $CuFe_2O_4$ nanoparticles has been reported to enhance the electrochemical sensing performance toward deoxyepinephrine detection.[33] Overall, the co-precipitation method remains a versatile and economically viable approach for tailoring ferrite nanostructures with tunable morphology and functional properties suitable for advanced sensor and energy-storage applications.

**2.2 Sol-Gel method**

The sol-gel technique has attracted considerable attention for the synthesis of spinel ferrite nanomaterials because of its lower processing temperature, outstanding compositional uniformity, and capacity to create ultrafine particles with regulated size and shape. In this approach, a uniform precursor solution (sol) experiences hydrolysis and polycondensation reactions to create a three-dimensional gel network, which is then dried and calcined to achieve the desired ferrite phase.[34] This method allows for precise control over particle size distribution, porosity, crystallinity, and structural uniformity, rendering it particularly effective for customizing ferrite materials for gas sensing and electrochemical energy-storage uses. Additionally, the ultimate microstructure and morphology can be greatly affected by synthesis parameters such as precursor makeup, pH level, calcination temperature, fuel-to-oxidizer ratio, and heating conditions, enabling adjustable electrochemical and surface characteristics.[35,36] Due to these benefits, sol-gel synthesis has been widely utilized for producing various spinel ferrites and their composites, including $MgFe_2O_4$, $CoFe_2O_4/SiO_2$, $MFe_2O_4/TiO_2$ (M = Zn, Co), and $NiFe_2O_4$-based nanomaterials for sensing and supercapacitor uses.[37–40] The increased surface activity through tailored morphology achieved via sol-gel techniques enhances charge-transfer properties and electrochemical efficacy.

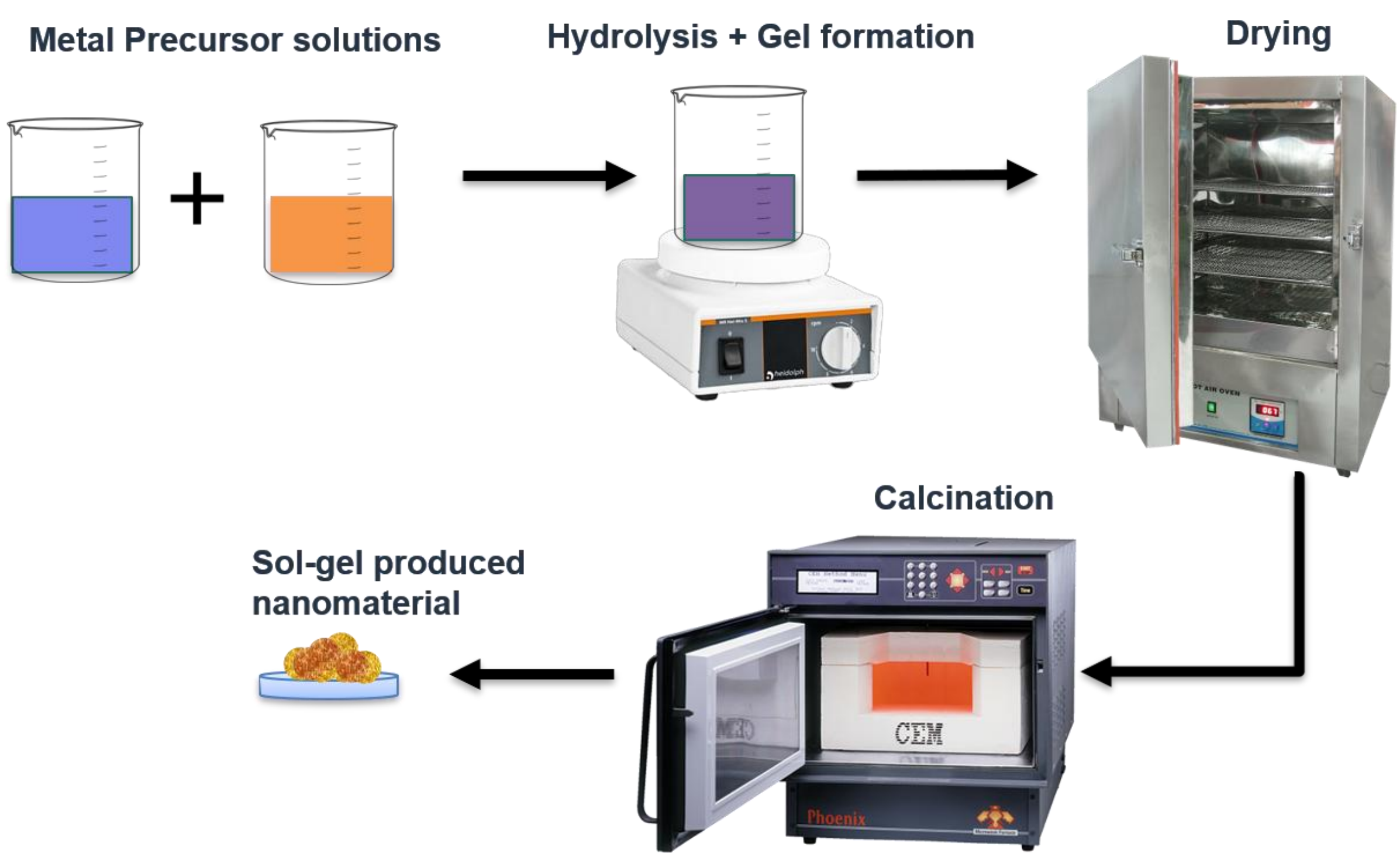

Fig. 5 Synthesis schematic of spinel ferrite nanomaterials via sol-gel method

Despite these advantages, the conventional sol-gel method generally involves prolonged synthesis duration and relatively high thermal treatment temperatures. To overcome these limitations, the sol-gel auto-combustion route has come out to be as an efficient modification, capable of reducing processing time and crystallization temperature through rapid exothermic reactions.[41,42] The combustion process also promotes rapid crystallization and frequently generates porous ferrite nanostructures with high surface area, which are advantageous for electrochemical applications. The combustion behaviour and resulting ferrite microstructure are strongly governed by factors such as fuel type, fuel-to-oxidizer ratio, pH, and heating conditions, which influence particle size, crystallinity, agglomeration, and phase purity.[36] The statements are supported by studies on $MgFe_2O_4$ synthesised using the citric acid-assisted sol-gel auto-combustion method demonstrating that crystallinity, lattice parameters, and cation distribution were strongly influenced by both pH and annealing temperature.[43] Similarly, investigations on $CoFe_2O_4$ and $NiFe_2O_4$ prepared through solution combustion synthesis revealed fuel-to-oxidiser ratio manipulating purity phase and crystalline size using glycine as the fuel.[44] These characteristics have encouraged the widespread use of sol-gel auto-combustion synthesis for developing ferrite-based electrodes. For instance, cubic spinel $ZnFe_2O_4$ synthesized through this route demonstrated improved particle-size control and enhanced phase purity suitable for electrochemical charge-storage applications.[45] Similarly, Nikam et al. optimized the composition of $Ni_xCo_{1-x}Fe_2O_4$ ferrites synthesized via an energy-efficient auto-combustion process resulted in nanocrystalline single phase, with cubic spinel ferrite for supercapacitor applications.[46] In another notable study, Gabal and co-workers developed a bio-templated solution combustion approach for synthesizing $MFe_2O_4$ (M = Co, Ni, Cu, Mg, and Zn) ferrospinels using egg white as a bio-fuel in place of conventional organic solvents. In this process, the metal ions acted as oxidizing agents while the egg white served as fuel, generating an exothermic redox reaction that facilitated ferrite formation.[47] Such environmentally benign and sustainable synthesis strategies demonstrate the continuing evolution of sol-gel-based methods as versatile platforms for designing tailored ferrite nanomaterials for advanced electrochemical applications.

## 2.3 Hydro(solvo)thermal synthesis methods

The thermal method, especially the hydrothermal and solvothermal approaches, is one of the most popular among the methods utilized in the synthesis of spinel ferrite nanoparticles. The hydrothermal approach requires the reaction to be carried out at high temperature and high pressure using the aqueous solution. On the other hand, in the solvothermal approach, the reaction is carried out under the similar condition of hydrothermal synthesis using non-aqueous solvents such as organic solution. Among the parameters required for these methods include temperature, pressure, time taken to complete the reaction, solvent used, and the amount of the reactant used. The most important advantage associated with the use of these methods includes the ability to control the morphology of the synthesized homogeneous ferrites. One of the most environmentally beneficial features of the hydrothermal technique is its capability to minimize waste generation.[48]

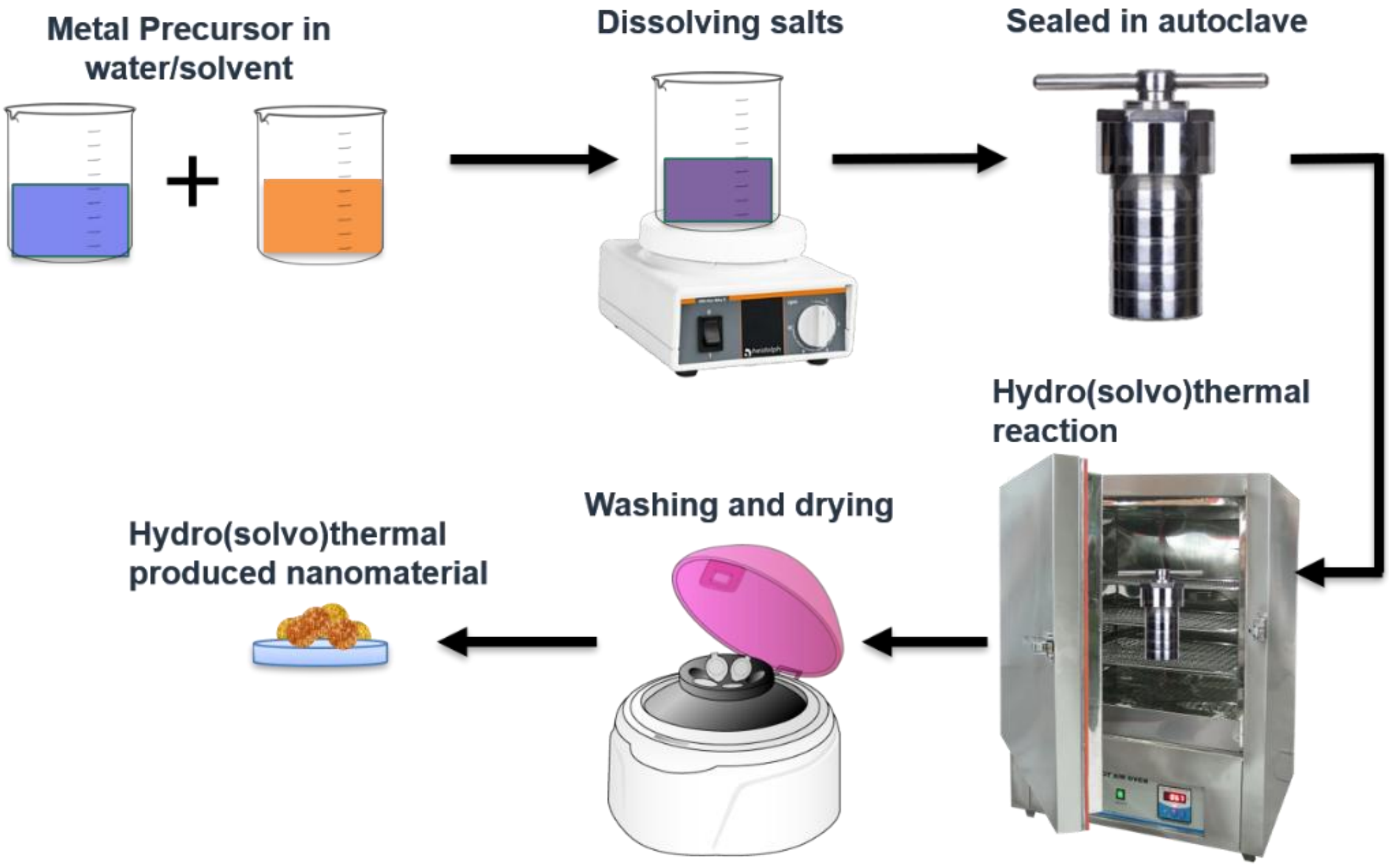

Fig. 6 Synthesis schematic of spinel ferrite nanomaterials via hydrothermal and solvothermal method

Even when based on similar synthesis principles, hydrothermal and solvothermal processing can produce ferrite nanostructures with significantly different structural and morphological characteristics due to variations in solvent environment and surfactant interactions. For instance, comparative studies on $CuFe_2O_4$, $MgFe_2O_4$, and $ZnFe_2O_4$ nanoparticles synthesized under identical reaction temperature and duration demonstrated notable differences between hydrothermal and solvothermal routes. In the hydrothermal process, trisodium citrate aqueous solution was used as a chelating agent, whereas ethylene glycol and polyethylene glycol were used as solvents in the solvothermal approach. The solvothermally synthesized samples exhibited comparatively larger crystallite sizes. Distinct morphology variations were also observed with hydrothermal synthesis producing cubical particles and solvothermal processing leading to porous spherical structures. Furthermore, residual polyethylene glycol and ethylene glycol could not be completely eliminated under identical thermal treatment conditions, signifying the major role of solvent chemistry in determining the final microstructure and surface characteristics of ferrite nanoparticles.[49] Similarly, solvent driven chemistry demonstrated distinct morphologies for $ZnFe_2O_4$ subjected to various solvents such as oxalic acid, HMTA, glycerol, and ascorbic acid exhibiting a diverse range of distinct ZFO nanostructures, including stacked rods, microparticles, spheres, and nanosheet-assembled flowers, respectively.[50]

To better illustrate the contribution of the present review, Table 1 compares representative research articles published in recent years with the current work. This review establishes a comprehensive correlation between synthesis methods, structural evolution, defect engineering, and electrochemical performance.

Table 1 Recent reports on hydrothermal and solvothermal method-based synthesis of $MFe_2O_4$

| System | Synthesis method | Solvent(s) used | Morphology | crystalline size/particle size | Thermal temperature | Red |
|---|---|---|---|---|---|---|
| $CuFe_2O_4$ nanoparticles | Solvothermal | Ethylene Glycol | core-shell nanoparticles | 5.5 nm w/o citric acid<br>3.5 nm with citric acid | 190°C, 10h | [51] |
| $ZnFe_2O_4$ | Hydrothermal | Double distilled water | Spherical nanoparticles | 40.11 nm | 165°C, 16 h | [52] |
| $MnFe_2O_4$ | Solvothermal | Ethanol | nanoflower | - | 180°C, 20h | [53] |

| System | Synthesis method | Solvent(s) used | Morphology | crystalline size/particle size | Thermal temperature | Red |
|---|---|---|---|---|---|---|
| $ZnFe_2O_4$ | Hydrothermal | Distilled water | Microcubes | - | 180°C, 22 h | [54] |
| ($MFe_2O_4$, M = Fe, Co, Ni, Mn, Cu, Zn) | Solvothermal | Ethylene glycol solvent and PEG 600 | Mesoporous spherical nanoassemblies | 20 to 35 nm (crystallite size) and 50 to 140 nm (particle size) | 200°C, 24 h | [55] |
| $MnFe_2O_4$ | Hydrothermal | $H_2SO_4$, $HNO_3$, and deionized water | $MnFe_2O_4$ nanoparticles decorating CNT network on 3D supported carbon cloth | - | 140°C, 6h | [56] |

| System | Synthesis method | Solvent(s) used | Morphology | crystalline size/particle size | Thermal temperature | Red |
|---|---|---|---|---|---|---|
| $CuFe_2O_4$ | Hydrothermal | Distilled water | spherical particles deposited over of rGO nanosheets | 35.5 nm | 160°C, 4 h | [57] |
| $MFe_2O_4$ (M= Ni, Co, Mn) | Solvothermal | Ethylene glycol | Agglomerated nanoparticles ($MnFe_2O_4$,$CoFe_2O_4$) and densely packed aggregates ($NiFe_2O_4$) | 13.72 nm ($MnFe_2O_4$), 20.72 nm ($CoFe_2O_4$), 11.86 nm ($NiFe_2O_4$) | 200 °C, 24 h | [58] |

## 2.4 Template synthesis Method

The template-assisted approach has surfaced as a versatile and widely adopted strategy for the synthesis of nanostructured materials, particularly with advances in nanotechnology. This method enables precise control over crystal nucleation and growth, hence, allowing fine tuning of morphology, pore structure, and overall architecture of the target material. The synthesis typically involves three key steps: These steps involve: (i) Template selection; (ii) Deposition or formation of the right material in or on the template by using processes like hydrothermal/solvothermal/co-precipitation/sol-gel process, etc.; and (iii) Extraction of the template to get the desired product. These templates can be broadly classified in two types: (a) Rigid templates like silica or anodic aluminum; and (b) Soft templates that consist of surfactants or polymers. Both these categories have their unique strengths concerning the control over the size and shape as well as porosity through molecular interactions, either intra or intermolecular. It is highly essential that the template should be extracted carefully due to this very reason.[59,60]

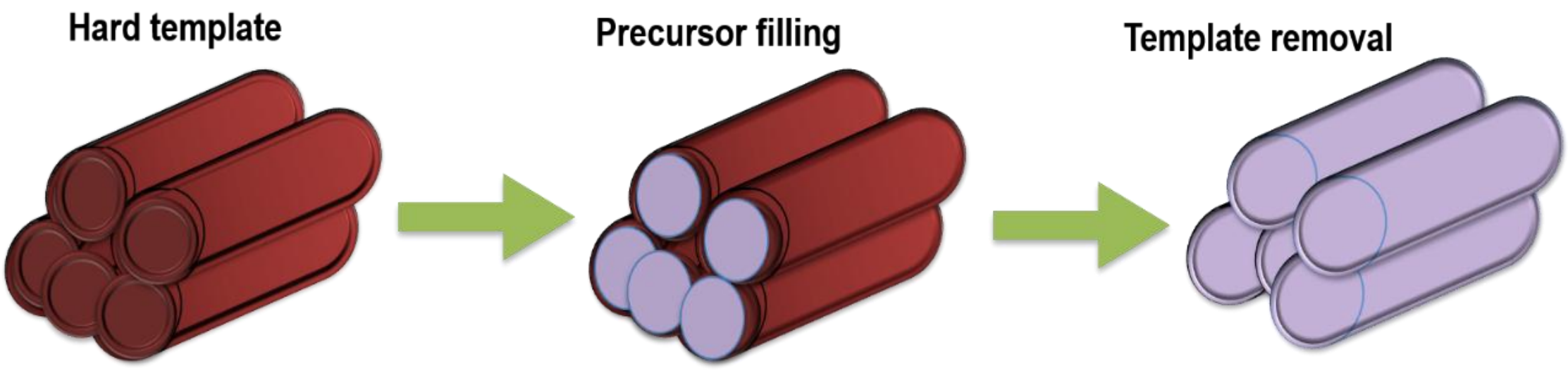

Fig. 7 Synthesis schematic of spinel ferrite nanomaterials via hard template-assisted method

The template method has been widely used in the production of ferrite nanostructures for gas electrochemical sensing and capacitors applications, where morphology and surface accessibility are critical to performance, where morphology and surface accessibility play a decisive role in performance. For instance, Huang et al. synthesised $ZnFe_2O_4$ pleated hollow microspheres for acetone sensing applications using NaCl as a sacrificial hard template. The cubic structure of NaCl provided confined growth sites as template for ferrite deposition and dissolved easily, leaving behind the formation of hollow porous architectures with enhanced gas diffusion and active surface area.[61] Zheng et al. fabricated $CoFe_2O_4$ nanowires using a double-pass porous alumina template, which used the well-ordered nanochannels of anodic alumina to direct anisotropic growth of one-dimensional crystals. The resulting nanowire arrays had improved aspects ratios and directional charge transport properties due to the template-based confinement.[62] García et al. utilized bacterial cellulose nanoribbons as a bio-template for synthesizing cobalt ferrite nanotubes. As a result, the interconnected fibrous network of cellulose provided a natural scaffold for ferrite deposition, while further removing template created porous nanotubular architectures with enhanced surface accessibility and structural interconnectedness. This method showcased the possibilities of sustainable and eco-friendly bio-templating techniques for shaping ferrite morphology.[63] Overall, template-assisted synthesis provides exceptional flexibility in tailoring ferrite nanostructures with controlled morphology and porosity, making it a powerful strategy for advanced electrochemical and sensing applications.

## 2.5 Electrospinning

In recent decades, electrospinning has emerged as a highly effective alternative for the fabrication of ferrite-based nanostructures, compared to conventional synthesis techniques such as hydrothermal, solvothermal, and co-precipitation methods. The principal advantage of electrospinning lies in the controlled synthesis of continuous one-dimensional (1D) nanofibers that can be tailored according to their functional and structural attributes. The technique works on basic mechanism using the high electric field to draw polymer-precursor solutions into extremely fine fibers under fairly mild processing conditions. This causes charged polymer jets to elongate and stretch in a strong electric field, promoting anisotropic growth and enabling the formation of continuous one-dimensional structures with a high aspect ratio. These properties are especially useful for gas sensing and electrochemical energy storage applications because they provide an abundance of active sites for gas adsorption while also facilitating efficient ion diffusion and electron transport. Furthermore, electrospinning has advantages such as cost effectiveness, scalability, environmental compatibility, and operational simplicity.[64,65]

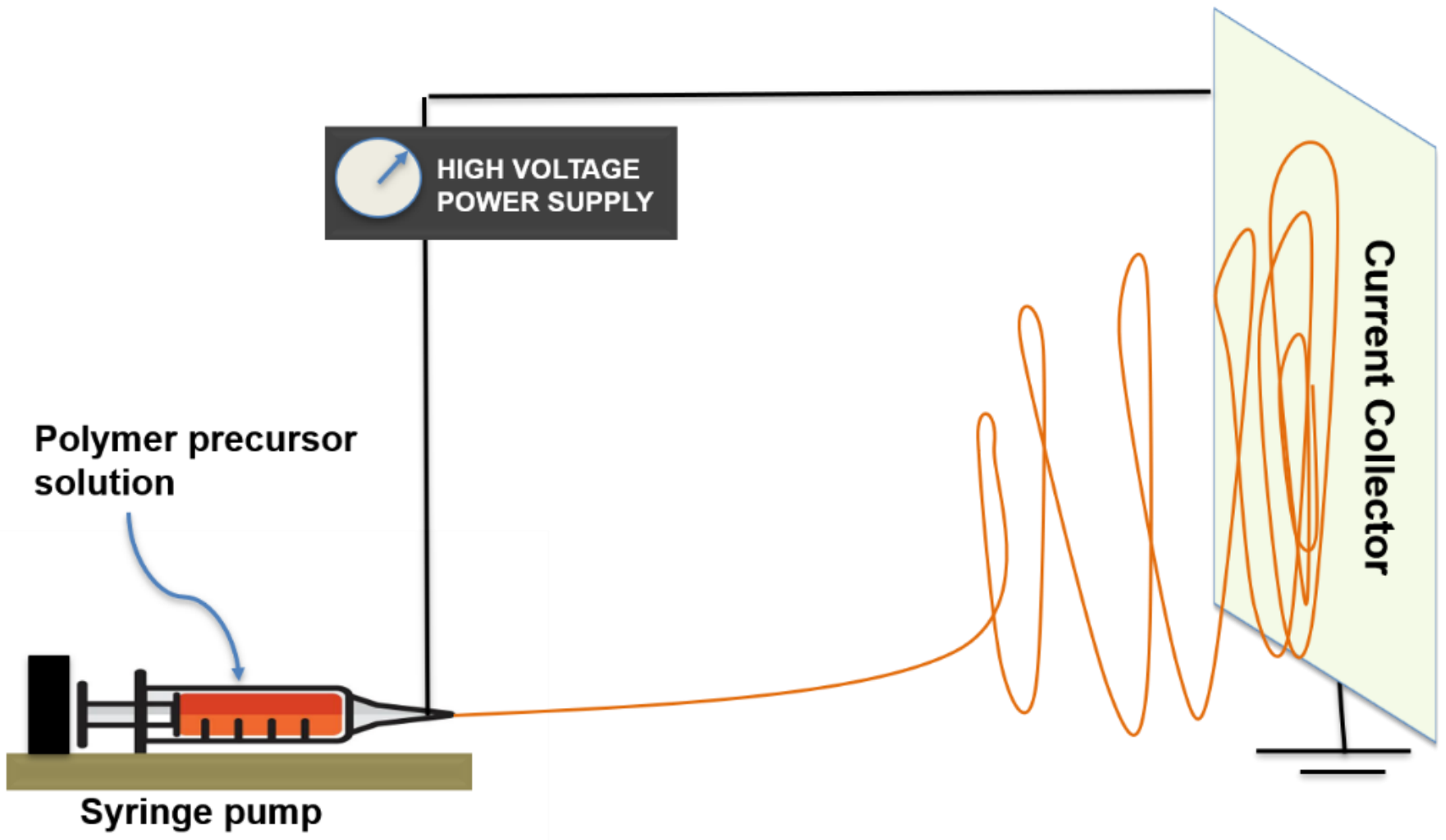

Fig. 8 Synthesis schematic of spinel ferrite nanomaterials via electrospinning method

Electrospinning has been investigated in the synthesis of one-dimensional nanomaterials for various multifunctional electrochemical devices. A few illustrations, such as, Yang et al. demonstrated that electrospun $CoFe_2O_4$ nanofibers synthesized through single-spinneret electrospinning exhibited distinct structural transitions from wire-in-tube morphology to belt-like and fiber-like architectures by the variations in the Co/Fe precursor ratio, exemplifying the strong influence of precursor composition on anisotropic growth behaviour and structural organization during electrospinning.[66] $CuFe_2O_4$ nanofiber fabricated via electrospinning, showed spider-like morphology influenced by optimised parameters. The electrospun ferrite has been investigated for gas sensing and exhibited strong detection of $NO_2$ and $H_2S$ gases.[67] Similarly, at varying electrospinning duration, diameter, grain size and porosity of the nanofiber and hence the selectivity for the gases were influenced in $NiFe_2O_4$ nanofibers.[68] Recently, Sakthinathan et al. fabricated $MnFe_2O_4$ microfibers through electrospinning using a PVP-assisted precursor, followed by calcination to obtain crystalline ferrite microfibers with interconnected porous morphology. The electrospun microfibers were subsequently integrated with reduced graphene oxide through a hydrothermal route to form $MnFe_2O_4$-mf@rGO composites for supercapacitor applications. The synergistic combination of conductive rGO sheets and porous one-dimensional ferrite microfibers facilitated enhanced charge transport and electrolyte accessibility, resulting in a high specific capacitance of 1797 F $g^{-1}$ and excellent cycling stability[69]. The ability of electrospinning to generate interconnected one-dimensional ferrite architectures with tunable porosity and continuous electron-transport pathways makes it a particularly promising strategy for next-generation flexible and high-performance electrochemical devices.

### 2.6 Microwave-Assisted Synthesis Method

Microwave (MW)-assisted synthesis has emerged as a rapid, energy-efficient, and environmentally favorable approach for the preparation of both organic and inorganic nanomaterials, including ferrite nanostructures. In this technique, electromagnetic radiation, typically operating at a frequency of 2.45 GHz, interacts with polar molecules and ionic species present in the reaction medium, resulting in direct volumetric heating through dipolar polarization and ionic conduction mechanisms. Unlike conventional heating processes that rely on conduction and convection, microwave irradiation enables rapid and uniform heating throughout the reaction medium, significantly reducing synthesis duration and energy consumption. The unique heating mechanism associated with microwave irradiation promotes accelerated nucleation, enhanced crystallization, and improved phase formation while often suppressing excessive particle growth and agglomeration. As a result, microwave-assisted synthesis is particularly effective for producing ferrite nanomaterials with controlled morphology, high crystallinity, narrow particle-size distribution, and enhanced surface properties. Furthermore, the technique offers advantages such as operational simplicity, shorter reaction time, reduced thermal gradients, and improved reaction efficiency, making it highly attractive for large-scale and sustainable nanomaterial synthesis.[70,71]

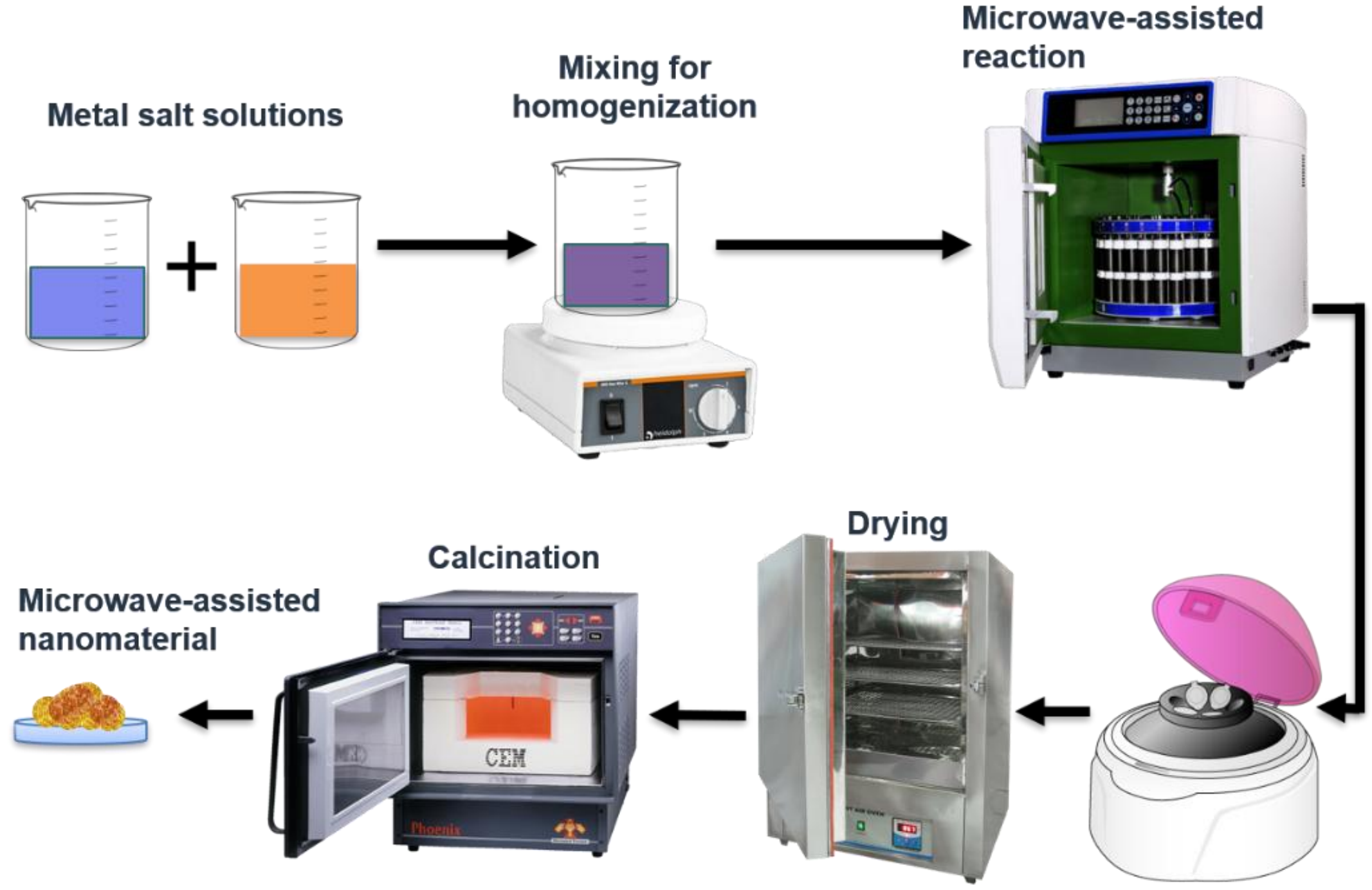

Fig. 9 Synthesis schematic of spinel ferrite nanomaterials via microwave-assisted method

Microwave irradiation has been successfully integrated with various synthesis routes including hydrothermal, solvothermal, sol-gel, and combustion methods for the fabrication of ferrite-based nanostructures tailored for sensing, catalytic, and electrochemical energy-storage applications.[72] Recent studies have demonstrated the effectiveness of microwave-assisted synthesis strategies for engineering ferrite nanostructures with enhanced electrochemical and functional properties. Microwave combustion synthesis of $MnFe_2O_4$ nanoparticles enabled rapid formation of spinel ferrite nanocrystals with reduced synthesis duration and improved electrochemical behavior for supercapacitor applications.[73] Similarly, zinc-doped $CuFe_2O_4$ nanostructures synthesized through microwave-assisted combustion exhibited improved supercapacitive performance, highlighting the role of microwave irradiation in promoting rapid crystallization and controlled nanostructure formation.[74] Microwave irradiation has also been integrated with co-precipitation approaches to achieve environmentally sustainable ferrite synthesis. For instance, microwave-assisted co-precipitation synthesis of $CoFe_2O_4$ and $MnFe_2O_4$ nanoferrites using biogenic coir extract demonstrated that rapid microwave heating can facilitate efficient nucleation and nanoscale morphology control while reducing processing time and energy consumption.[75]

These findings collectively indicate that microwave-assisted synthesis not only accelerates ferrite formation but also provides improved control over crystallinity, morphology, and electrochemical activity, making it a promising strategy for advanced sensing and energy-storage applications.

After the brief discussion on various synthesis routes, it is established that each synthesis technique presents advantages and limitations and has been summarized in Table 1. Therefore, careful selection and optimization of the synthesis method are essential to achieve the desired material morphology, which in turn can significantly influence the overall performance of the ferrospinel applications.

Table 2 Benefits and drawbacks of various methods of synthesis

| Method | Advantages | Disadvantages |
|---|---|---|
| Hydrothermal/Solvothermal | High crystallinity, ultra-fine material, environmentally friendly | Extreme experimental conditions |
| Co-precipitation | Homogeneity, high purity, scalable | Possible presence of impurities |
| Sol-gel | Ultra-fine particles, homogeneous particles, control of parameters such as size | High cost, toxic reagents |
| Template-assisted | Control of the nanomaterial growth, porosity | Precise experimental conditions regarding the template |
| Electrospinning | Particles with high porosity and large surface area, environmentally friendly | Number of parameters that must be controlled-complex proces |
| Microwave-assisted | Controlled morphology, high crystallinity, fast duration, reduced energy cost | Not all materials can be used, expensive, industry-level unscalable as of now |

## 3. Influence of Synthesis Approaches and Material Characteristics on Electrochemical Performance

### 3.1 Crystallinity and phase purity

A crystalline material refers to atoms arranged in an ordered pattern within the grain boundaries, while phase purity resembles the homogeneity of the intended chemical composition of spinel ferrites without any secondary impurity phases. Both the characteristics are heavily influenced by synthesis conditions. For instance, calcination temperature governs the crystal growth and phase formation of the ferrite. Increasing the temperature generally improves crystallinity of the spinel structure.[76] However, excessive heat treatment can induce agglomeration of particles directly affecting the surface area and thus, the electrochemical performance.[77]

Phase purity is also affected if any inhomogeneous solvent is not eliminated or misappropriate oxidisation which leads to secondary phases. pH stimulates nucleation growth and therefore influences phase purity.[76] For electrochemical applications, higher crystallinity improves electrical conductivity, whereas optimised nanocrystalization and defect-rich structures can offer abundant active sites for gas/electrolyte adsorption and electrochemical reactions.

### 3.2 Morphology and porosity

Morphology of spinel ferrites plays an important role in accessibility of the electrochemically active sites to the ions (from electrolyte in supercapacitors or gas/air in gas sensors). Various synthesizing routes and factors can produce distinct nanostructures, including, nanoparticles, nanorods, nanowires, nanosheets, hollow spheres, nanoflowers, etc.[78] As an example, Co-precipitation commonly yields nanoparticles, whereas hydrothermal and solvothermal route offers greater control over crystal growth and often produce complex morphologies such as nanoflowers, nanocones, and urchin-like structures.[21]

The dimensionality of the ferrite nanostructures also affects the electrochemical performance. The 0D particles have highest volume-to-surface area due to highest quantum confinement. 1D nanowires, nanorods provide elongated channels for electron transport, while 2D nanostructuring provides large, exposed surface area for the electrochemical reactions. The 3D hierarchical structures combine large surface area with interconnected diffusion pathways for fast charge transfers.[79]

To enhance electrochemical activity on surface sites, scientists have also explored porous architecture in electrode materials such as microporous, mesoporous, microporous as well as hierarchical meso-micro, micro-macro, macro-meso,etc. architectures. Porous structures allow faster penetration of electrolyte ions and gas molecules while also enhancing the exposure of redox-active sites hence, the size, shape and orientation of the pore play a part. Additionally, porous electrode design are great technique to not only mitigate volume expansions but also heat dispersions during the cycles thereby, lessening the structural degradation and improving the cycle life.[80]

### 3.3 Defects engineering with oxygen vacancies and cation distribution

With spinel ferrite structure of metal cations at tetrahedral (A) and octahedral (B) sites, oxygen vacancies and cation substitution become intrinsic features for electrochemical applications. The introduction of oxygen vacancies in perfectly crystalline spinel ferrites is due to the absence of oxygen anions from lattice on the surface of spinels possessing ample localized electrons as well as cationic reformation also induces defect in the spinel structure by rearrangement of metal cations at octahedral and tetrahedral sites. These defects can be generated by two approaches- (i) direct defect during the synthesis (ii) post-preparation by heat treatment.[81]

Manipulating with structure via substitution or doping of aliovalent metal cation or much rare earth elements with much larger ionic radii creates defects. In heteroatom doping, aliovalent M cation has higher affinity towards A-site (octahedral) while Fe cation can arrange in either of A or B sites. This creates disparity in charge neutrality and thus, creates oxygen vacancies. And even when oxygen vacancies aren't feasible, cationic defect also leads modulation in density of states, which results in tuned bandgap and electrical conductivity. The oxygen vacancies and lattice defects serve as active centres for adsorption and redox processes. These sites encourage better electron transport, improve carrier mobility, create additional pathways for ion diffusion and increase catalytic selectivity. Furthermore, another approach of elemental substitution or doping causes lattice distortion and vacancy formation, thereby tailoring the electronic structure of the ferrite. And so, the chemical and physical properties of spinel structures are strongly dependent not only on the nature of the constituent cations but also their distribution among the available lattice sites, termed as cation site occupancy or cation distribution.[82] Hence. during synthesis, precursor composition, and doping strongly affects cation occupancy and defect concentration in the spinel structure. Post-synthesis, heat treatment is the most common and straight-forward way to induce defects.[83]

Such modifications extensively improve gas adsorption behaviour in sensing applications and boost the pseudocapacitive contribution in supercapacitor electrodes. Henceforth, controlling defect chemistry has become an important strategy for enhancement of the electrochemical performance of spinel ferrites.[84]

### 3.4 Structure–Property–Performance Relationship

The electrochemical performance of spinel ferrites is fundamentally governed by the intimate relationship between synthesis conditions, structural characteristics, and charge-transport behaviour. Variations in synthesis parameters directly influence crystallinity, particle size, morphology, porosity, cation distribution, and defect concentration, all of which determine the density of electrochemically active sites and the efficiency of electron and ion transport. Therefore, understanding the structure-property-performance relationship is essential for rational design of high-performance ferrite-based electrochemical materials.

Highly crystalline spinel ferrites generally exhibit improved electrical conductivity owing to reduced grain-boundary resistance; however, excessive crystal growth often decreases the specific surface area and limits the availability of active sites. Conversely, nanostructured and defect-rich ferrites provide abundant adsorption sites and shorter diffusion pathways, leading to enhanced electrochemical activity despite slightly lower crystallinity. Similarly, oxygen vacancies and cation redistribution modify the electronic structure by introducing localized energy states near the Fermi level, thereby facilitating electron hopping and improving charge-transfer kinetics.

Morphology also plays a decisive role in electrochemical performance. Hierarchical porous architectures provide interconnected pathways for electrolyte penetration and gas diffusion while simultaneously accommodating structural strain during repeated electrochemical cycling. Consequently, rational control of synthesis conditions enables simultaneous optimization of crystallinity, defect chemistry, morphology, and electronic conductivity, ultimately leading to enhanced sensitivity in gas sensors and improved capacitance and cycling stability in supercapacitors.

Keeping these concepts in mind, below are few examples that emphasizes the effect of synthesising routes and parameters on spinel ferrite structure, their morphologies for various electrocatalytic applications -

Table 3 Spinel ferrite systems influenced by synthesis conditions

| System | Synthesis Method | Key Synthesis Parameters | Morphology | Material Characteristics | Applications | Ref |
|---|---|---|---|---|---|---|
| $Cu_xMn_{1-x}Fe_2O_4$ | Combustion | Calcination at 400-700 °C | Highly crystalline nanoparticles | Improved crystallinity | Methyl blue adsorption | [85] |
| $Mg_{0.5}Zn_{0.5}Fe_2O_4$ | Hydrothermal | Hydrothermal treatment at 80 °C; calcination at 600 °C | Hollow spherical structures | Surface area ≈ 49 $m^2\ g^{-1}$, pore size ≈ 12 nm | Methane gas sensing | [86] |
| $CoFe_2O_4$ | Hydrothermal | pH 7–14; temperature 120–220 °C | Spherical/agglomerated nanocrystallites | Particle size increased with temperature | Ethanol gas sensing | [87] |
| $CoFe_2O_4$ / $NiFe_2O_4$ | Thermal decomposition | Calcination at 600 °C | Flake-like (CFO) and flower-like (NFO) morphologies | Crystallite size ≈ 20 nm | Supercapacitor electrode | 88 |
| $MFe_2O_4$ (M = Co, Mn, Ni) | Hydrothermal | Metal-ion variation; pH 9 | Sphere-like, quasi-cubic, and octahedral nanoparticles | Different degrees of spinel inversion | Chloramphenicol sensing | 89 |
| $ZnFe_2O_4$ / $MgFe_2O_4$ / $CoFe_2O_4$ nanofibers | Electrospinning | Polymer-assisted spinning and annealing at 800 °C | Porous nanofibers (~104-125 nm diameter) | Mesoporous texture and interconnected 1D pathways | Li-ion battery anode | 90 |
| $NiFe_2O_4$ | Co-precipitation / Sol-gel / Microwave- | Variation in synthesis route | Agglomerated nanoparticles; MW-assisted synthesis | Crystallite size reduced from ~56 nm (CP) to ~26 nm (MW) | Supercapacitor electrode | 91 |
| | assisted combustion | | produced more uniform particles | | | |
| $MnFe_2O_4$ | Sol-gel | pH 12; calcination at 600 °C | Mesoporous interconnected nanoparticle clusters | Oxygen vacancies, surface area ≈ 36.6 $m^2\ g^{-1}$ | HER electrocatalyst | 92 |
| $CoFe_2O_4$ microflowers | Hydrothermal | Hydrothermal temp at 180 °C | Hierarchical flower-like microstructures | Crystallite size ≈ 14 nm, high surface area | Carbendazim sensing | 93 |
| $CoFe_2O_4$ thin film | Chemical bath deposition | annealing at 723 K | Porous interconnected nanoflakes | Low charge transfer resistance (~4.59 Ω) | Supercapacitor electrode | 94 |
| $CoFe_2O_4$ nanoparticles | Electrochemical synthesis | Temperature-controlled electrosynthesis | Temperature-dependent nanoparticle morphology | Higher temperature improved phase | Structure-property study | 95 |
| | | | | purity and stoichiometry | | |
| $NiFe_2O_4$ | Citrate-assisted sol-gel combustion | Citrate-assisted combustion and calcination | Sponge-like mesoporous structure | Surface area ≈ 73 $m^2\ g^{-1}$, pore size 3–30 nm | Glucose electrooxidation | 96 |
| $Co_{1-x}Mg_xFe_2O_4$ | PEG-assisted gel synthesis | Mg substitution and annealing at 700 °C | Nearly spherical nanoparticles | Cation redistribution and modified lattice parameter | Pseudocapacitive energy storage | 97 |
| $NiFe_2O_4$ | Hydrothermal | pH variation 7–13, hydrothermal at 160 °C, 10 h | pH 7: 10–25 nm nano-spheres<br>pH 12: nanorods (length ~1 µm, diameter 50–60 nm) | inverse/mixed spinel and phase/morphology controlled by pH | Structure-property study | 98 |
| | | | pH 13: nano-octahedra (side length ~150 nm or ~60 nm) | | | |

## 4. Application of spinel materials as electrode materials

Spinel ferrites ($MFe_2O_4$) have become intriguing multifunctional electrode materials for diverse electrochemical applications by virtue of their tunable electronic structure, multiple oxidation states, structural stability, and rich redox activity. The coexistence of transition-metal cations within tetrahedral and octahedral lattice sites facilitates efficient electron hopping and reversible Faradaic reactions, making these materials highly attractive for electrochemical sensing and energy-storage devices. Furthermore, morphology tunability, their compositional flexibility, and compatibility with conductive composites enable the development of high-performance ferrite-based electrodes with enhanced charge-transfer kinetics and electrochemical stability.[99] The sensing mechanism primarily involves adsorption of gas molecules onto the ferrite surface, followed by charge-transfer interactions with surface oxygen species that induce measurable changes in electrical resistance or current response. The presence of multiple transition-metal ions within the spinel lattice provides abundant electroactive sites that facilitate rapid electron-transfer kinetics and enhanced interaction with analyte molecules. Consequently, ferrite-based sensors often exhibit improved sensitivity, response-recovery characteristics, and operational stability toward toxic and volatile gases.[100,101]

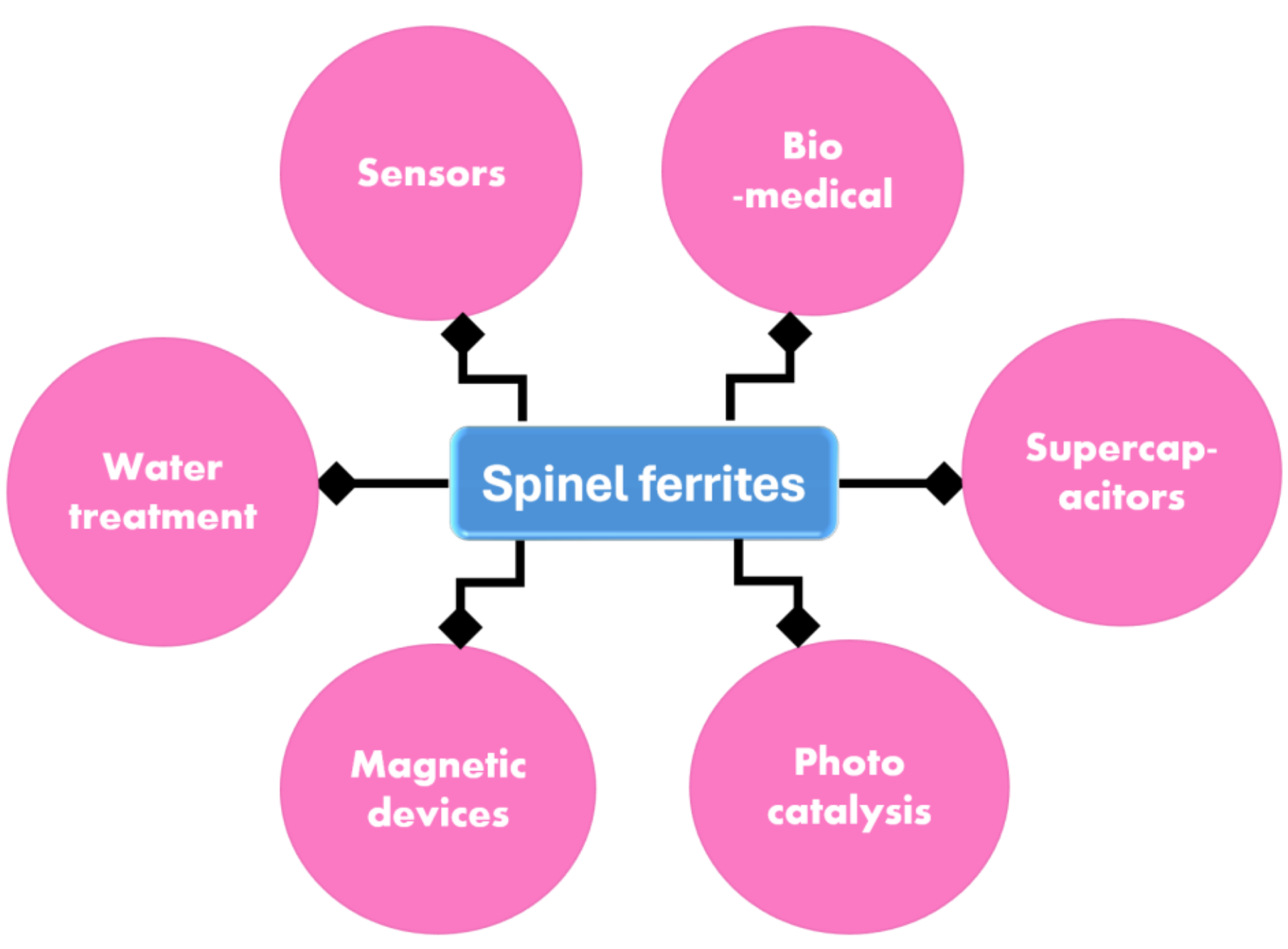

Fig. 10 Application of spinel ferrites in various fields

Owing to their rich redox chemistry and multiple accessible oxidation states, spinel ferrites have attracted significant attention as pseudocapacitive electrode materials for supercapacitor applications. The electrochemical charge-storage mechanism is primarily due to reversible Faradaic redox transitions including transition-metal ions such as $Fe^{2+}/Fe^{3+}$, $Co^{2+}/Co^{3+}$, and $Mn^{2+}/Mn^{3+}$ at the electrode-electrolyte interface. These rapid surface and near-surface redox reactions contribute to enhanced specific capacitance and energy density compared to conventional electric double-layer capacitors. Furthermore, the interconnected spinel architecture promotes efficient electron transport and ion diffusion, resulting in improved charge-discharge kinetics and cycling stability.[102] Beyond supercapacitors, spinel ferrites have also been investigated as electrode materials for rechargeable battery systems because of their structural stability and favourable ion-transport characteristics. The robust spinel framework can withstand repeated ion intercalation and deintercalation processes while maintaining structural integrity, which contributes to long-term cycling stability. Moreover, the three-dimensional interconnected pathways present within spinel structures allows the diffusion of charge carriers such as $Li^{+}$, $Na^{+}$, and $Zn^{2+}$ ions, making ferrite-based electrodes promising for high-rate energy-storage applications.[103,104] While, spinel ferrites are popular amongst various applications, in this review, our focus remains on ferrites-based electrode materials for gas sensor and supercapacitor applications.

### 4.1 Application of spinels in Gas sensors

Gas sensors have wide application in numerous areas for monitoring flammable, hazardous and toxic gases. For instance, Hydrogen is used in propellants and fuels for aerospace and electric vehicle applications and needs supervision for its concentration in the atmosphere as hydrogen gas is explosive when high concentration is exposed in air.[105] Thus, a gas sensor detects gases in the surrounding environment as a preventative measure against the effects of harmful gases such as methane, carbon monoxide, or volatile organic compounds (VOCs) such as ethanol and methanol in health care and medicine. This leads the scientists to investigate high-performance gas sensors.

The gas-sensing behaviour of metal oxides is governed by complex interfacial interactions occurring at the solid-gas interface, although a fully unified sensing mechanism has not yet been established due to the diversity of ferrite compositions and operating conditions. Nevertheless, the widely accepted sensing mechanism primarily involves surface adsorption and charge-transfer processes occurring upon exposure to ambient air. Oxygen molecules are initially adsorbed onto the metal oxide electrode surface, where they extract electrons from the conduction band to form chemisorbed oxygen species such as $O_2^{-}$, $O^{-}$, and $O^{2-}$, depending on the operating temperature, creating depletion layer in n-type metal oxide and accumulation layer for p-type metal oxide. The following equations govern the chemisorption of oxygen molecules on the surface of the metal oxide material[106] –

$$O_{2(gas)} \leftrightarrow O_{2(adsorbed)}$$

At T<150°C:

$$O_{2(adsorbed)} + e^- \leftrightarrow O^-_{2(adsorbed)}$$

At 150°C<T<400°C:

$$O_{2(adsorbed)} + e^- \leftrightarrow O^{2-}_{(adsorbed)}$$

At T>400°C:

$$O_{(adsorbed)} + e^- \leftrightarrow O^{2-}_{(adsorbed)}$$

This space-charge layer and hence the grain boundaries of the electrode material have certain resistance which changes when the gas to detected is exposed to the surface. The oxygen ions react with the gas influencing the charge carriers on the surface, changing the depletion/accumulation layer and hence, resistance on the grain boundaries. This modulation in local resistivity of the material due to shrinking or widening of space-charge layer when connected to a transducer provides a signal for the detection of the gas molecules in question.[107]

In n-type semiconducting metal oxides, at an operating temperature, electrons are abundant charge carriers. The oxygen adsorption usually removes electrons from the surface. This electron removal leads to the formation of an electron depletion layer near the surface which results in an increase in the electrical resistance as the fewer electrons are available. When exposed to reducing gases, the adsorbed oxygen species react with target as molecules and release the trapped electrons back into the conduction band. The following equations formulate the mechanism by[106]

$$\mathrm{Oxide}^{-}_{(\mathrm{adsorbed})} + \mathrm{Gas} \leftrightarrow \mathrm{Gas\ Oxide} + \mathrm{e}^{-}$$

Consequently, the depletion layer thickness decreases, decreasing the resistance and leading to a measurable reduction in sensor resistance.

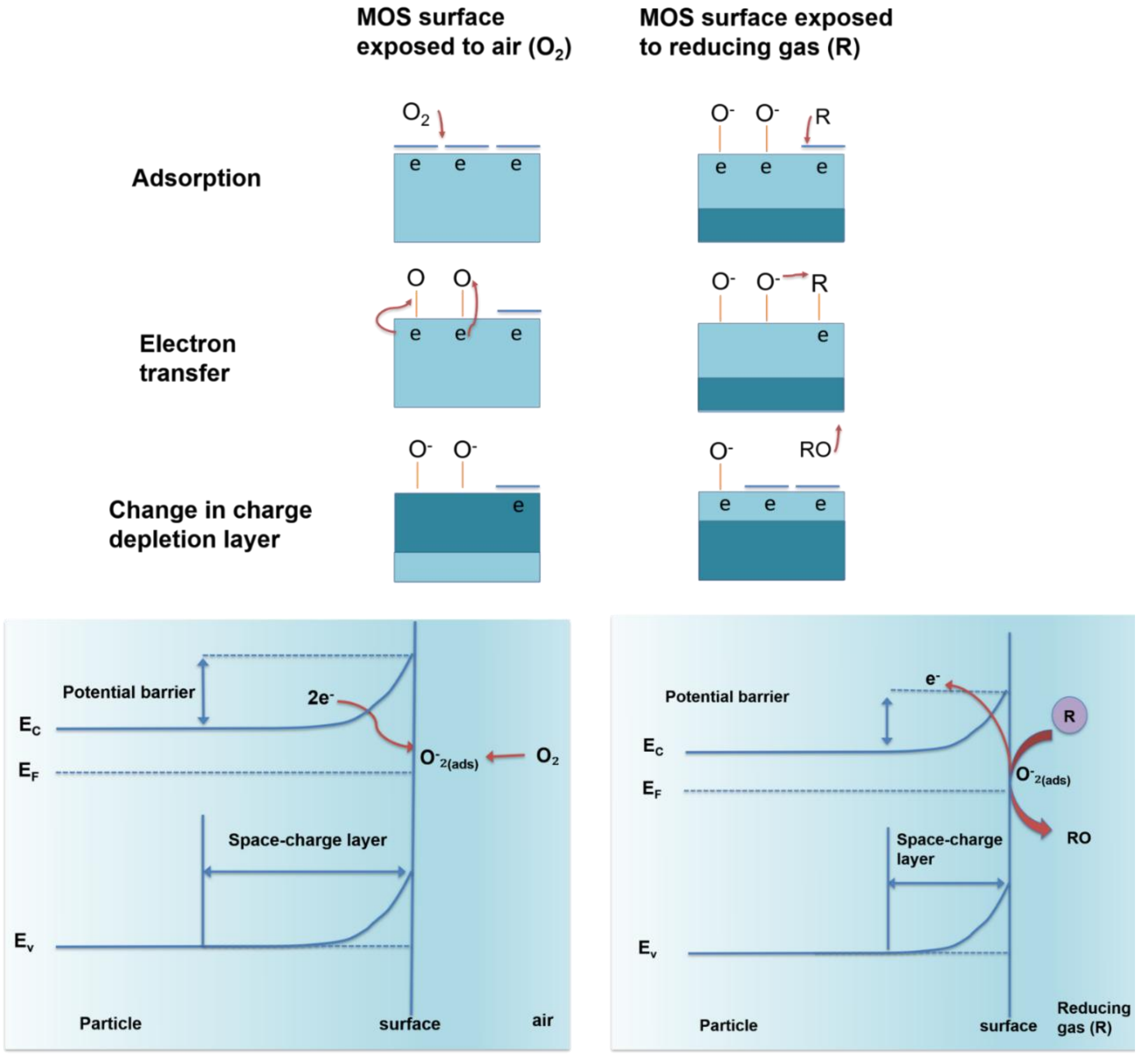

Fig. 11 Schematic diagram of the sensing mechanism of n-type semiconducting metal oxide nanostructures for reducing gas and affect on energy bandgap by adsorption of oxygen in air and interaction with reducing gas

Whereas, in p-type semiconducting metal oxides, holes are the dominating charge carriers. So, removal of electrons with adsorption of oxygen leads to formation of accumulation layer leaving behind abundant holes. This lowers the resistance by accumulating a wider channel for holes and when exposed to target gas molecules, the accumulation layer shrinks as the oxygen bonds with the target gas molecules. The recombination of electrons back from oxygen and holes at the surface, leaving the layer with fewer holes leads to increase in resistance. This resistance variation induced by gas adsorption and surface redox reactions constitutes the fundamental operating principle of metal oxide-based gas sensors. The sensitivity of the MOS-based gas sensors is generally defined as $R_a/R_g$ (reducing gas), $R_g/R_a$ (oxidizing gas), where, $R_a$ is the resistance of the sensor exposed to air or the background gas, $R_g$ is the resistance of the sensor exposed to the target gas, ΔR is the change in resistance of the sensors after exposure to the target gas).[108]

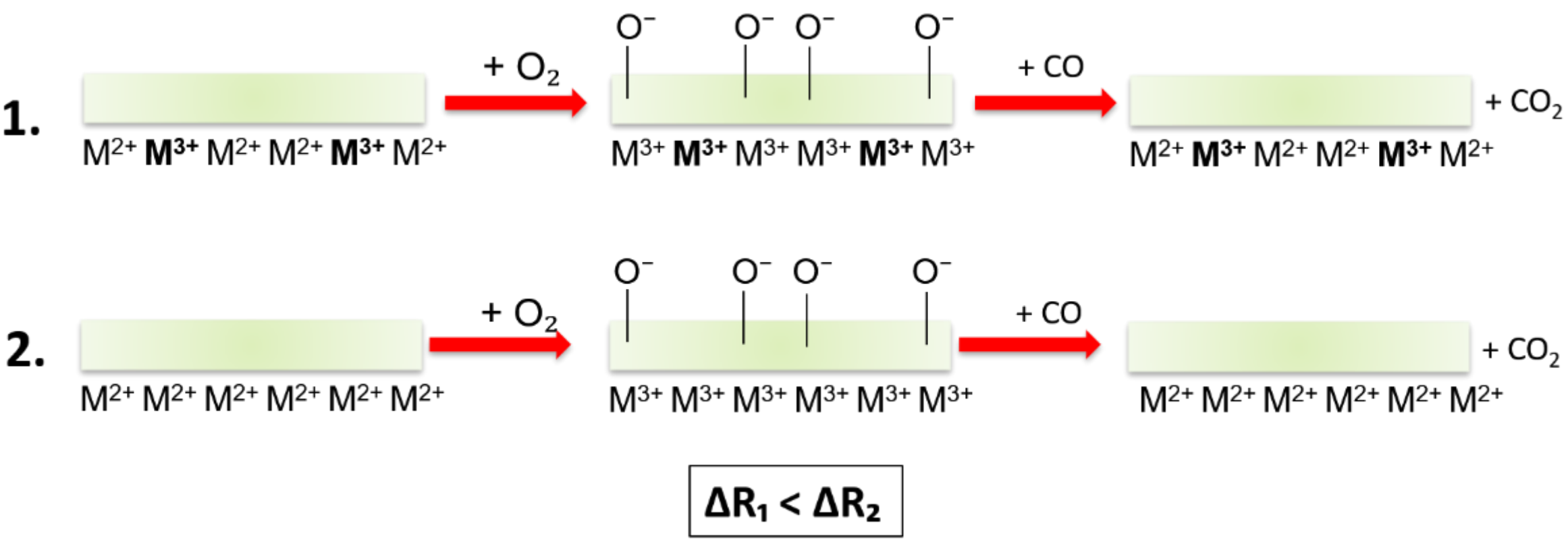

Fig. 12 divalent metal cations at the surface interacting with oxygen molecules exhibits larger change in resistance than trivalent valent oxidation states of metal cations $MFe_2O_4$ spinels are particularly attractive for gas sensing due to their distinctive structural features, various stable oxidation states and oxygen vacancies at tetrahedral and octahedral sites. The divalent cations of the transition metals change to trivalent cations of the same element when an adsorbing oxygen molecule removes an electron. Reversibly, when the testing gas is exposed to the surface, the chemisorbed oxygen bonds with the gas molecules, transferring back the electron to the trivalent cations to change to divalent cations of the same species given by $\mathrm{Me^{3+} + e^- \leftrightarrow Me^{2+}}$ and $\mathrm{Fe^{3+} + e^- \leftrightarrow Fe^{2+}}$. This electron hopping of the transition metals result in modulation of electrical conductivity and thus, resistance. The gas sensing requires for excess of $Me^{2+}$ or $Fe^{2+}$ on the surface to chemisorb more oxygen molecules and therefore, more change in electrical resistance which enhances the sensitivity of the sensor as shown in figure with CO gas detecting Ni-ferrite as an illustration. The catalytic surface exposure of structure due to their valence states and oxygen defect, the thermal stability and the cost effectiveness of the spinel ferrites motivates researchers to explore the material for various gas sensing applications.[107]

For instance, cation engineered $MFe_2O_4$ with various M=Fe, Co, Ni, Zn, each cation modulating the morphology, microstructural properties and gas sensing kinetics for detecting acetone gas in diabetic patients' breath as a low-cost, rapid and effective material.[109]

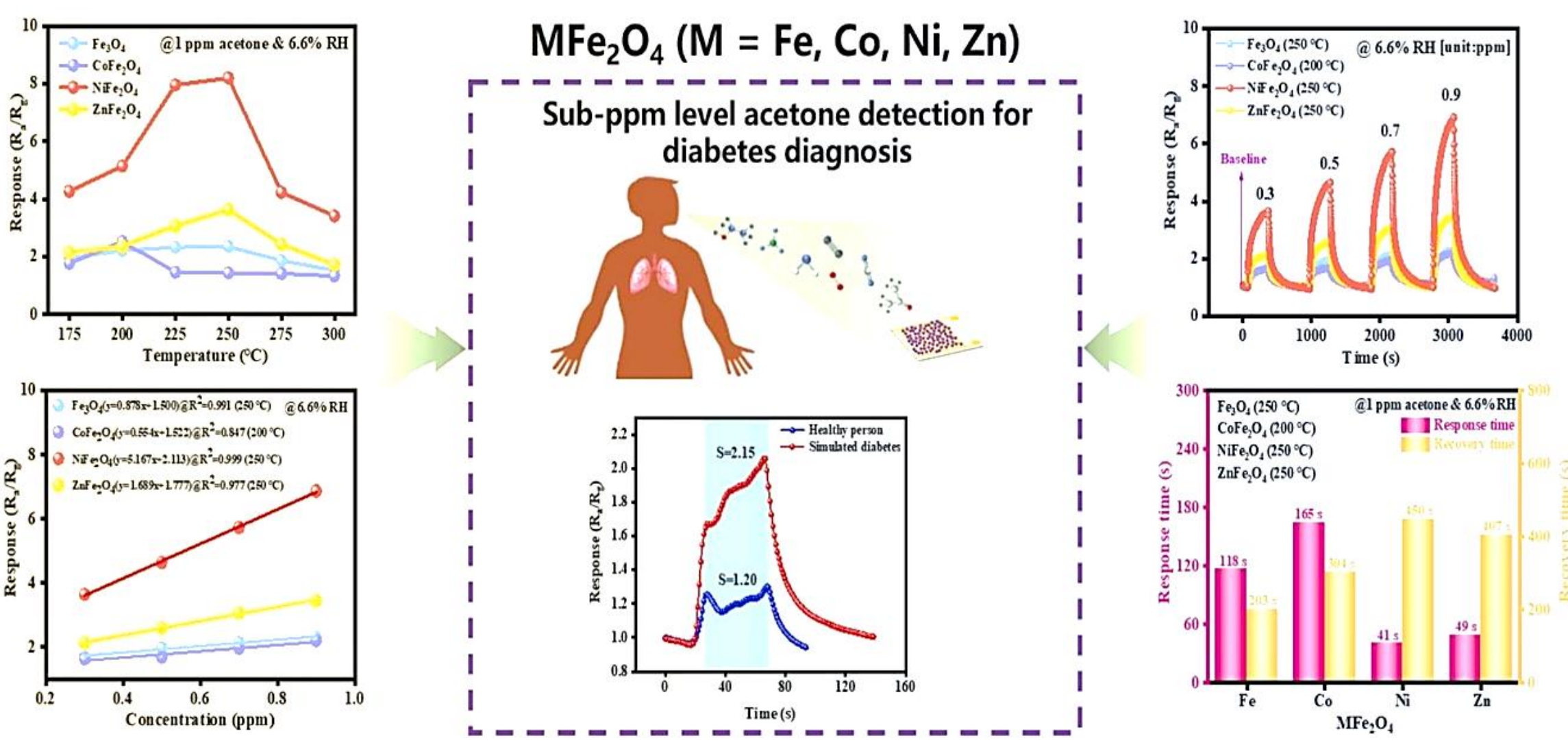

Fig. 13 Acetone sensing characteristics of $MFe_2O_4$ (M = Fe, Co, Ni, and Zn) spinel ferrites for breath-based diabetes diagnosis[109]

Thus, spinel ferrites provide a versatile sensing platform where cation distribution, morphology, and defect chemistry collectively govern sensitivity, selectivity, and operating stability.

### 4.2 Application of spinel ferrites in Supercapacitors

Supercapacitor plays a vital role in energy storage and utility applications. It bridges the gap between high-power density capacitors and high-energy density batteries for optimum energy storage as well as longevity for efficient energy use. Unlike batteries, which store energy via slow chemical reactions, supercapacitors store and release energy by building electrical charge on their surface. This allows them to charge and discharge incredibly quickly, making them suitable for the applications that require short bursts of power, such as regenerative braking systems in hybrid electric vehicles.

Supercapacitors adhere capacitance based on two mechanisms- (a) Electrical double layer capacitance (EDLC) and (b) pseudocapacitance (PC) which results in high-performance SCs. Electrical Double Layer Capacitance store energy by forming an electric double layer at the electrode-electrolyte interface. EDLC is simply produced when voltage is applied at the electrodes, these electrodes polarise to pertain as anode and cathode and to counter-compensate, ions from electrolyte accumulate on the electrode surface. This small-ranged distance between electrode charges and anion/cation of electrolytic layer creates an electric field which stores energy (potential energy) as $E = \frac{1}{2}CV^2$. Therefore, this leads to capacitance due to the electric double layer effect as $C = \frac{\varepsilon A}{d}$. EDLC shows rapid charge-discharge rate since it is purely electrostatic, and no faradaic reactions are involved. This limits SCs to enhance their energy density. Thus, another mechanism, termed as Pseudocapacitance, was coined in 1960s for other electrochemical applications but made its place in supercapacitors as well. Pseudocapacitance involves fast and reversible faradaic reactions, specifically redox reactions of the electrode material, which enhances energy density of supercapacitors. The term 'pseudo' came into picture as it mimics the linear relationship between charge and voltage found in ideal capacitors, despite involving chemical bonds and charge transfer in its mechanism. The pseudocapacitive property of the material revolves around its valence electrons and oxidation-reduction states. It primarily involves three mechanisms- (a) underpotential deposition (b) redox pseudocapacitance (c) Intercalation. The underpotential deposition involves sub-monolayer deposition of a foreign metal on electrode surface at potentially above bulk deposition potential where the deposition-desorption. The redox pseudocapacitance, reversible redox reactions take place at the surface-active sites of electrode coupling valence electrons with the help of electrolytic cations. Finally, the insertion of electrolytic ions inside the electrode layer results in intercalation pseudocapacitance. This mimics the which in-reality are just the overlapping of many redox reactions resulting in a smooth capacitor-like Q-V curve. These, together with EDLC, enhance the capacitance of electrode of the supercapacitor.[110,111] The electrode fabricated of Transitional Metal Oxides (TMOs) generally go through redox reactions to produce pseudocapacitance. It is important to note that pseudocapacitance does not exclude capacitance due to EDL. As the voltage is supplied, EDL is formed and extra pseudo-capacitance is offered by redox reaction of electrode spinels involving the electrolyte ions which can be given by the following equation[112]

$$M^{n+} + OH^{-} \leftrightarrow M^{(n+1)+}OH + e^{-}$$

where, M is metal (Mn, Ni, Fe, Co, etc.) and OH is generalised ion from the electrolyte participating in this reversible redox reaction.

In the interest of ferrospinel as supercapacitor electrode material, the metal cations in spinels transition to their possible stable oxidation states ($Fe^{2+}$/$Fe^{3+}$, $Mn^{4+}$/$Mn^{3+}$/$Mn^{2+}$, etc.) results in pseudocapacitive behaviour of the spinel ferrites. The fast electron exchange i.e. redox reactions at the electrode-electrolyte surface stores energy in the supercapacitor via pseudocapacitance. The $MFe_2O_4$ consisting of positive metal ions- Fe and M and negative oxygen ions forms an interconnected structure, providing more active sites for ions from the electrolyte to diffuse onto the electrode surface.[113]

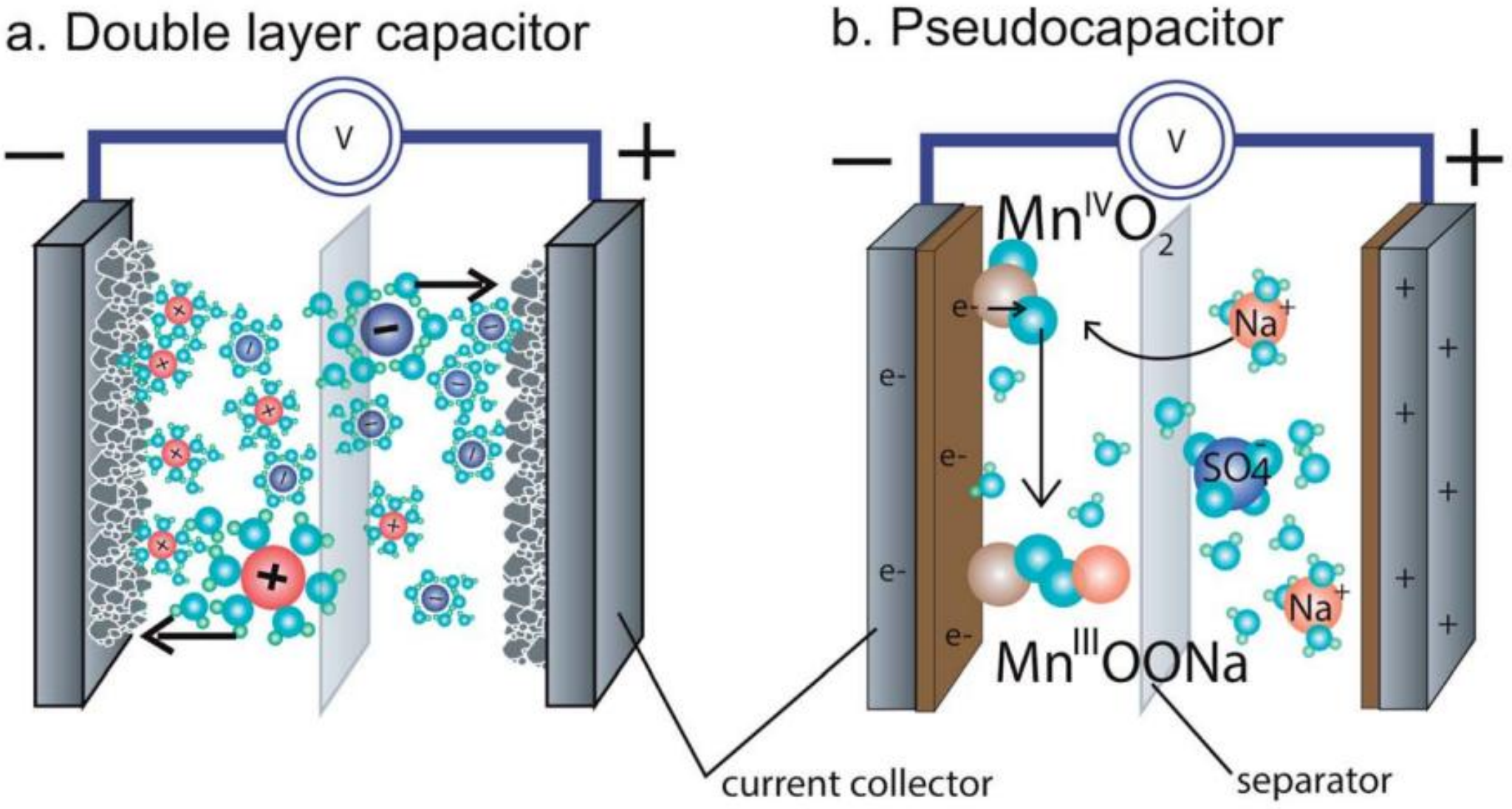

Fig. 14 Fundamental mechanisms of supercapacitors (a) Double layer capacitance via electrostatic charge storage at electrode-electrolyte interface (b) Pseudocapacitance via rapid redox reactions at electrode-electrolyte interface[113]

The electrochemical properties of the electrode material have a significant role in determining supercapacitor performance, including specific capacitance, rate capability, and cycling stability. In contrast, sensitivity and selectivity for the target species are essential considerations in the evaluation of electrochemical sensors. The physiochemical properties of the material, including as electrical conductivity, redox reactions, and surface chemistry, are critical in influencing reaction kinetics and performance in both scenarios. When treating these parameters fundamentally, it is predicted that the electrode surface plays an important influence in analyte adsorption in the sensor and ion diffusion pathways in the supercapacitor. This is heavily influenced by the dimensions, morphology, and geometric structure of the ferrites. The high porosity materials, especially, offer increased surface area and numerous active sites, improving gas adsorption and catalysis in sensor technologies and efficient charging/discharging processes and fast ion transportation in electrochemical energy storage devices.

Henceforth, conscious control over the composition, shape, and porosity of spinel ferrites used in energy storage and sensing applications is necessary.[100,114–117]

Therefore, the next section discusses how to utilise and enhance the electrochemical properties for gas sensors and supercapacitors among various multi-applications of spinel ferrites.

## 5. Techniques to enhance spinal ferrite electrochemical properties

Although the potentials of ferrospinels are being explored in variety of functions like water splitting, hydrogen evolution reduction, catalysis and photo-catalysis, along with the above-mentioned electrode-based applications, few challenges such as volume expansion, sluggish reaction kinetics, and limited electrical conductivity may affect the long-term performance of spinel electrodes.[118] To address these issues, researchers have developed various techniques such as nanoscale architectures such as nanowires, nanosheets, hollow spheres, and porous structures of spinel ferrites.[104] Also, ferrite nanostructures integrated with conductive materials such as graphene derivatives and carbon nanostructures exhibit synergistically enhanced electrochemical performance.[119] Additionally, cation redistribution due to transition metal and rare earth doping modifies the electrical and chemical properties of spinel structure of ferrites in order to enhance the electrochemical performance for the applications of gas sensors and supercapacitors.[120,121]

### 5.1 Nanostructured Spinel Ferrite Materials

Nanostructured materials have a central role in sensing and energy storage technologies due to their high surface-to-volume ratio. As the size of the spinel atoms decrease, the surface reactivity significantly drives the electrochemically active sites and interfacial charge kinetics at the electrode. In comparison to their bulk form, nanostructured ferrites exhibit tuned electronic structure, enhanced surface reactivity, shorter ion-diffusion pathways and improved adsorption capability, all of which contribute to high-performing electrochemical applications.[122,123]

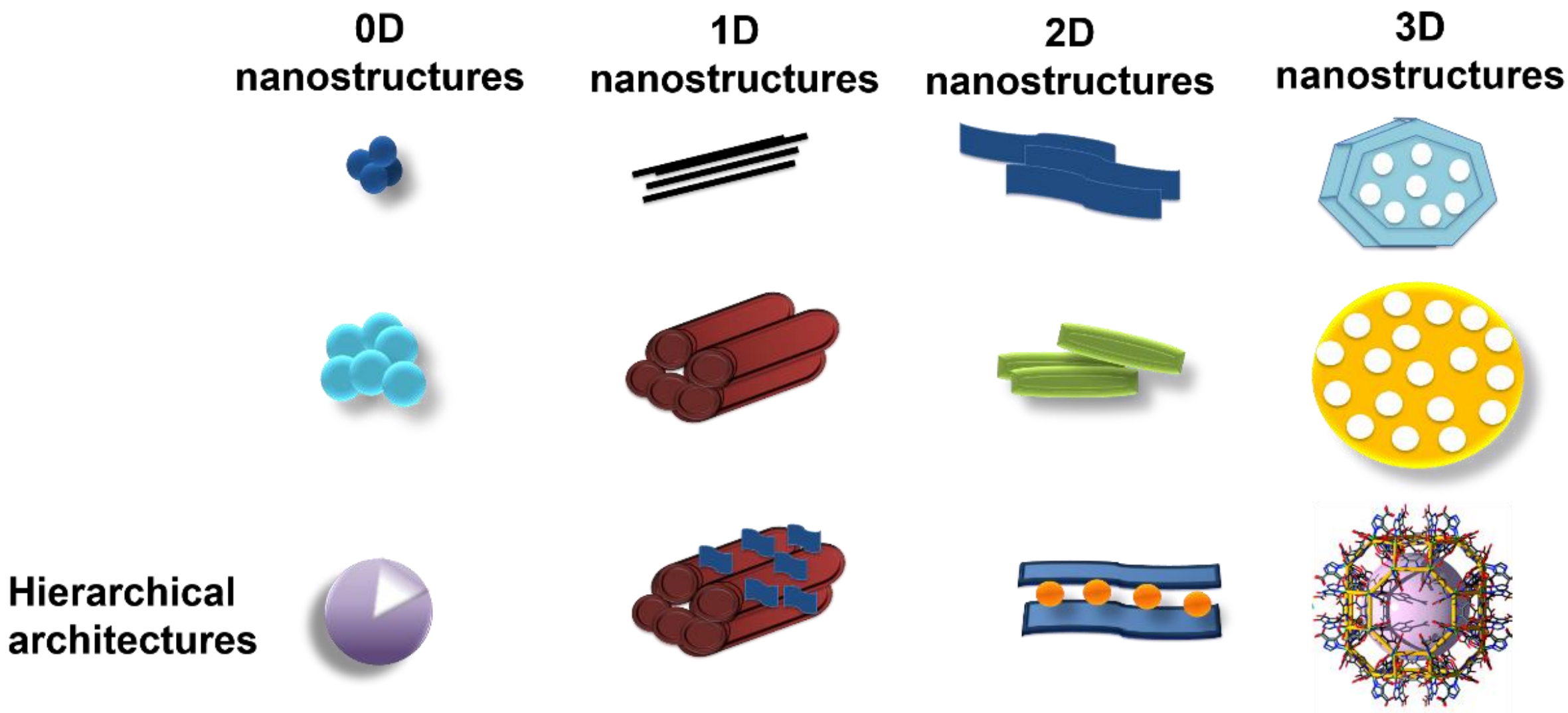

Fig. 15 Various nanostructures and hierarchical architectures of nanomaterials

Due to reduced size, the grain boundaries, the cation distribution often display tuned electronic and optical properties. In nanocrystalline $CoFe_2O_4$, diminished grain sizes increase electrical conductivity owing to expansion of grain boundary density and structural distortions due to reduced crystallite size.[124] Similarly, the bulk $ZnFe_2O_4$ maintains a normal spinel structure at room temperature, however, at nanoscale, it displays cation inversion, with $Zn^{2+}$ and $Fe^{3+}$ ions exchanging tetrahedral and octahedral sites. This cation distribution allows the material for tunable energy bandgap and thus, better electrical conductivity. This enables size-dependent tuning of electrochemical characteristics.[9,125]

Many advanced applications are made possible by such nanoscale alterations. Through improved surface reactivity due to elevated specific surface areas (SSAs) increase gas sensing sensitivity of the target gas due to rapid charge transfer kinetics at electrode-air/gas surface.[126,127]

In gas sensors, more oxygen molecules adsorbed due to the enhanced sensitivity elevates the overall performance as various alterations take place at nano-dimension. For instance, $ZnFe_2O_4$ nanospheres at controlled tuning of its particle size, boosted high surface area from 9 to 19 $m^2/g$. The nano-scaling delivered tuning of energy bandgap due to the quantum confinement of the nanoparticles. One more advantage could be observed higher concentration of $Fe^{2+}$ at the surface instead of $Fe^{3+}$ which indicates better electrochemical performance. As the bulk transformed to nanometre, it exhibited higher pore volume and created oxygen vacancies enhancing the gas sensing of $CO_2$.[128]

As stated in various synthesis techniques described in previous sections such as co-precipitation, hydrothermal, and sol-gel method, spinel ferrite can be synthesized by different morphological structures. This is due to their specific characteristics of varying particle size, shape, and distribution towards achieving the nanometric structure that governs efficiency of the spinel ferrites. As an example, microwave-assisted $MnFe_2O_4$ NPs exhibited a high response of 1.70 to 5 ppm of n-butanol at the operating temperature 120 °C, indicating that reducing the ferrite size reflects positively on high sensitivity and offer acceptable response and recovery time.[129] $CoFe_2O_4$ NPs exhibited that 6nm average particle size yields better $CO_2$ detection at room-temperature than 60 nm. The resistance response is shown from low concentration as 250 ppm covering wide range to 4000 ppm in 30% of room humidity. The synthesized gas sensor also has a response time of ~11.25 min and recovery time of about 12.8 min demonstrating a high-performance spinel ferrite-based gas sensor.[130]

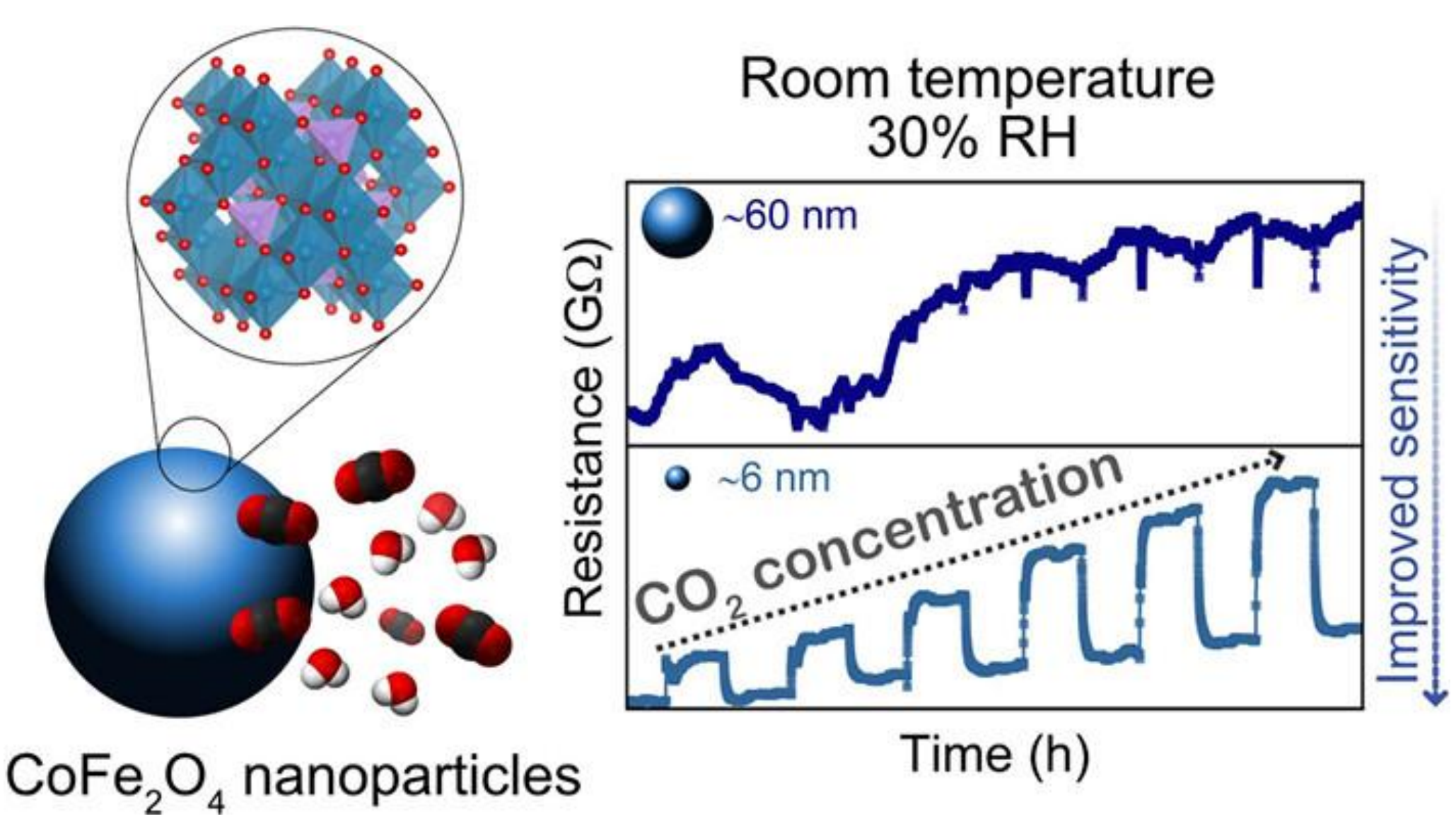

Fig. 16 $CoFe_2O_4$ Nanoparticles in gas sensors detect $CO_2$ gas enhances with decreasing particle size from 60 nm to 6 nm[130]

In energy storage devices, nanosized spinel ferrites exhibit superior electrochemical performance including multiple redox states, stability, and facile ion transport which surpasses the bulk materials in supercapacitors through sublattice-mediated pseudocapacitance.[131] Hydrothermally synthesized $NiFe_2O_4$ delivered 2-dimensional nanosheets of crystallite size 10.25 nm, along with mesopores and micropores favour multiple channels for electrons from metal cations and electrolyte ion diffusion pathways leading to fast redox reactions. Nanostructuring thus expands their utility as the 2D porous of $NiFe_2O_4$ plays huge role in enhancing the electrochemical properties of the supercapacitor.[132] Shahid et al. synthesized $NiFe_2O_4$ via different routes leading to distinct morphologies from each. The work inferred that sol-gel formed $NiFe_2O_4$ showed highest surface area and interconnected pathways for ion diffusion.[133]

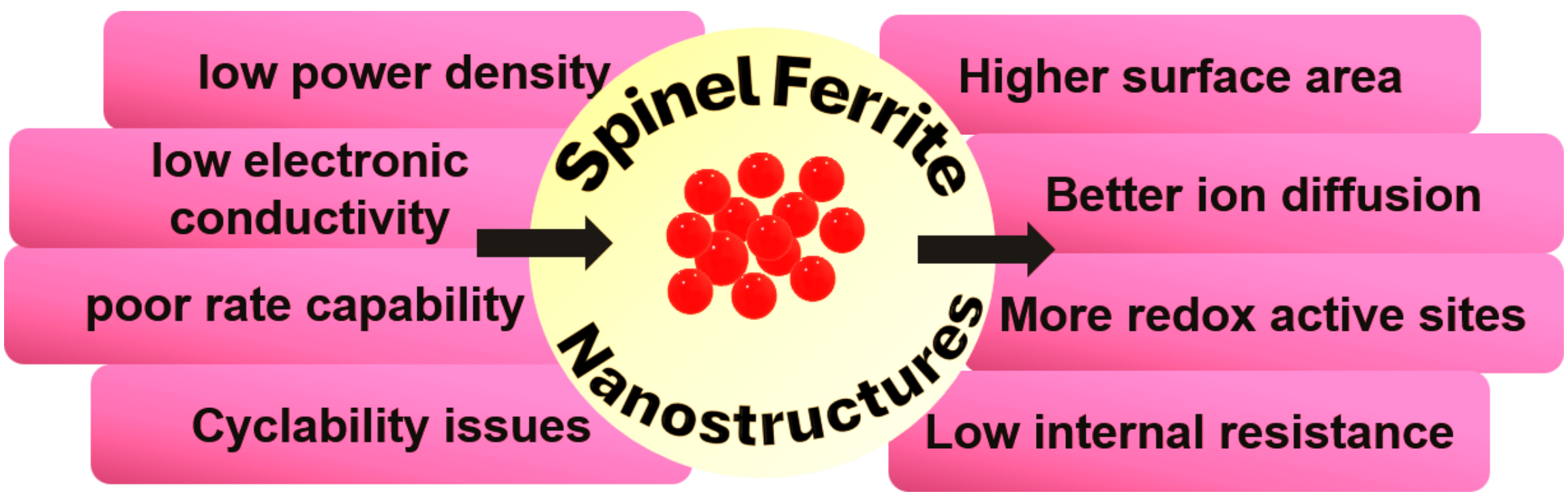

Fig. 17 Limitations of bulk spinel ferrites modified by nanostructuring the spinel structure particles

Optimizing synthesis techniques towards tailoring these structural properties towards achieving enhanced performance is an important step as extremely nano-reduction of particles can lead to extensive agglomeration due to magnetic interactions between the metal cations when grouped too close. Similarly, optimising porosity of the electrode material is crucial otherwise, pore size lesser than the size of ions interacting creates resistance. Hence, to deliver effective gas sensing and energy storage applications, optimised tuning of the nanostructures is a fundamentally vital step. Nanostructured spinel ferrite technology has become one of the extensively studied techniques to improve electrochemical performance in various modern technologies.[134,135]

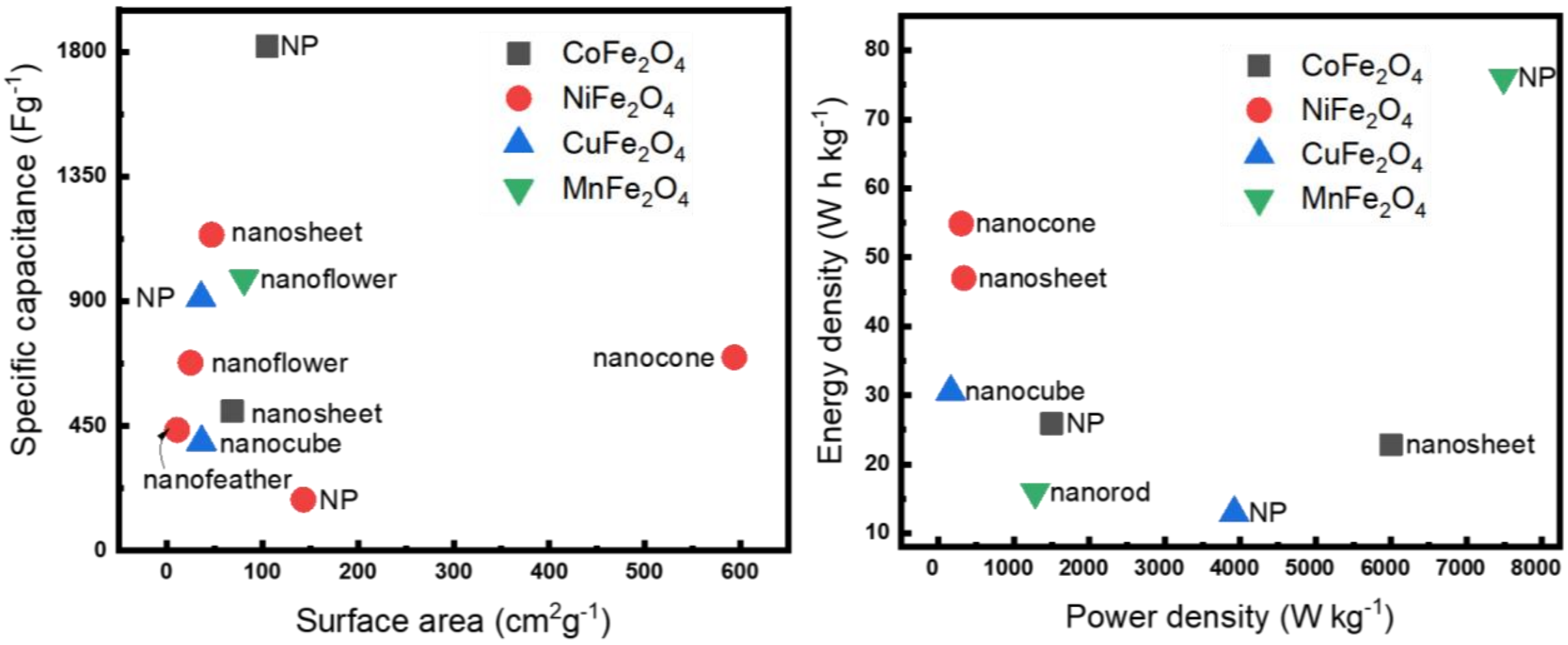

Fig. 18 (a) Power density vs. Energy density for various morphologies of spinel ferrites (b) Surface area vs. specific capacitance plot for various morphologies of spinel ferrites[53,136–144]

## 5.2 Graphene as Carbonaceous Conductive Hybridization

Graphene and its derivatives have emerged as highly effective conductive platforms for enhancing the electrochemical performance of spinel ferrites in both gas-sensing and energy-storage applications. Owing to its two-dimensional honeycomb lattice, exceptional carrier mobility, high theoretical surface area, excellent mechanical stability, and superior electrical conductivity, graphene facilitates rapid electron transport and efficient interfacial charge transfer within ferrite-based electrochemical systems.[145,146]

Among graphene derivatives, graphene oxide (GO) and reduced graphene oxide (rGO) are particularly attractive due to their scalable synthesis, high dispersibility, and compatibility with metal oxides like ferrites. The abundant oxygen-containing functional groups present in GO improve hydrophilicity and promote strong interfacial interaction with ferrite nanoparticles, while reduction of GO partially restores the π-conjugated structure and improves electrical conductivity in rGO. Consequently, graphene derivatives provide conductive frameworks capable of suppressing ferrite agglomeration, improving electrolyte accessibility, and enhancing electrochemical stability.[146,147]

Table 4 Comparison table of theoretical surface area and conductivity of Graphene, GO and rGO

| **Property** | **Graphene** | **Graphene Oxide (GO)** | **Reduced Graphene Oxide (rGO)** |
|---|---|---|---|
| Theoretical Surface Area ($m^2/g$) | 2630 | ~2400 (practical <500) | ~2600 (practical 400-500) |
| Conductivity (S/m) | $10^6$–$10^8$ | $10^{-3}$–$10^1$ | $10^2$–$10^4$ |

In gas-sensing applications, graphene incorporation improves sensing performance by enhancing the charge-transfer kinetics by improving electron flow. This fast charge transfer facilitates the adsorption of oxygen ions as well as the target gas molecules at electrode-gas surface. The adsorption of gas molecules onto graphene-based conductive networks induces measurable modulation in carrier concentration and electrical resistance, thereby improving sensor sensitivity and response characteristics. Furthermore, the large surface area and defect-rich interfaces generated in ferrite/graphene heterostructures provide abundant adsorption-active sites and promote rapid diffusion of analyte gases.[148]

Shoeb et al. explored that when arc furnace dust treated in nitric acid and citric acid via combustion method, it was indicative of Zn ferrite. His team fabricated ferrite@rGO electrode which was able to detect heavy metal ions like $Pb^{2+}$, $Cd^{2+}$, and $Hg^{2+}$ 100-folds more against 11 common metals, due to high binding affinity of graphene to soft Lewis acids. The sensor met World Health Organization for standard limit detection range (1.01 – 1.13 ppb) for drinking water. It yielded linear response ($R^2 > 0.995$) throughout 10-150 ppb referring to fast transfer kinetics of ferrite electrons and demonstrated a sustainable and effective usage of industry waste.[149]

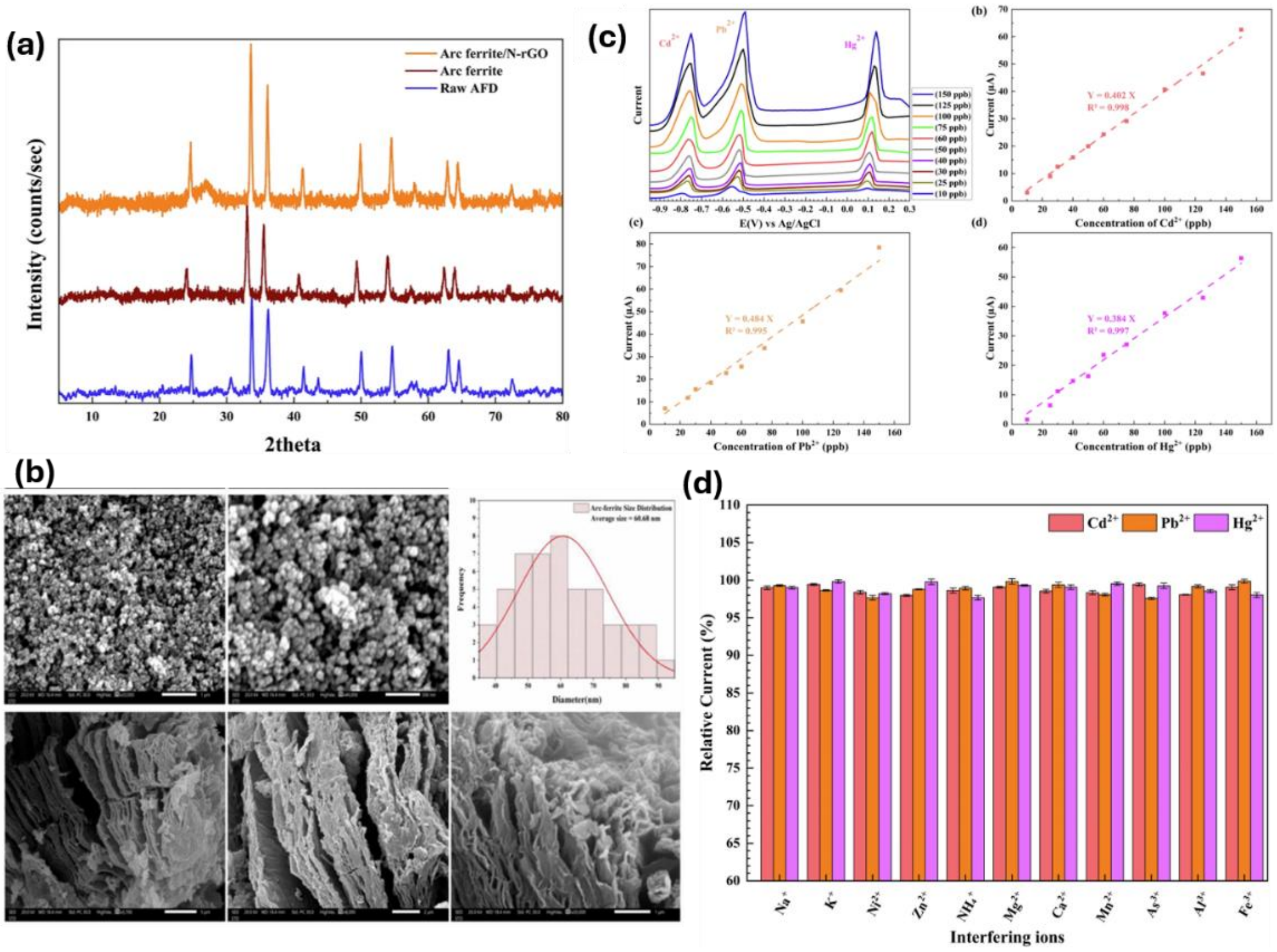

Fig. 19 (a) X-ray diffraction for Furnace dust ferrite and its composite with rGO (b) TEM images of pristine furnace dust ferrite and composite with rGO (c) $Cd^{2+}$, $Pb^{2+}$ and $Hg^{2+}$ detection at various ppb and linear response to these gases (d) 100-fold sensitivity to $Cd^{2+}$, $Pb^{2+}$ and $Hg^{2+}$ against 11 common alkali and earth metals[149]

Mesoporous $ZnFe_2O_4$ hollow octahedron anchoring on reduced graphene oxide detected $NO_2$ gas (50ppb), along with $SO_2$ (500 ppb), $NH_3$ (5,000 ppb), acetone (5,000 ppb), and acetic acid (5,000 ppb). The highest detection of $NO_2$ gas is due to interaction with $O_2^-$ whereas other gases interacted with $O^-$ ions resulting in their lower sensitivity at room temperature. Thus, $ZnFe_2O_4$ /rGO shows a high performance chemiresistive flexible gas sensor with sensitivity of 219.4 $ppm^{-1}$.[150]

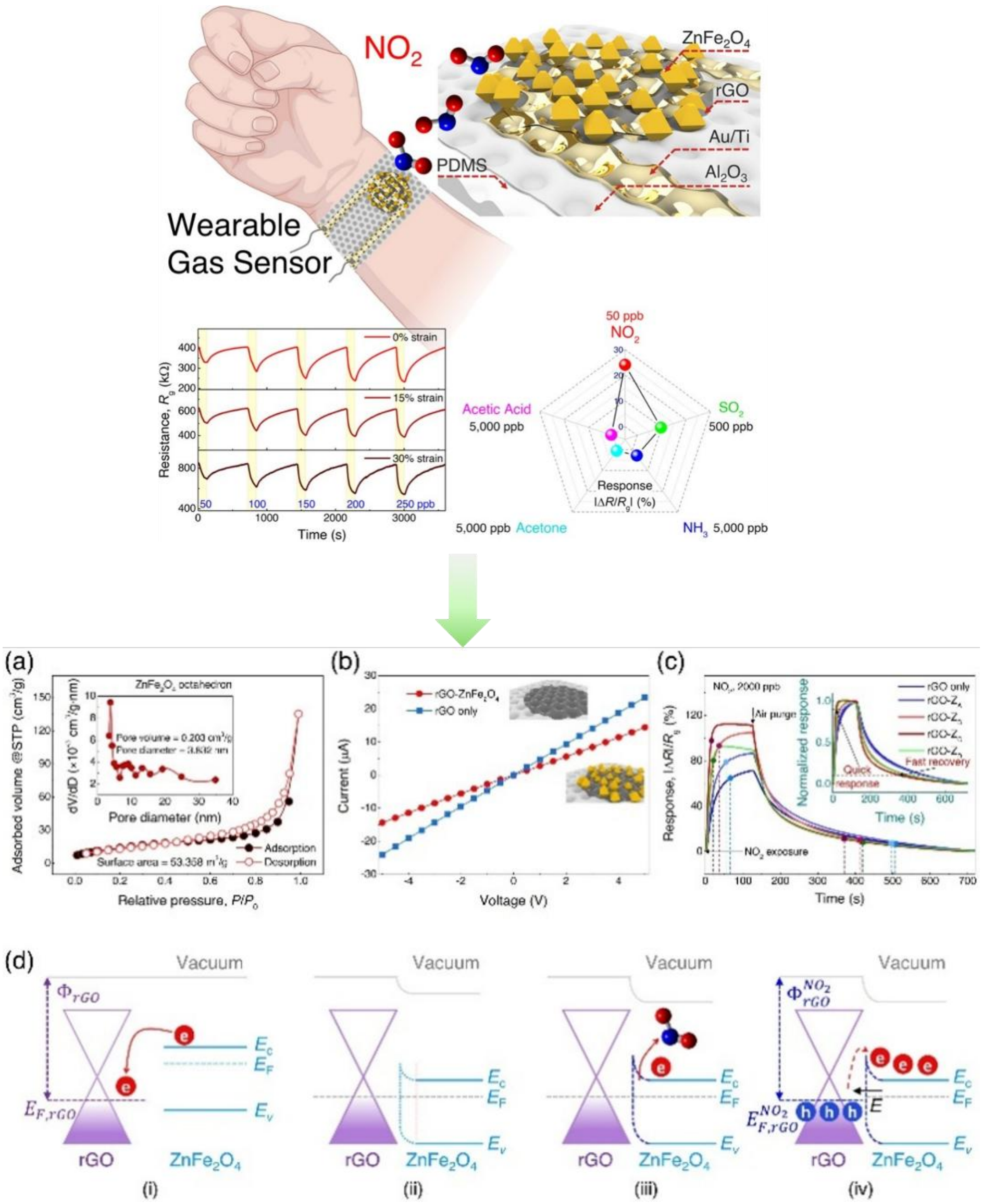

Fig. 20 Flexible $NO_2$ gas sensor based on mesoporous $ZnFe_2O_4$ hollow octahedra anchored on reduced graphene oxide (rGO). The figure illustrates the wearable sensing platform and its gas selectivity together with the structural and sensing characteristics of the $ZnFe_2O_4$/rGO heterostructure. (a) $N_2$ adsorption–desorption isotherms and pore-size distribution confirming the mesoporous nature of $ZnFe_2O_4$. (b) Current–voltage (I–V) characteristics comparing rGO and rGO-$ZnFe_2O_4$. (c) Dynamic response of the sensor toward various $NO_2$ concentrations and corresponding response/recovery behavior. (d) Schematic illustration of the energy-band alignment and charge-transfer mechanism at the rGO/$ZnFe_2O_4$ heterojunction during $NO_2$ sensing[150]

Similarly, in supercapacitor systems, graphene-based hybridization effectively addresses the intrinsic limitations of spinel ferrites, including low electrical conductivity, sluggish charge transport, particle agglomeration, and structural degradation during repeated charge-discharge cycles. The conductive graphene framework facilitates rapid electron mobility and shortened ion-diffusion pathways, while simultaneously buffering volume changes and improving mechanical integrity. Additionally, the synergistic interaction between ferrite nanoparticles and graphene sheets enhances electroactive surface exposure and accelerates reversible Faradaic reactions.[151,152] Several studies have demonstrated substantial electrochemical enhancement through graphene integration. Tudu et al. synthesized $MnFe_2O_4$/rGO nanocomposites through a one-pot solvothermal approach, where $MnFe_2O_4$ nanospheres were homogeneously distributed across wrinkled rGO sheets. The conductive rGO network effectively prevented nanoparticle agglomeration and created interconnected ion-transport channels with improved electrolyte accessibility. The synergistic interaction between $MnFe_2O_4$ and rGO significantly enhanced electron-transfer kinetics and electroactive surface exposure, resulting in a specific capacitance of 253 F $g^{-1}$ compared to 133 F $g^{-1}$ for pristine $MnFe_2O_4$. Furthermore, the composite exhibited remarkable cyclic stability with 96% capacitance retention after 5000 cycles, highlighting the structural stabilization effect of rGO incorporation.[153]

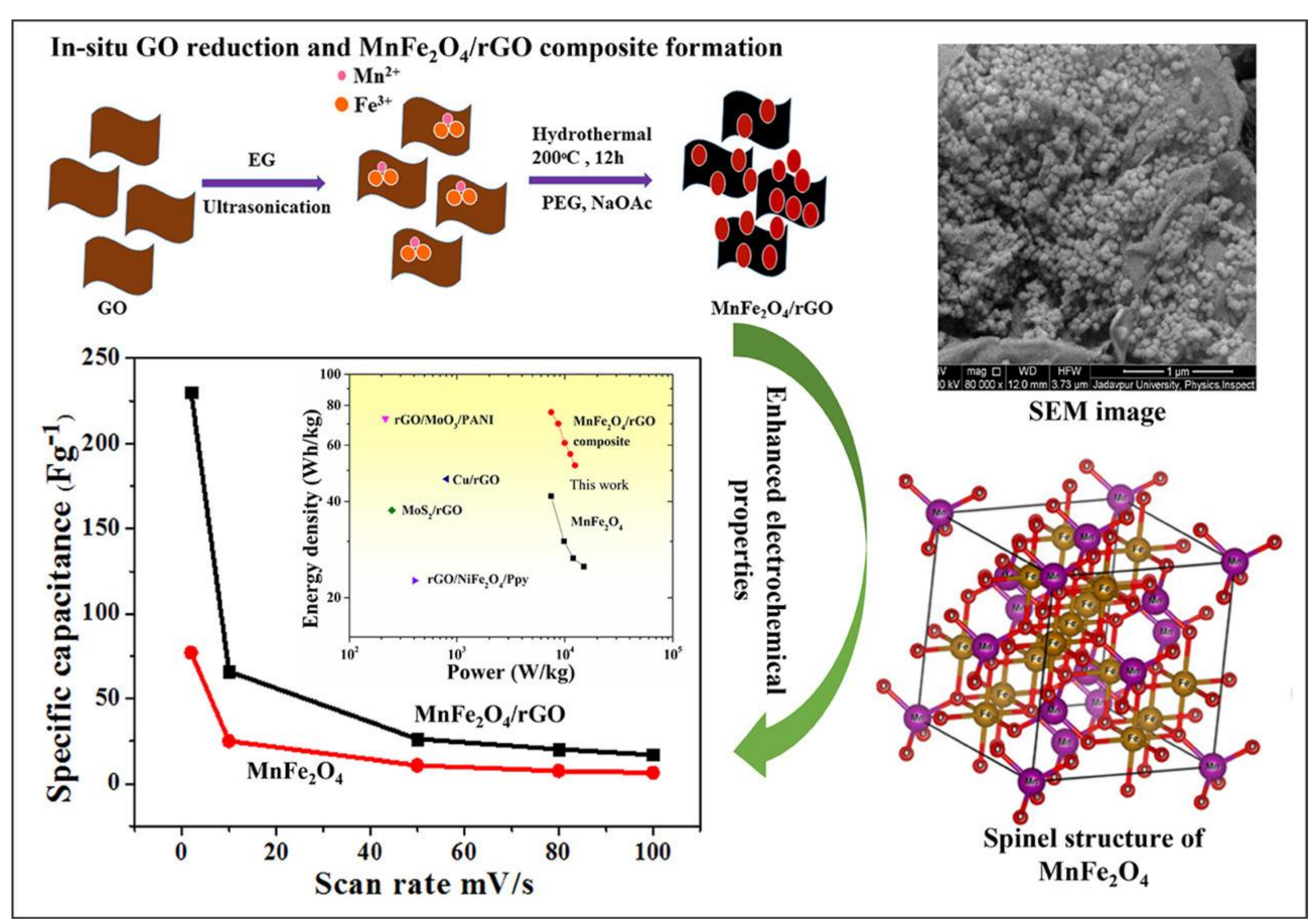

Fig. 21 In-situ graphene oxide reduction and $MnFe_2O_4$ spherical agglomerated nanocomposite formation, resulting in enhancement in specific capacitance for pristine $MnFe_2O_4$ and $MnFe_2O_4$@rGO[153]

Masoudpanah et al. investigated the influence of varying rGO content on $CoFe_2O_4$ electrochemical behavior and observed that integration of rGO produced optimal electrochemical performance. The uniform dispersion of $CoFe_2O_4$ nanoparticles over conductive rGO sheets increased the specific surface area from 85 to 105 $m^2\ g^{-1}$ and significantly reduced charge-transfer resistance from 34 to 1.9 Ω. Such improvements facilitated rapid $OH^-$ ion diffusion and enhanced redox accessibility at the electrode–electrolyte interface. Consequently, the $CoFe_2O_4$-10/rGO composite exhibited a remarkably high specific capacitance of 1820 F $g^{-1}$ at 1 A $g^{-1}$ while maintaining excellent rate capability and cycling stability.[137] Kumar et al. reported $NiFe_2O_4$/rGO nanocomposites synthesized through a sol-gel auto-combustion route, where incorporation of rGO substantially enhanced the electrochemical activity of pristine $NiFe_2O_4$. The conductive rGO framework improved electron mobility and facilitated rapid charge-transfer processes, while simultaneously increasing electroactive surface exposure. Electrochemical measurements revealed intensified redox peaks corresponding to Ni-O/NiOOH and Fe-O/FeOOH transitions, indicating improved pseudocapacitive behavior. The optimized composite achieved a high capacitance of 1709.87 F $g^{-1}$ at 1 A $g^{-1}$ along with 91% capacitance retention after 4000 cycles, demonstrating the beneficial role of graphene-mediated conductive pathways and structural stabilization.[154] Jennifer et al. synthesized hydrothermally derived $NiFe_2O_4$/rGO nanocomposites consisting of uniformly dispersed $NiFe_2O_4$ nanorods anchored onto interconnected rGO sheets. The porous conductive framework generated by rGO provided efficient ion-diffusion channels and minimized internal resistance, while simultaneously suppressing nanorod agglomeration. The synergistic interaction between $NiFe_2O_4$ redox-active sites and the conductive graphene network resulted in enhanced capacitance, energy density, and cycling durability, with the composite retaining 91.8% of its capacitance after 5000 cycles.[155]

Despite these advantages, graphene-based systems still face challenges such as graphene-sheet restacking, limited intrinsic pseudocapacitance, and interfacial stability issues. To overcome these limitations, heteroatom doping and hybridization with transition-metal oxides, conducting polymers, and chalcogenides have been extensively explored to further improve electrochemical activity and charge-storage capability.[156–159]

Overall, graphene-based conductive hybridization has emerged as a highly effective strategy for simultaneously improving electrical conductivity, ion diffusion, active-site accessibility, and electrochemical stability in spinel ferrite-based gas sensors and supercapacitor electrodes.

### 5.3 Transitional Metal Doping

Transition-metal doping has emerged as one of the most effective approaches for improving the electrochemical performance of spinel ferrites by overcoming intrinsic limitations such as low electrical conductivity, sluggish charge-transfer kinetics, and insufficient electroactive surface accessibility. The substitution of transition-metal ions within the spinel $ABFe_2O_4$ lattice modifies the cation distribution between tetrahedral (A) and octahedral (B) sites. This affects the electronic structure, defect concentration, magnetic interactions, and electrochemical activity of the host ferrite. Thus, such modifications often induce oxygen vacancies on the surface due to lattice distortions or modified energy bandgap, etc., which collectively enhance carrier mobility and facilitate rapid electron-transfer processes.[160–162]

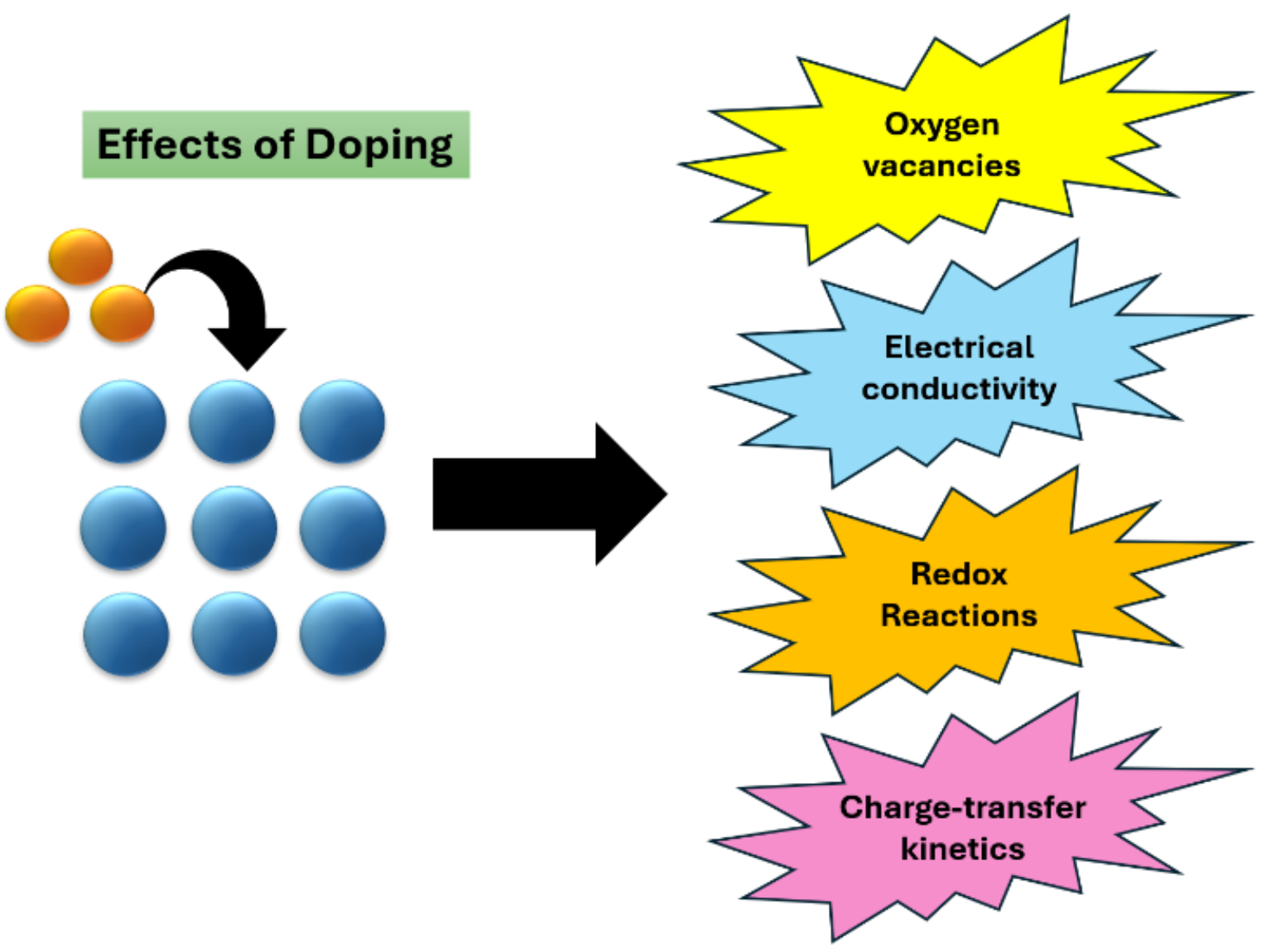

Fig. 22 Effects of doping in electrochemical performance

In gas-sensing systems, transition-metal doping plays a crucial role in tuning adsorption behaviour, carrier concentration, and surface catalytic activity. Dopant atoms possess partially filled orbitals that strongly interact with adsorbed gas molecules and surrounding lattice atoms, thereby modifying the Fermi-level characteristics and charge-transfer behaviour of metal oxide semiconductor (MOS) sensors. Such interactions improve adsorption-desorption kinetics, sensitivity, and selectivity toward target gases. Depending on the synthesis strategy, doping may occur either through bulk substitution within the crystal lattice or through surface decoration of ferrite nanostructures. In both cases, dopants significantly influence electron depletion-layer modulation and gas-response characteristics by facilitating electron exchange between the adsorbed analyte molecules and the ferrite surface.[64] Manoharan et al. investigated Cu dopped $NiFe_2O_4$ exhibiting highest selectivity for Liquid Petroleum Gas (LPG) monitoring response and recovery times are 46 and 39 seconds respectively at 75 ppm LPG. The following equation showcases the reaction occurred at the electrode-gas interface[120]

$$C_3H_8 + 5O_2^- \leftrightarrow 3CO_2 + 4H_2O + 5e^-$$

$Ni_{1-y}Zn_yFe_2O_4$ nanoparticle exhibited excellent selectivity ∼600% at 20 ppm at the operating temperature 140C to detect $H_2S$ gas. The transition metal doping changed the degree of inversion of the spinel structure resulting in $Zn^{2+}$ at the A-sites, while the B-sites filled by Ni and Fe ions. This structure leads to lesser hopping of electrons and thus, high resistance. Additionally, the high porosity achieved due to Zn doping enhanced the gas adsorption resulting in high change in resistivity.[163] Similarly, $ZnFe_2O_4$ and Cu-doped $ZnFe_2O_4$ spinel nanoferrites were compared in isopropyl gas sensing performance. The $ZnFe_2O_4$ sensor's response and recovery times at room temperature are 42 and 38 seconds, respectively, whereas the $ZnCuFe_2O_4$ sensor's response and recovery times are 30 and 28 seconds. This enhancement was seen due to the doping of Cu, improving the electrical conductivity and as well as creation of oxygen vacancies, thus, resulting in a high-performance hetero-transition metal composition for sensing VOCs gases.[164]

Transition-metal doping similarly enhances pseudocapacitive performance in supercapacitor electrodes through the introduction of multiple redox-active centers and improved electrical conductivity. The coexistence of various oxidation states within doped ferrites enables rapid reversible Faradaic reactions, while structural modifications such as increased porosity, reduced charge-transfer resistance, and enlarged electroactive surface area contribute to enhanced ion diffusion and charge-storage capability. Consequently, transition metal substituted ferrites often exhibit higher capacitance, improved rate capability, and superior cycling stability compared to pristine ferrites.

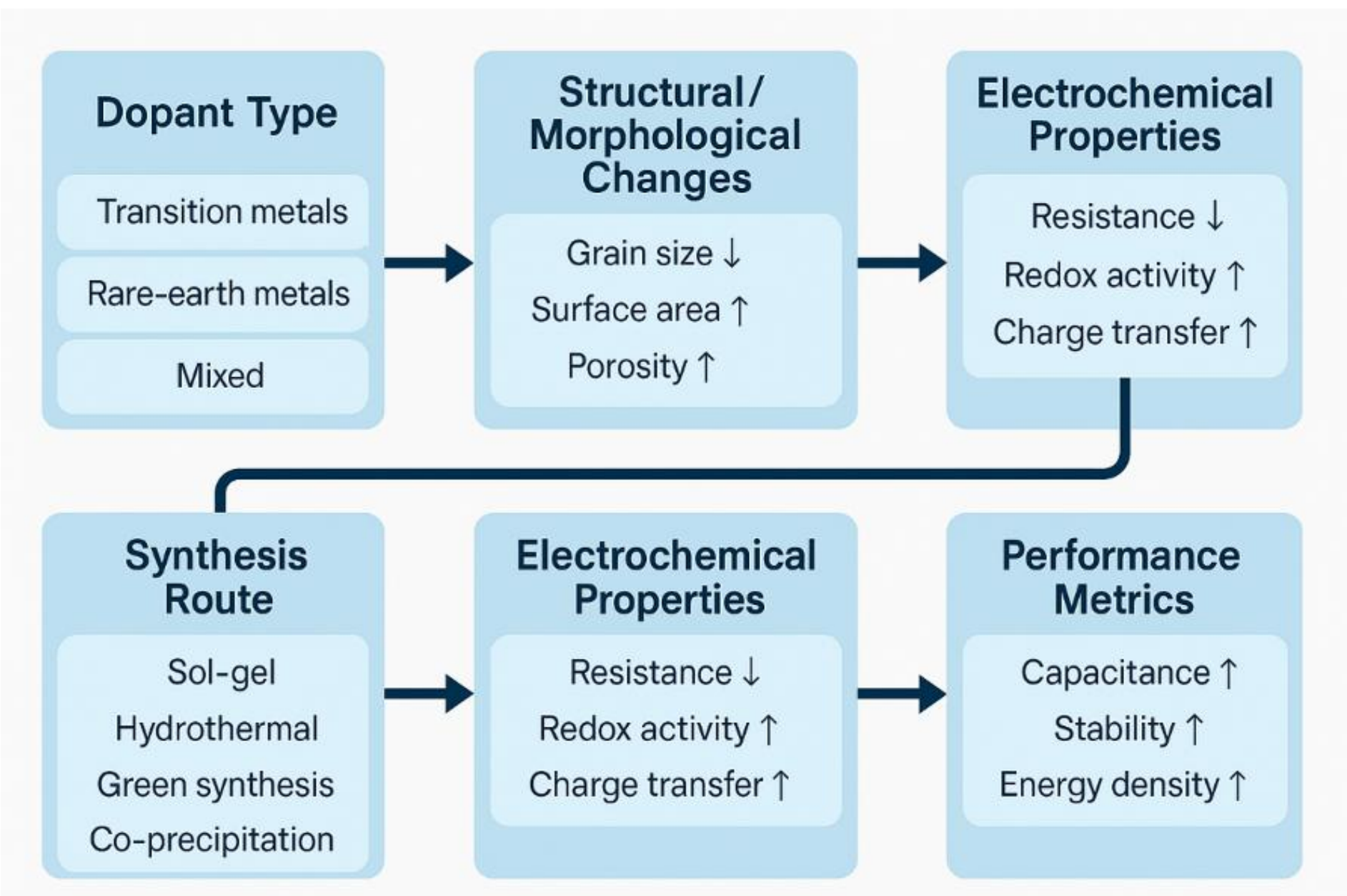

Fig. 23 Effect of dopant- transition metal and rare earth element in structural and electrochemical properties[165]

Several illustrations displayed the beneficial role of transition metal substitution in ferrite-based electrochemical systems. Tran et al. investigated Mn-doped $CoFe_2O_4$ nanocomposite and observed that increasing Mn concentration expanded the lattice parameter and promoted redistribution of $Fe^{3+}$ ions between tetrahedral and octahedral sites. Such cation redistribution altered the electrical and electrochemical behavior of the ferrite structure by modifying the local electronic environment and improving charge-transport characteristics.[166] Similarly, Angadi et al. explored Zn-doped $CuFe_2O_4$ spinel ferrites synthesized through a dual-fuel combustion approach. Zn incorporation induced lattice expansion, highly porous morphologies, and band-gap narrowing, resulting in improved electrochemical conductivity and enhanced pseudocapacitive performance. The optimized $Cu_{0.75}Zn_{0.25}Fe_2O_4$ composition exhibited excellent electrochemical stability and an areal capacitance of 11.07 mF $cm^{-2}$, highlighting the synergistic role of band-gap engineering and morphology tuning in multifunctional ferrite electrodes.[167] Ni-Co mixed ferrite system $Ni_{1-x}Co_xFe_2O_4$ synthesized via citrate-assisted auto-combustion, the optimized $Ni_{0.5}Co_{0.5}Fe_2O_4$ composition exhibited enhanced ion transport and improved electrochemical conductivity due to the coupled $Ni^{2+}/Ni^{3+}$ and $Co^{2+}/Co^{3+}$ redox transitions. The work demonstrated the advantages of multi-redox-center engineering in spinel ferrites.

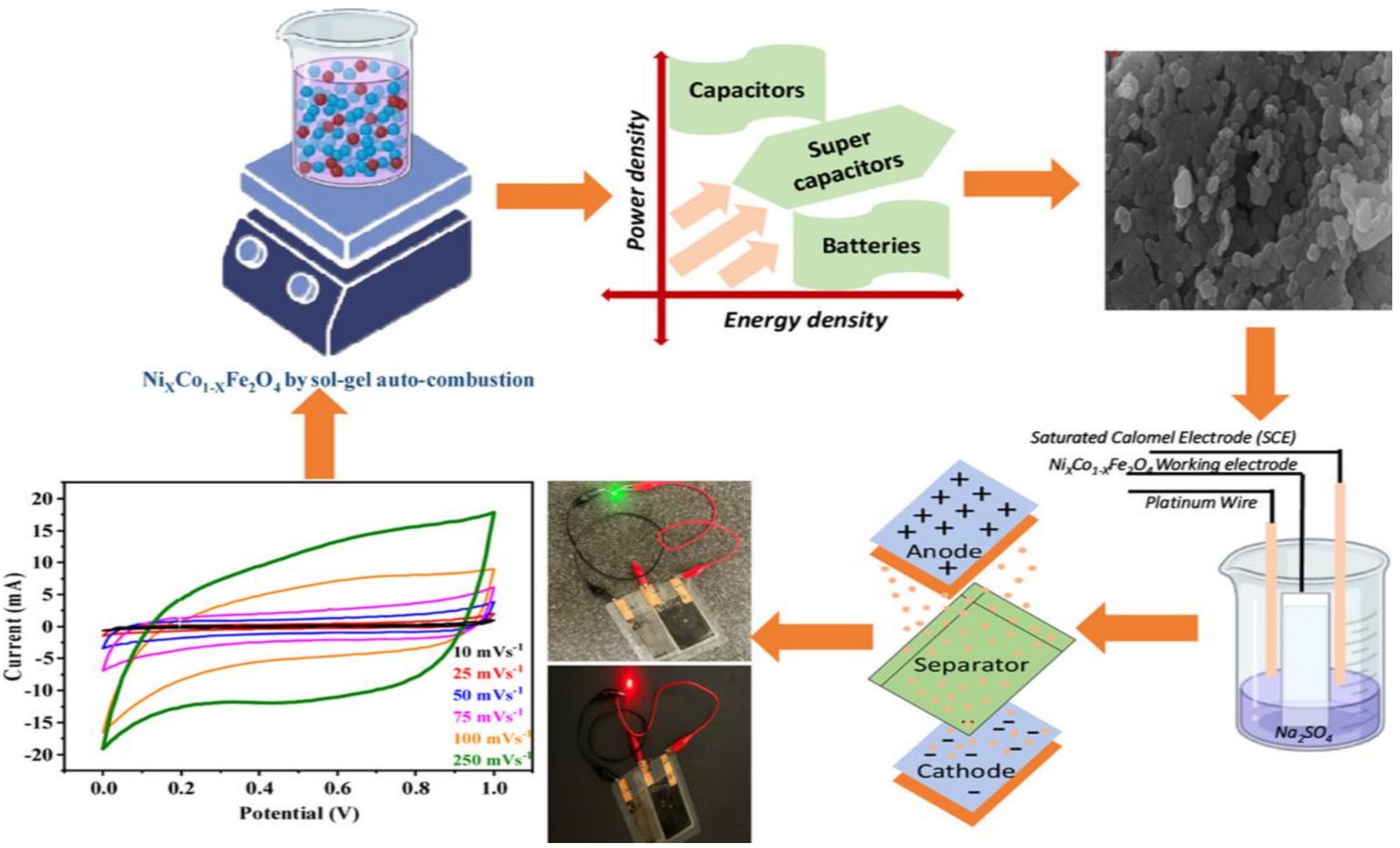

Fig. 24 Ni1-x$Co_xFe_2O_4$ spinel ferrite nanoparticles synthesized via sol-gel auto-combustion method utilised in high-performance supercapacitor electrode powering LED[167]

Likewise, Mn-doped $NiFe_2O_4$ nanosheet electrodes revealed enhanced electrolyte-ion transport and reduced charge-transfer resistance owing to the increased density of redox-active sites generated by Mn incorporation. The optimized Mn-doped ferrite electrode achieved high specific capacitance and improved electrochemical stability in alkaline electrolyte systems.[168] Overall, transition-metal doping represents a highly versatile strategy for tailoring the electronic structure, redox activity, defect chemistry, and interfacial charge-transfer behavior of spinel ferrites. Through appropriate dopant selection and compositional engineering, transition metal substituted ferrites can achieve significantly enhanced electrochemical performance for both gas-sensing and energy-storage applications.

Table 5 Recent studies for transitional heteroatom spinel ferrite systems on supercapacitor applications

| System | Synthesis Method | Morphology | Specific Capacitance ($Fg^{-1}$) | Cycles/Retention | Surface area ($m^2g^{-1}$) | Ref |
|---|---|---|---|---|---|---|
| **$Ni_{1-x}Co_xFe_2O_4$** | Hydrothermal | Nanoneedles (50-100 nm) | 996 @ 5mV/s | 5000 cycles, 90% | - | [168] |
| **Zn-Mn-FO** | Solgel-Combustion | Porous NPs (~25 nm) | 282 @ 0.25 A/g | 1000 cycles, 84% | 15.53 | [169] |
| **$Ni_xCo_{1-x}Fe_2O_4$** | Sol-gel auto-combustion | Spherical (14 nm) | 1700 @ 10 mV/s | 5000 cycles, 91% | 55.45 | [170] |

| System | Synthesis Method | Morphology | Specific Capacitance ($Fg^{-1}$) | Cycles/Retention | Surface area ($m^2g^{-1}$) | Ref |
|---|---|---|---|---|---|---|
| **Mn-$NiFe_2O_4$** | precipitation | Nanocrystals (~20nm) | 913 @ 5 mV/s | 5000 cycles, 84.3% | - | [171] |
| **$CuCoF_2O_4$** | Sol-gel | Thin films | 221 @ 5 mV/s | 1000 cycles | - | [172] |
| **$NiCuFe_2O_4$** | Microwave-assisted combustion | hexagonal-shaped nanoparticle (27nm) | 220.6 @ 5 mV/s | 5000 cycles, 79.63% | - | [173] |

Collectively, these findings stress on the critical role of transition metal doping in optimizing the electrochemical properties of spinel ferrites for energy storage applications.

### 5.4 Rare-Earth Elements Doping

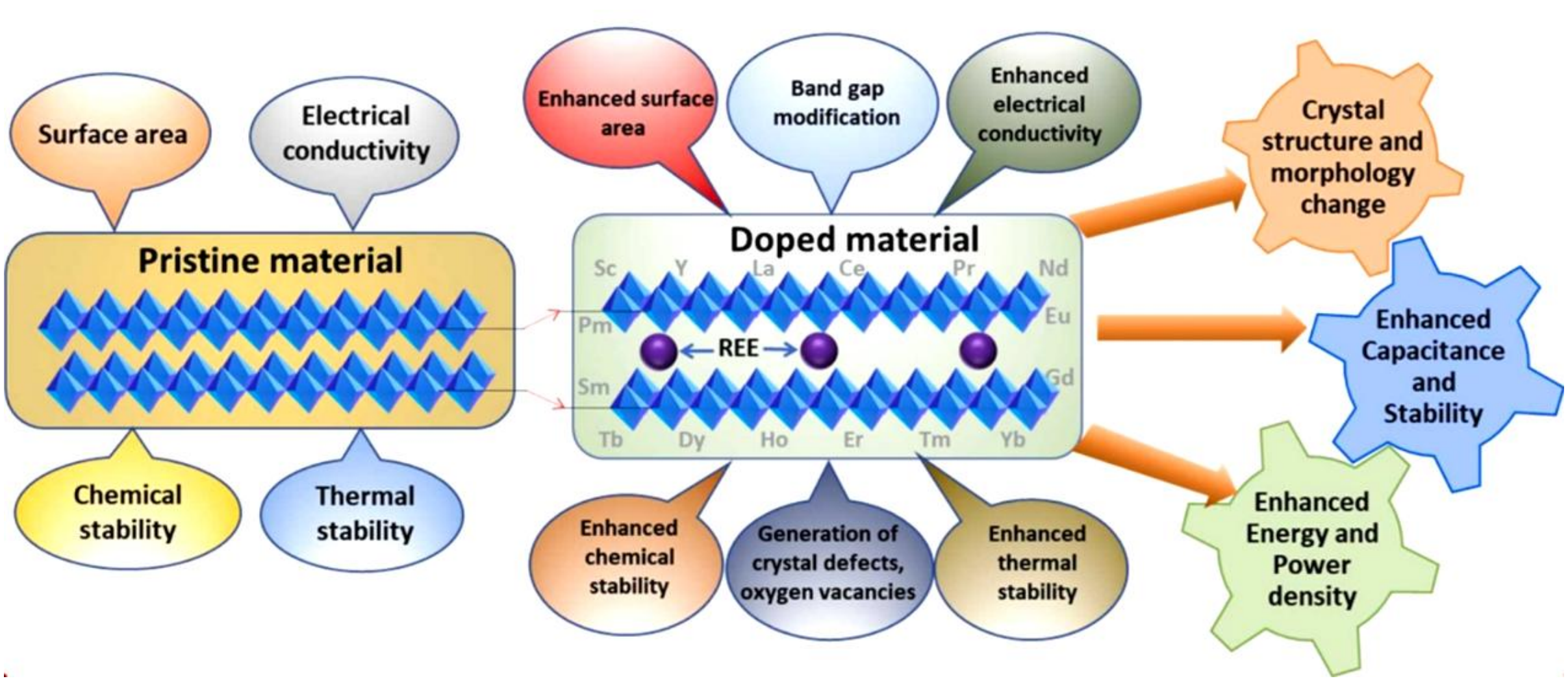

Fig. 25 Rare earth Elemt doping effects on electrochemical applications[174]

Rare-earth element (REE) doping has emerged as a strategic approach for tailoring the structural, electronic, and electrochemical properties of spinel ferrites for advanced sensing and energy-storage applications. Due to their larger ionic radii and 4f electronic configurations, the REE ions typically induce significant lattice distortions, defect formation, oxygen vacancies, and localized electronic states when incorporated into the spinel lattice. Such modifications strongly influence charge-transfer kinetics, band-gap characteristics, redox activity, and ion-transport behaviour in ferrite-based electrochemical systems. REE doping is commonly achieved through co-precipitation, hydrothermal, sol-gel, combustion, and solid-state synthesis methods, where the dopant ions are generally incorporated at octahedral B-sites within the spinel framework.[175] The incorporation of REE ions generally leads to lattice strain arising from ionic-radius mismatch between the host and dopant ions. As the ions generates oxygen vacancies and defect-rich interfaces, this improves the electrochemical activity of the spinel ferrite. These defects act as additional redox-active centres and facilitate rapid ion diffusion and electron transport. Furthermore, localized impurity states introduced by REE ions narrow the electronic band gap and improve conductivity, while morphological refinement and enhanced porosity provide improved electrolyte ion or gas ion accessibility. Consequently, REE-doped ferrites often exhibit enhanced electrochemical surface activity compared to pristine ferrite systems.[174]

In supercapacitors, several findings have demonstrated the beneficial influence of REE doping on ferrite delivers enhanced capacitance, improved rate capability, superior cycling stability. Ahmad et al. synthesized Ce-doped $SnFe_2O_4$ through a hydrothermal approach and observed that incorporation of $Ce^{3+}$ ions significantly reduced particle agglomeration, increased specific surface area, and enhanced electrochemical activity. The larger ionic radius of $Ce^{3+}$ generated lattice strain and oxygen-vacancy defects that promoted proton transport and facilitated rapid charge-transfer processes. Consequently, the Ce-doped ferrite achieved a high specific capacitance of 1216 F $g^{-1}$ along with substantially improved energy density and electrochemical stability compared to pristine $SnFe_2O_4$.[176]

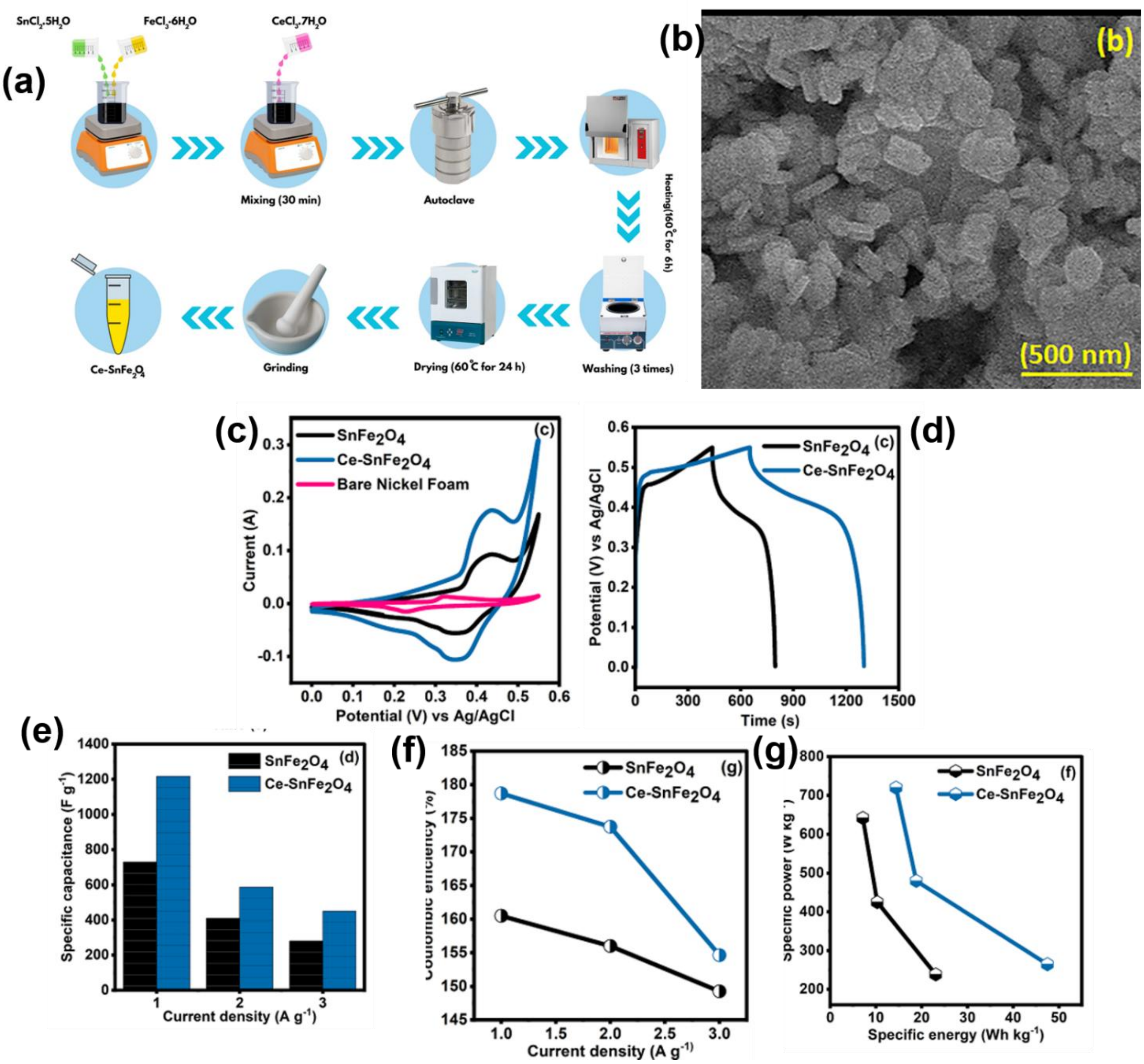

Fig. 26 Effect of Ce doping on the structural and electrochemical properties of $SnFe_2O_4$. (a) Schematic illustration of the hydrothermal synthesis procedure for Ce-doped $SnFe_2O_4$ (Ce-$SnFe_2O_4$). (b) SEM image showing the morphology of the synthesized Ce-$SnFe_2O_4$ nanoparticles. (c,d) Cyclic voltammetry (CV) curves and galvanostatic charge-discharge (GCD) profiles comparing pristine $SnFe_2O_4$ and Ce-$SnFe_2O_4$ electrodes. (e) Specific capacitance at different current densities, (f) Coulombic efficiency, and (g) Ragone plot illustrating the energy and power density characteristics of the electrodes. Ce incorporation enhances the electrochemical performance through improved conductivity, increased active surface area, and additional redox-active sites[176]

Similarly, Tewari et al. investigated La-doped $ZnFe_2O_4$ synthesized via an auto-combustion route and reported progressive reduction in crystallite size and strain with increasing La incorporation. The resulting rugby-shaped nanoparticles exhibited improved electrochemical behavior owing to enhanced surface accessibility and defect-assisted charge transport.[177] In Sr-doped Co-Ni ferrites synthesized by solution combustion, Basavegowda et al. observed that incorporation of $Sr^{2+}$ ions induced lattice distortion and increased porosity, thereby improving surface-to-volume ratio and electrochemical accessibility. The addition of gum arabic further enhanced capacitance retention and rate capability by improving conductive interfacial pathways.[178] Sm-doped cobalt ferrites additionally demonstrated that REE incorporation can significantly enhance porosity and reduce internal resistance, thereby facilitating improved ionic and electronic transport. The optimized Sm-doped composition exhibited nearly threefold enhancement in capacitance compared to pristine cobalt ferrite along with excellent long-term cycling stability.[179]

In sensors, the doping of Yttrium in Nickel ferrite nanofibers showed detection of triethylamine at a concentration of 100 ppm at 125 °C. The system had amazingly rapid response time of only 2 seconds. Due to its larger ionic radius, $Yt^{3+}$ replaces $Fe^{3+}$ at octahedral site rather than substituting at tetrahedral site which created lattice distortion and oxygen vacancies for the oxygen and target VOCs to be adsorbed at the electrode surface and hence, a high yield of chemiresistive sensing.[180] The rare earth doping elevates the resistance modulation, rapid response and recovering time and thus, better selectivity and sensitivity of target gases.

Table 6 Recent studies for rare earth spinel ferrite systems on gas sensor applications

| System | Target gas | Optimum temp (°C) | Gas concentration | Response value at optimum | Response time | Recovery time | Ref |
|---|---|---|---|---|---|---|---|
| Erbium-doped $ZnFe_2O_4$ nanofibers (E-ZFO NFs) | n-Butanol | 175 | 100 ppm | 30.3 (Er-ZFO), ~3× (pure ZFO) | 9 s | 94 s | [181] |
| Yttrium-doped $ZnFe_2O_4$ nanofibers (Y–ZFO NFs) | n-Butanol | 160 (Y-doped) vs 175 (pure ZFO) | 100 ppm | 52.2 (Y-ZFO), ≈ 3× (pure ZFO) | 6 s | ~280–300 s | [182] |
| Lanthanum-doped $ZnFe_2O_4$ nanoparticles (La–ZFO NPs) | Xylene | 250 (La-doped) vs 300 (pure) | 100 ppm | 118 (La-ZFO) vs 38 for pure | 69 s | 87 s | [183] |
| Cerium-doped $ZnFe_2O_4$ hollow microspheres (Ce–ZFO) | Acetone | 160 | 5 ppm | 7.08 | 10 s | 160 s | [184] |
| Lutetium-doped $ZnFe_2O_4$ nanoparticles (Lu–ZFO) | CO | RT | 20,000 ppm | 80% | - | - | [185] |

| System | Target gas | Optimum temp (°C) | Gas concentration | Response value at optimum | Response time | Recovery time | Ref |
|---|---|---|---|---|---|---|---|
| Samarium-doped $ZnFe_2O_4$ nanoparticles (Sm–ZFO) | $H_2S$ | 400 | 30-40 ppm | 10 | ~7.5-35 s | ~90-170 s- | [179] |

The synergistic integration of REE doping with conductive carbonaceous frameworks and has also shown remarkable promise. Mirza et al. investigated Pr- and Nd-doped $CoFe_2O_4$/graphene oxide composites and observed that the oxygen functional groups of graphene oxide promoted strong interfacial interaction with the rare-earth-doped ferrite nanoparticles, thereby improving hydrophilicity, conductivity, and electrochemical accessibility. The dual-doped PrNd-$CoFe_2O_4$/GO composite delivered exceptionally high capacitance and excellent cyclic retention due to the combined effects of REE-induced electronic modulation and graphene-mediated conductive pathways.[186] Likewise, $Tb^{3+}$-doped cobalt ferrites synthesized through sol-gel auto-combustion exhibited enhanced porosity, reduced crystallite size, and narrowed band-gap characteristics that collectively improved electron transport and redox activity. The nanoscale crystallites provided abundant electroactive sites and facilitated rapid electrolyte-ion diffusion, resulting in significantly enhanced specific capacitance and electrochemical stability.[187]

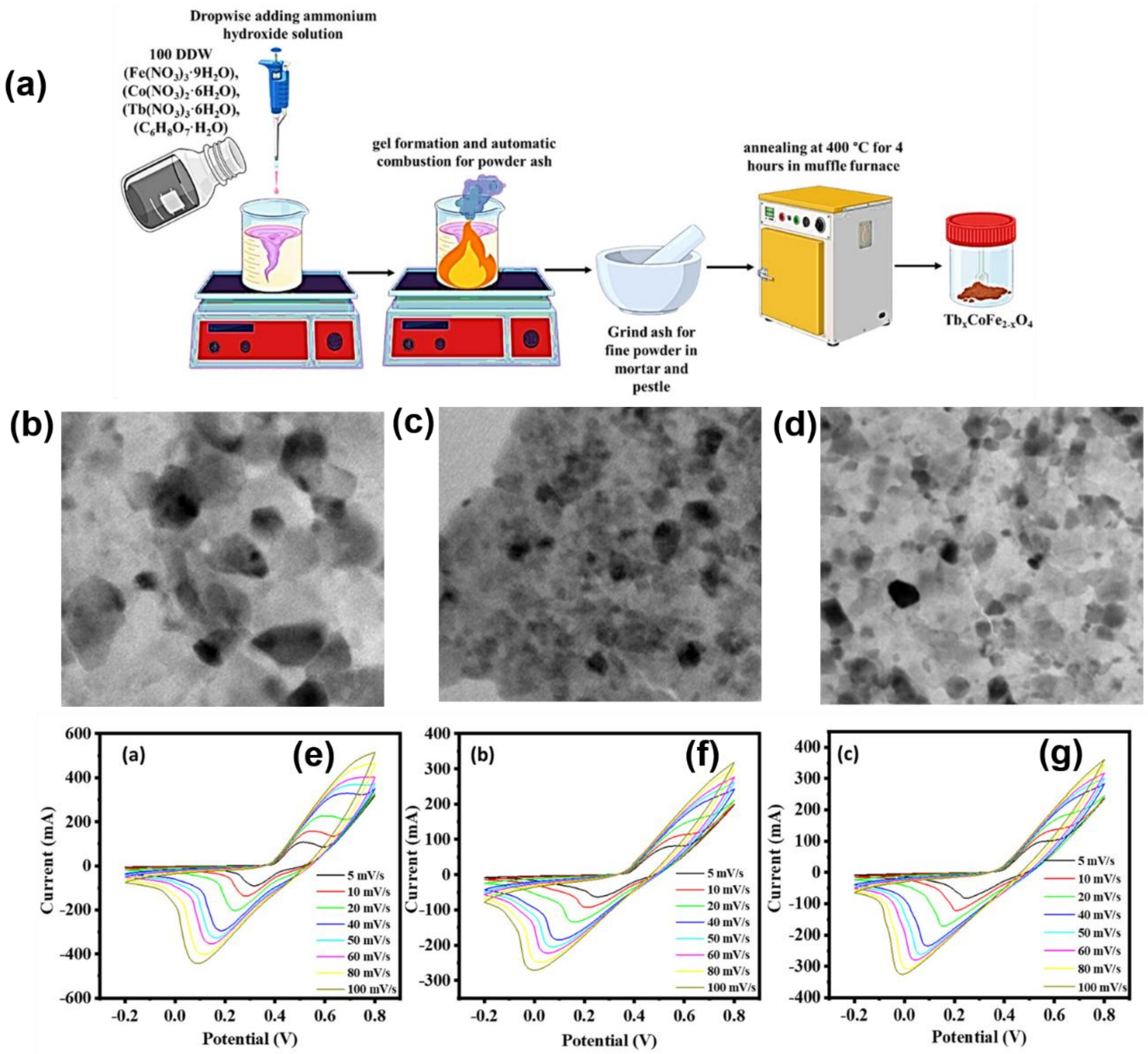

Fig. 27 Tb-doped cobalt ferrite ($Tb_xCoFe_{2-x}O_4$) synthesized via the sol-gel auto-combustion route. (a) Schematic illustration of the synthesis process. (b–d) TEM images showing the morphology of ferrites with varying Tb content at x=0, x=0.75 and and x = 0.15 (e–g) Cyclic voltammograms recorded at different scan rates, demonstrating the influence of Tb doping on the redox behavior and capacitive response of the electrodes. The enhanced electrochemical performance is attributed to structural distortion, reduced crystallite size, and increased electrochemically active surface area induced by $Tb^{3+}$ incorporation

Recent studies on Dy-doped Cu–Zn ferrite/rGO composites further indicate that incorporation of rare-earth ions such as $Dy^{3+}$, $Ce^{3+}$, $La^{3+}$, and $Gd^{3+}$ effectively modulates electronic states and defect chemistry, thereby improving ion transport, conductivity, and pseudocapacitive behavior in multifunctional ferrite-based electrochemical devices.[188]

Overall, rare-earth doping represents a highly promising strategy for engineering the structural defects, electronic structure, conductivity, and redox-active interfaces of spinel ferrites. Through appropriate dopant selection and compositional optimization, REE-substituted ferrites can achieve substantially improved electrochemical performance for both gas sensing and supercapacitor applications.

### 5.5 Formation of junction in sensors

In the quest to enhance gas sensing properties, researchers have turned their attention to the creation of p–n, n–n, and p–p heterojunction contacts at the nanoscale, Through the amalgamation of many metal oxide semiconductors possessing distinct bandgaps. This approach involves fusing two semiconductors, either p-type, n-type, or a combination thereof, to create an interface where Fermi energy levels cross across or move closer to one another. The result is a depletion layer forming and an increase in charge transfer. The effectiveness of heterojunction materials largely depends on two key factors: Chemical uniformity and great structural integrity. In a metal oxide semiconductor (MOS), heteroatoms displace the host metal oxide when they are incorporated into the bulk, effectively lowering the source material's grain size. This becomes particularly significant thus an electron depletion layer covers the whole grain when the particle size is smaller than twice the Debye dimension, hence improving the sample's gas sensing capabilities. Additionally, the material's surface area and porosity are changed when foreign heteroatoms are added, increasing the rate at which gas molecules are adsorbed and diffused.[189]

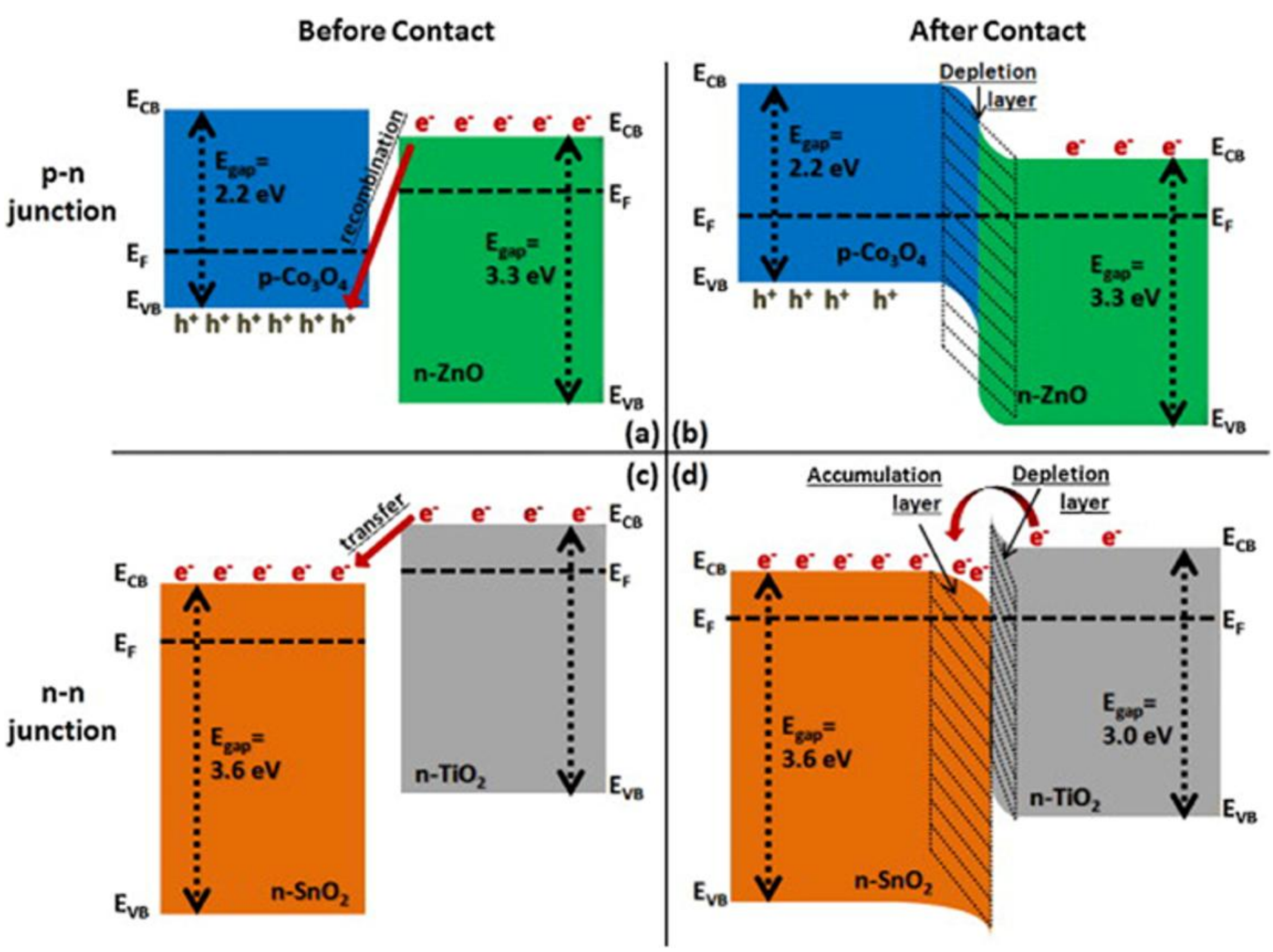

Fig. 28 Schematic diagram of formation of junction between n-type and p-type metal oxides[190]

The interaction between the Fermi energies of two metal oxide semiconductor materials causes high-energy electrons to flow over the junction interface to the vacant low-energy states until the Fermi energies balance out, which is a unique feature of the heterojunction process. This process is similar to the recombination of electrons and holes seen in p-n junctions. This causes the heterojunction interface to become a depleted region of charge carrier formation. Differences in the MOS materials' initial Fermi energy levels cause band bending, which might result in an energy barrier at the heterojunction contact. Electrons can cross the contact only after they have passed through this potential energy barrier, which increases resistance. For instance, the lower-energy valence band states (holes) are produced when p-type MgO and n-type $MgO_2O_4$ create a p-n junction promoting the transfer of electrons across the interface, establishing Fermi energy equilibrium.[191]

Similarly, when n-type α-$Fe_2O_3$/$CuFe_2O_4$ combine to produce a n–n junction, band bending occurs at the junction contact.[192] In this case, electron The MOS experiences an accumulation layer as opposed to a depletion layer as a result of migration to the lower-energy conduction band. The quick depletion of the accumulation layer on the MOS surface is caused by the high rate of oxygen adsorption, leading to an increase at the junction interface's potential energy barrier and a subsequent rise in resistance, ultimately influencing the sensor's response to analyte gases.

Likewise, in p-p junctions, band bending also occurs. For instance, the detection of formaldehyde has been done using NiO/$NiFe_2O_4$ with a p-p junction material.[193] The formation of p-p junctions, like other junction types, influences the electronic properties of the material and its sensitivity to specific gases. To sum up, the formation of junctions in gas sensors, whether p-n, n-n, or p-p is an effective method for modifying metal oxide semiconductor characteristics to maximize gas detecting abilities. Understanding the mechanisms involved in these junctions is crucial for developing sensors that respond effectively to a wide range of analyte gases.

## 6. Future Directions

The review underlined the potential of spinel ferrite-based materials in the development of electrochemical devices, specifically supercapacitors and sensors. Future research can look into the following directions to fully exploit the scope of possibilities and address existing challenges:

## 6.1 Spinel Ferrite and other Advanced Nanomaterials

Beyond the topics covered in the review, advanced nanomaterials such as 2D materials, quantum dots, polymers, and other developing nanocomposites are being investigated. These materials can result in innovative designs and increased performance. Recent research has shown that the combination of spinel ferrite nanostructures and two-dimensional nanomaterials like Mxenes, transition metal dichalcogenides like $MoS_2$, and others can considerably improve the overall charge transport characteristics and surface area.[194]

MXenes are a class of 2D transition metal carbides/nitrides and recently, a promising two-dimensional material for the advancement electrochemical devices. The general compositional formula of this 2-D material is $M_{n+1}AX_n$, where M is an initial transition metal, A is primarily an element from group IIIA or IVA (i.e., 13 or 14), X is C and/or N, and n is either 1, 2, or 3. MXenes offers high surface area and electrical conductivity but limiting due to sheets restacking.[195] Typically, spinel ferrite nanoparticles are decorated on the MXene sheets, which mitigates structural degradation of ferrospinels. With the same architectural foundation, asymmetric supercapacitor fabricated of $Mn_xZn_{1-x}Fe_2O_4$@MXene||AC used as power source for a calculator and delivered specific capacitance 646.9 F/g at a 5 mV/s along with energy density of 47 $W{\cdot}h{\cdot}kg^{-1}$, the highest power density of 4937.1 W/kg, a coulombic efficiency of 99%, and capacitance retention of 128.9% after 6000 continuous duty cycles.[196]

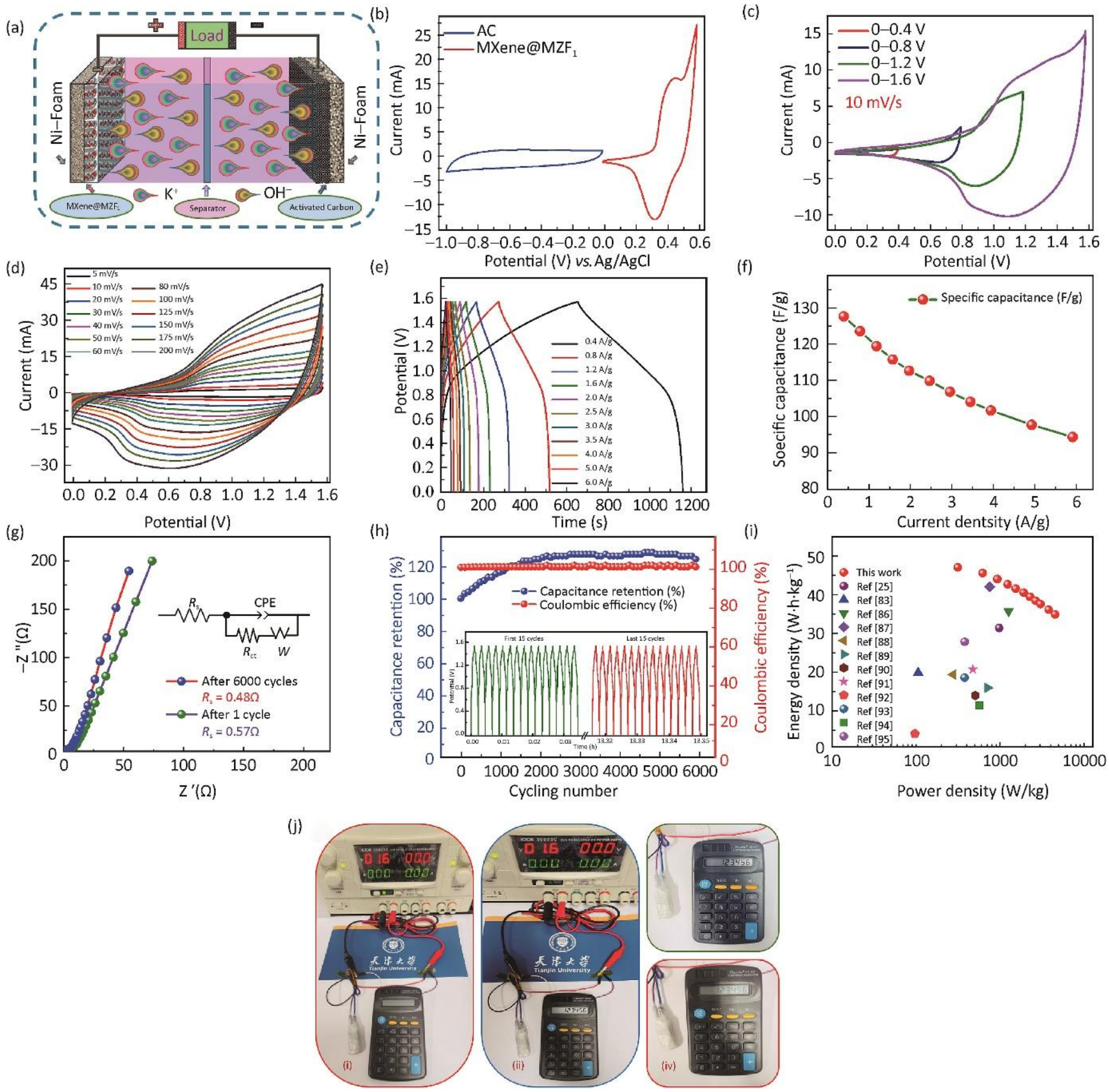

Fig. 29 Electrochemical performance of the MXene–spinel ferrite composite electrode and the assembled asymmetric supercapacitor device. (a) Schematic illustration of the MXene@$MZF_1$//activated carbon asymmetric supercapacitor. (b) Comparative cyclic voltammetry (CV) curves of activated carbon and MXene@$MZF_1$ electrodes. (c) CV profiles at different potential windows (d) CV curves measured at various scan rates. (e) Galvanostatic charge–discharge (GCD) profiles at different current densities. (f) Variation of specific capacitance with current density. (g) Nyquist impedance plots and equivalent circuit fitting. (h) Cycling stability and Coulombic efficiency during repeated charge–discharge cycling. (i) Ragone plot comparing the energy and power densities of the device with previously reported systems. (j) Practical demonstration of the assembled supercapacitor powering an electronic device- a calculator[196]

In another work, $NiFe_2O_4$ sandwiched in $Ti_3C_2T_x$ MXene nanosheets provided interlayer spacing between the sheets allowing better ion diffusion. MXene conductivity and hydrophilicity and Cobalt ferrite's redox activity as well as structural support creates synergy for electrochemical applications.[197]

An alternative burgeoning class of materials are Metal Organic Frameworks (MOFs). MOFs are a classification of coordinate polymers where metal cations or clusters coupled by organic ligand linkers. MOFs offer various fascinating features such as huge surface area for ion accessibility, resulting in abundance of active sites due to their extended organic ligands and hence, high-performance electrochemical applications.[198] These structures also work as template to metal oxides, which once etched, provides porous structure required for electrochemically active sites. For instance, MOF-derived bimetallic spinel ferrites $MFe_2O_4$ (M = Co, Ni, Cu and Zn) were investigated and as observed, Co, Ni and Cu ferrites had inverse spinel structure whereas, Zn was normal spinel ferrite. Prussian blue MOF constitutes of $Fe^{2+}/Fe^{3+}$ attached with linkers, which can be tuned structurally for various metal ions. Upon further research, Co and Ni were found to be p-type gas sensor whereas Cu and Zn to be n-type. Ni and Cu ferrite strongly detected harmful gases acetone and hydrogen sulfide respectively. Further, this work implied gas sensors applications in food monitoring and breath analyzer for diabetic patients.[199] As an example of high-performance supercapacitor, $CoFe_2O_4$ synthesized via MOF template, exhibited specific capacity of 312.8 F/g at a current density of 0.5 A/g with excellent energy density of 90.3 Wh/kg along the 88.4 % retention after 10,000 cycles.[200]

Among various 2D transitional dichalcogenide materials, $MoS_2$ is one of the most prominent materials in electrochemical applications due to its impressive properties such as, remarkable surface area, high electron mobility and ion diffusion pathways leading to superior performance of devices. DFT calculations were worked on $MoS_2$ nanosheets wrapped $CuFe_2O_4$ nanotubes synthesized via electrospinning and hydrothermal routes. $CuFe_2O_4$@$MoS_2$ exhibited 20-28% higher response value than pristine $CuFe_2O_4$ nanotubes. DFT calculations revealed more free electrons were available in the nanocomposite than the pristine ferrite.[201]

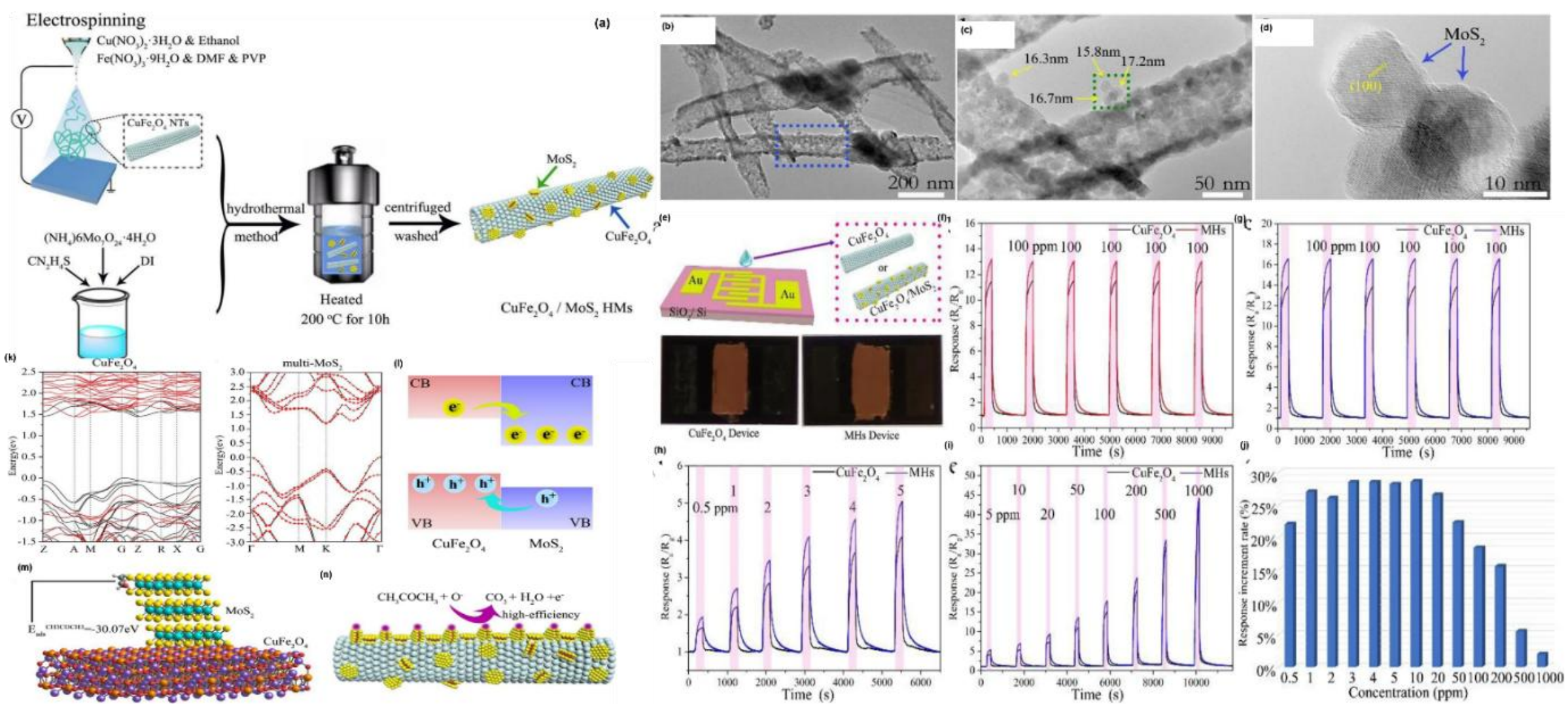

Fig. 30 $CuFe_2O_4$/$MoS_2$ heterostructure for high-performance acetone sensing (a) Electrospinning-assisted synthesis followed by hydrothermal growth of $MoS_2$ on $CuFe_2O_4$ fibers (b–d) TEM and HRTEM images confirming the formation of the $CuFe_2O_4$/$MoS_2$ heterostructure (e) Schematic and photographs of the fabricated sensing devices (f–g) Dynamic response–recovery behavior toward repeated exposure to 100 ppm acetone (h–i) Concentration-dependent sensing responses at low (0.5–5 ppm) and high (10–1000 ppm) acetone concentrations (j) Relationship between sensor response and acetone concentration (k–m) Electronic structure and interfacial charge-transfer analysis of the $CuFe_2O_4$/$MoS_2$ heterojunction (n) Schematic illustration of the enhanced acetone-sensing mechanism[201]

One of the early detections of diabetes is sourced by high acetone concentration range varies in a healthy patient and a diabetic patient. With the efforts to aim for the specific range 1.8 to 3.7 ppm, $ZnFe_2O_4$@$MoS_2$ has been investigated as gas sensor exhibiting a high response of 40.0-100 ppm acetone at 200 °C. The hollow microspherical $ZnFe_2O_4$ incorporated with $MoS_2$ nanosheets facilitates large active sites for oxygen and acetone adsorption with a response/recovery times of 31.7/103.3 s, respectively.[202]

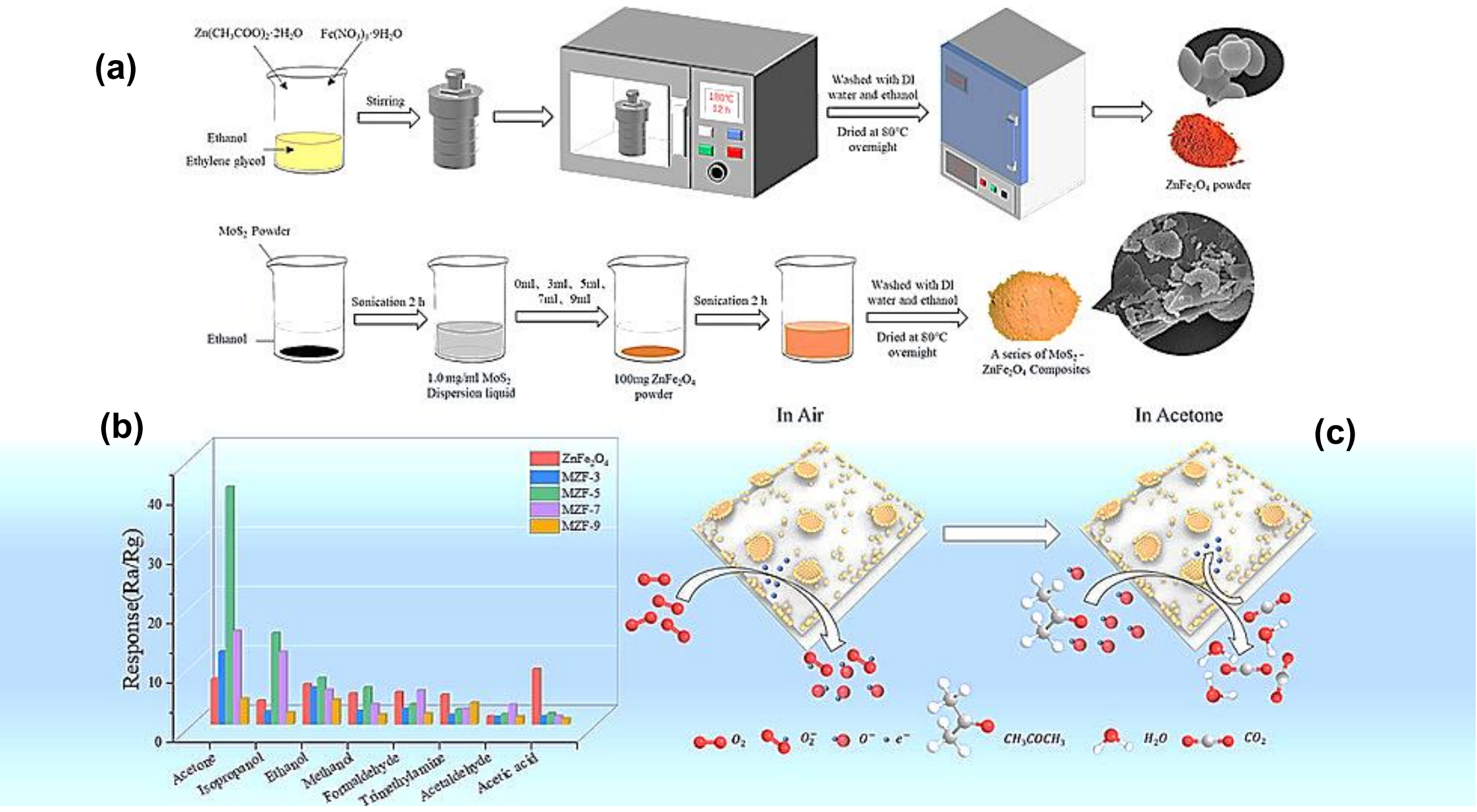

Fig. 31 $ZnFe_2O_4/MoS_2$ heterostructure for acetone sensing (a) schematic representation of the synthesis route and morphology of $ZnFe_2O_4$ and $ZnFe_2O_4/MoS_2$ composites (b) comparison of sensor responses toward various gases demonstrating superior selectivity toward acetone (c) proposed gas-sensing mechanism showing adsorption of oxygen species in air and their reaction with acetone molecules at the $ZnFe_2O_4/MoS_2$ surface, resulting in charge transfer and resistance modulation[202]

$MoS_2$ is emerging in applications of supercapacitors as well. Almessiere et al. $MoS_2@CoFe_2O_4$ hybrid nanocomposite exhibiting specific capacitance of 528.60 F $g^{-1}$ is reached. The supercapacitor displayed incredible energy density of 66.94 Wh $kg^{-1}$ and maintained a brilliant capacitance retention of 98.88 % along 10,000 cycles.[203] Ganesh babu and his team explored $MnFe_2O_4/MoS_2@MWCNT$ composite for supercapacitor electrode. The ternary nanocomposite electrode displayed energy density of 47 Wh $Kg^{-1}$.[204]

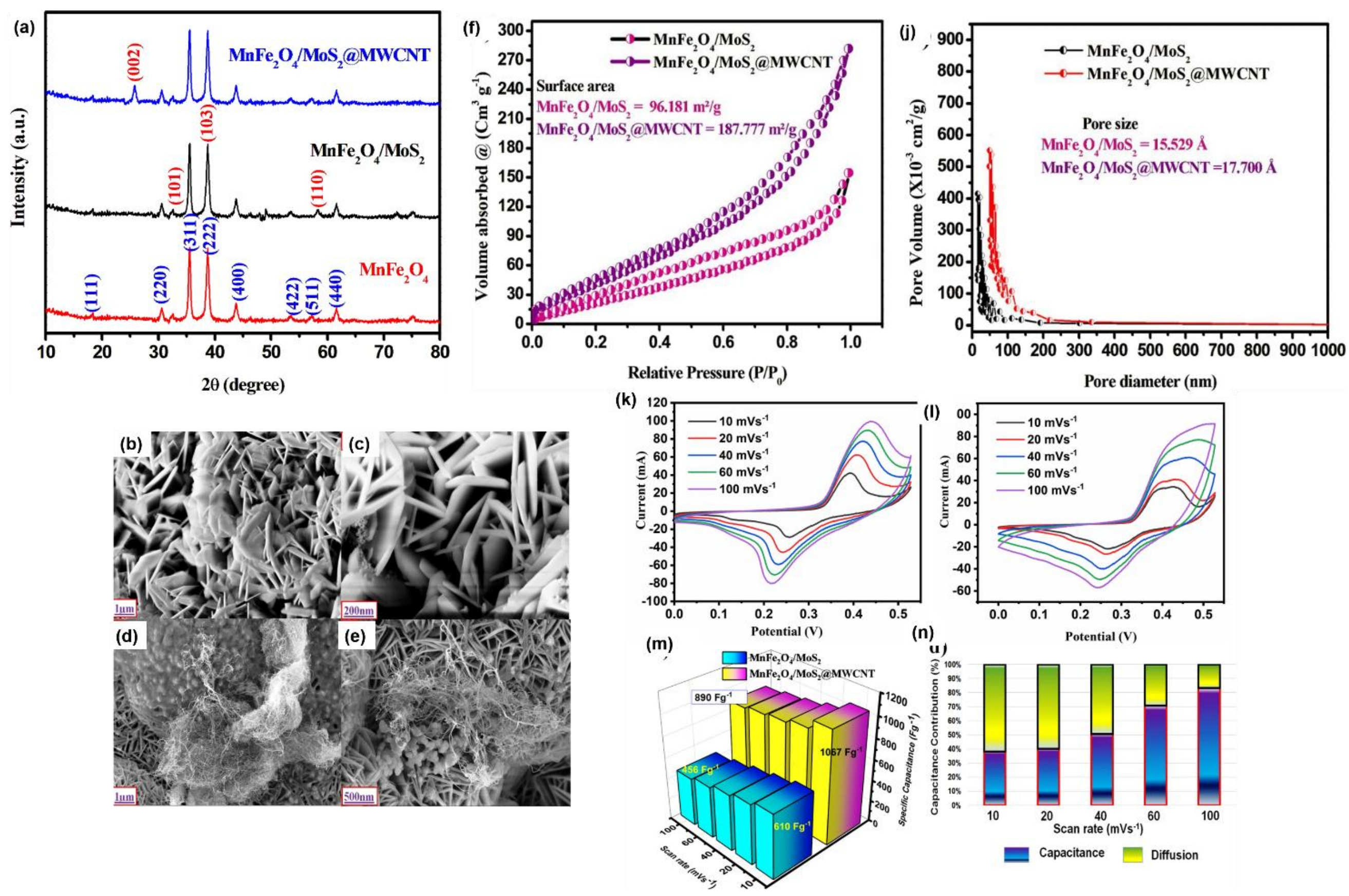

Fig. 32 Characterization and electrochemical performance of $MnFe_2O_4/MoS_2$@MWCNT composites (a) XRD patterns of $MnFe_2O_4$, $MnFe_2O_4/MoS_2$, and $MnFe_2O_4/MoS_2$@MWCNT (b–e) SEM images illustrating the morphology of the synthesized materials (f) $N_2$ adsorption–desorption isotherms and (j) pore-size distribution curves showing enhanced surface area and porosity after MWCNT incorporation (k,l) Cyclic voltammograms measurements at different scan rates for $MnFe_2O_4/MoS_2$ and $MnFe_2O_4/MoS_2$@MWCNT electrodes (m) Comparison of specific capacitance values and (n) capacitive versus diffusion-controlled charge-storage contributions at different scan rates[204]

## 6.2 Spinel Ferrites in Biomedical Applications

Recent studies in nanotechnology have proved spinel ferrite as versatile nanomaterials in biomedical applications. More recently, the studies have also begun to uncover the potential that these materials have for being used to detect disease, drug delivery, biocompatibility, etc., and their potential is significant. In one of the studied examples, $ZnFe_2O_4$, the experimental results suggest that zinc ferrites are not only readily interactable magnetically and chemically but are also readily functionalized to bind to proteins or to bind to DNA without the complex labelling processes needed to prepare tools with other biomolecules. Thus, devices based on ferrospinels can detect disease earlier. Scientists are probing the behaviour of ferrites for biocompatible and magnetic properties too, not just surface-related electrochemical devices.[205]

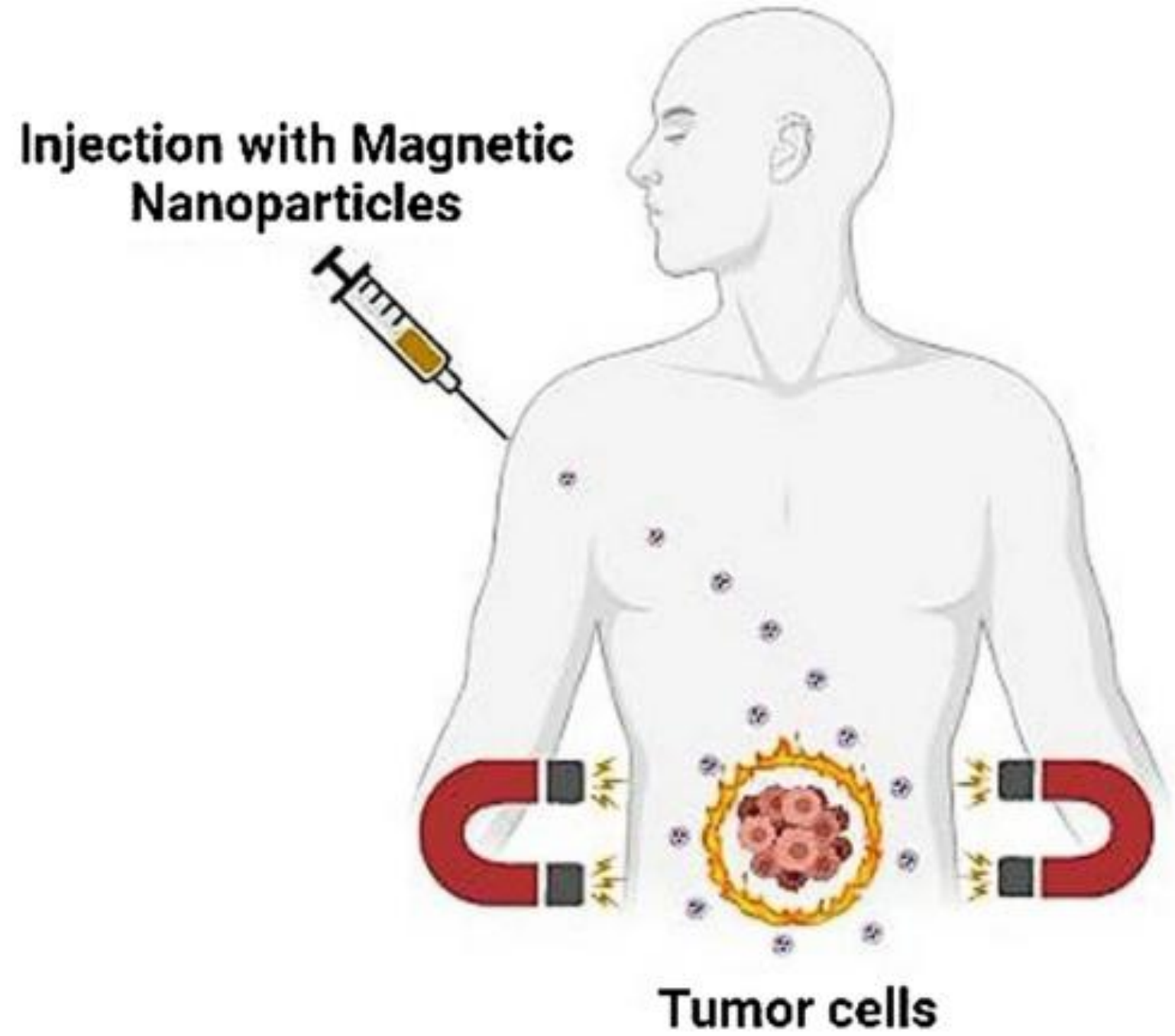

Fig. 33 Spinel ferrite magnetic nanoparticles for targeted cancer therapy through magnetic hyperthermia, where externally applied magnetic fields localize nanoparticles at the tumor site and induce therapeutic heating[205]

An investigation on magnetite and Mn, Co and Ni ferrite yielded approximately 9nm pure spinel nanoparticles where their magnetic property weakens when heated. The Mn-based variations were tried in tumor cells of the human cervix and were found to be relatively harmless with cytotoxic effect at various doses; however, the Co and Ni samples were toxic with an increase in doses. This work validated the previously discussed fact that compositions hold utter importance to be useful in their applications.[206] In another work, spinel ferrites were examined as a source of breast cancer treatment and detection. $Ni^{2+}$-doped $MgFe_2O_4$ NPs via the solvothermal reflux method were tested not only in breast cancer treatment applications but also in photocatalysis as a dual functional material.[207]

Some spinel ferrite NPs, such as cobalt, zinc, nickel, and Mg-doped ferrites, show antiproliferative activities against breast cancer cells, even at an external magnetic field, while maintaining acceptable biocompatibility, thus highlighting their potential for dual applications in diagnostics and therapy.[208]

In looking forward, the next steps should be focused on the integration of the spinel ferrite-based sensors in realistic and viable biomedical workflows. By linking electrochemical performance to biological safety and relevance, spinel ferrite-based devices have the potential to extend beyond proof-of-concept and start to make a real impact in the field of personalized and point-of-care medicine.[209]

### 6.3 Spinel Ferrites in Environmental Monitoring

Spinel Ferrites have been increasingly being explored not only in electrochemical sensors, energy storage materials, but also as active materials for environmental monitoring. Recent studies have also shown that spinel ferrite nanostructures have the ability to adsorb, convert, and decompose different kinds of pollutants in the system. These qualities make them particularly effective at absorbing and degrading dye pollution. Spinel ferrites' catalytic characteristics allow for effective oxidation and degradation of organic pigments.[210] For instance, Spinel ferrite nanostructures have been shown to adsorb and convert nitroaromatics and heavy metals in aqueous media. Nitroaromatics are synthetic industrial disposed chemicals, used in the manufacture of explosives, dyes, insecticides, medicines, and polymers and Spinel ferrites are magnetic, making them easy to separate from treated water.[211]

In addition to that, the nanocomposite of $ZnFe_2O_4$@Carbon were analysed by Rajani et al. showed high potential in the coordination of toxic hexavalent chromium in aqueous media, implying that spinel ferrites have the potential to act as active materials in the removal and coordination of heavy metals in aqueous media.[212]

Hydrothermally synthesized zinc ferrite nanoparticles have been reported to serve as fertilizer carrier nanomaterial for zinc and iron deficiency in soil, which enhanced photosynthesis process and fight against drought. Prashantha and team observed growing tomatoes with the fertilizer as well as fabricated pH and nutrient sensing Zn ferrite, implying that ferrite-based systems could be combined into smart agri-environmental systems capable of simultaneously monitoring soil chemistry and providing nutrients. In addition to all these applications, $ZnFe_2O_4$ supercapacitor also exhibited specific capacitance 360 $Fg^{-1}$ and 93.1% retention to 2400 cycles.[52]

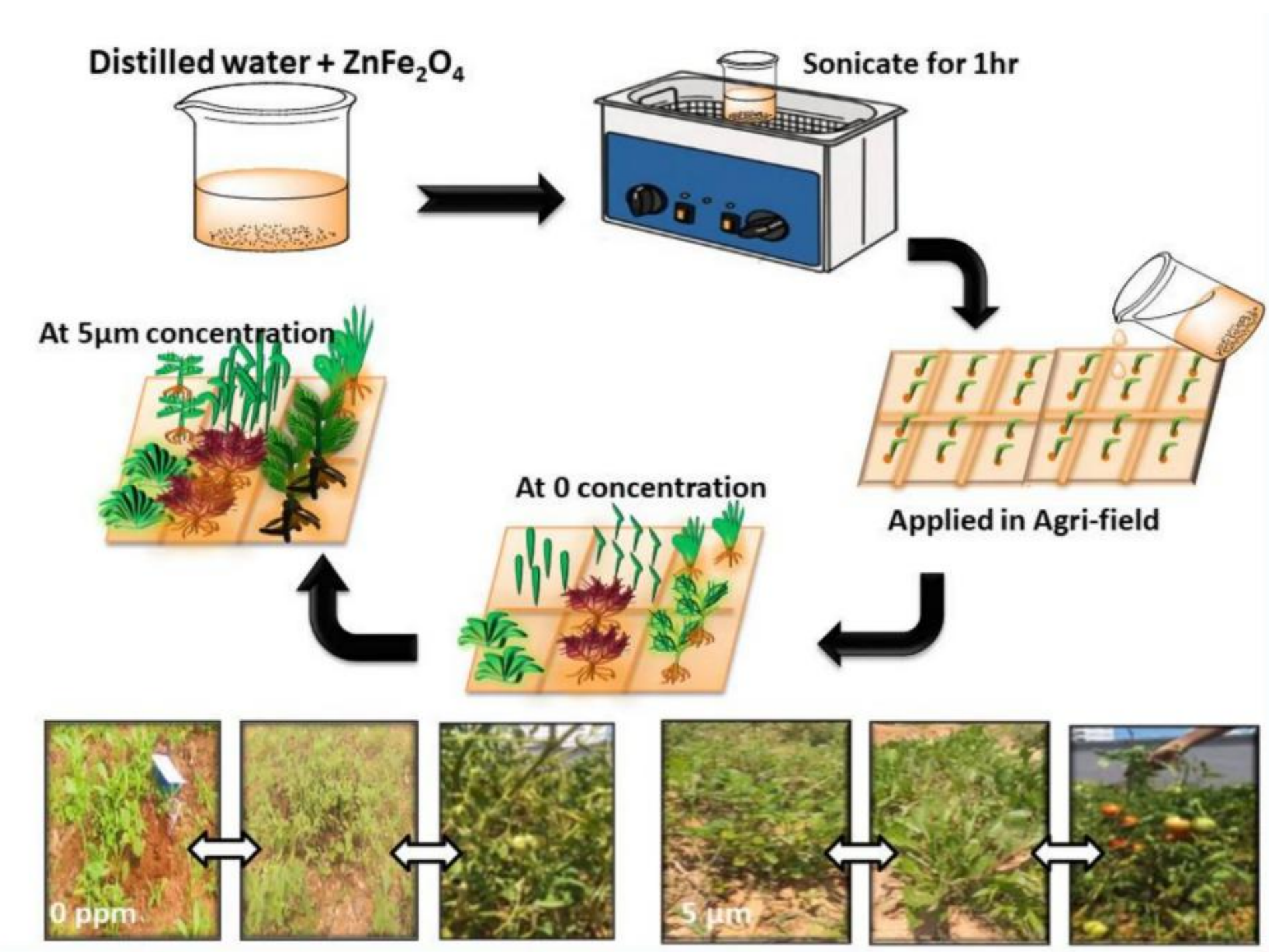

Fig. 34 $ZnFe_2O_4$ suspension is dispersed by ultrasonication and applied to agricultural fields, where it supplies essential Zn and Fe nutrients, promoting tomato plant growth[213]

Collectively, these advances suggest that spinel ferrites will have a multifunctional role in environmental science, as remediation agents and as low-cost, magnetically recoverable sensing surfaces to monitor pollutants in water and soil in real-time.

### 6.4 Cost-effective Fabrication Strategies

Even after numerous applications of spinel ferrites, scalability and cost issues persist. The development of cost-effective and scalable fabrication methods for spinel ferrite-based devices should be part of the research objectives. As discussed earlier, synthesis routes like microwave-assisted production take relatively less time and power. More techniques such as inkjet printing, roll-to-roll manufacturing, or additive manufacturing should be explored. Sanketh et al. reviewed recent trends of inkjet printing in chemiresistive gas sensors and underscored the advantages like precise material deposition, controlled nanostructuring and the ease to function under normal ambience over other methods. These advantages make this process environmentally sustainable as well as cost effective.[214] Recently, many researchers have also focused on green synthesis or solvent-free synthesis of ferrites, discarding the toxic chemicals being used in the processing. Roque et al. synthesized suberin polymer naturally from potato peels and cork powder as self-assembling substrate for detection of VOCs.[215] Another innovative approach would be micro-electromechanical systems (MEMS). The miniaturised form requires less materials and less power, making them eligible for biomedical implants and flexible wearable devices.[216] Further, more concrete tactics and techniques should be bookmarked to fabricate ferrospinels at a suitable cost.

### 6.5 Integration with AI smart technology

Combining spinel ferrite-based sensors with artificial intelligence (AI) and machine learning algorithms has become crutial for improved data analysis and decision-making. This technical advancement allows studying real time data and locate patterns in order to optimise the system. Patil and his team used the hybrid DNN-WSO algorithm to effectively calculate and optimize gas sensing performance, which is significant for advancing efficient gas sensor development. They proposed synthesis of $Ni_{0.6-x}Mg_xCo_{0.4}Fe_2O_4$ nanoparticles sol-gel method with various stoichiometries. The algorithm predicts x= 0.2 showed superior results in sensing $NO_2$ and $NH_3$ gases.[217]

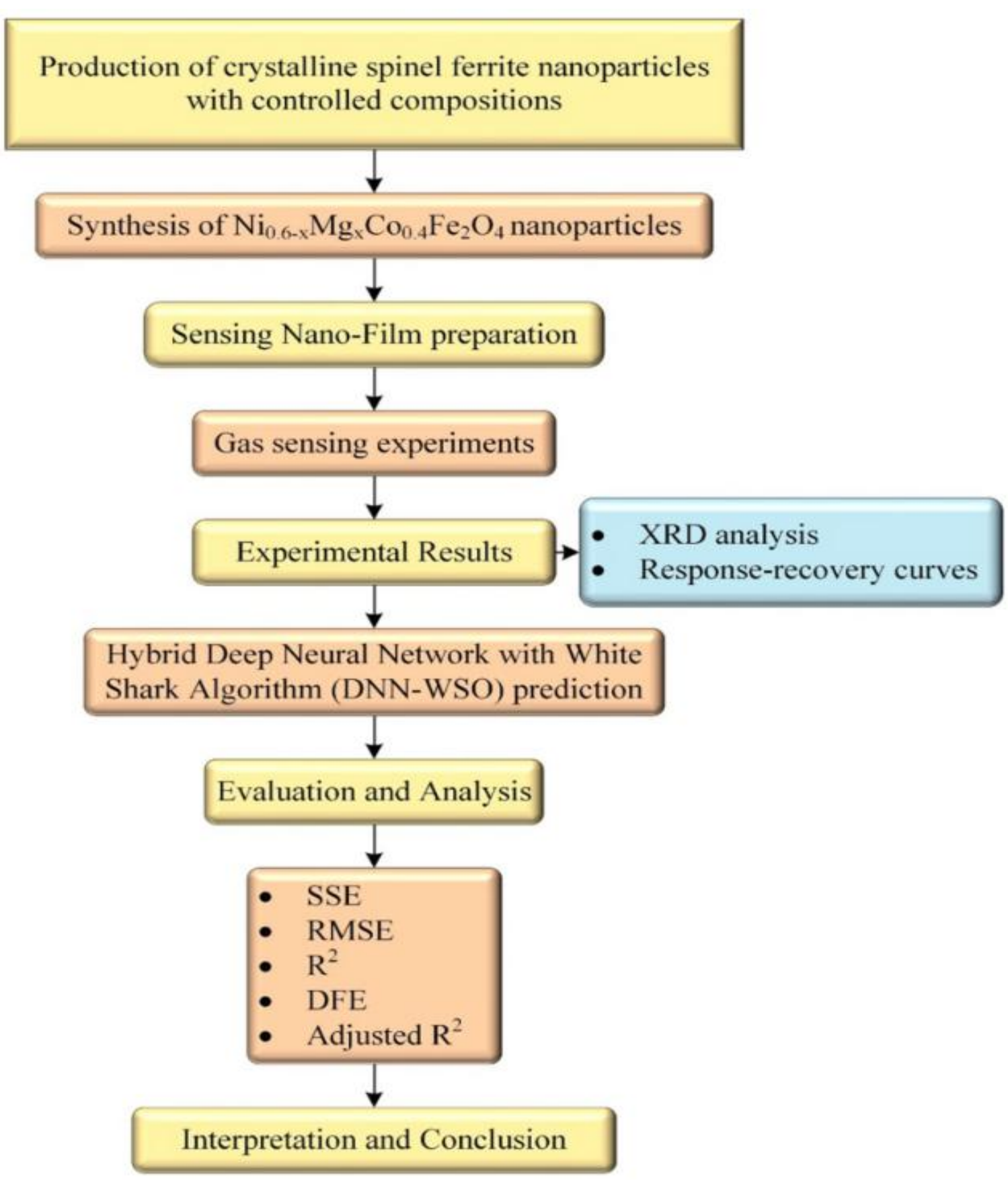

Fig. 35 Schematic representation of an AI-assisted gas sensing framework based on $Ni_{0.6-x}Mg_xCo_{0.4}Fe_2O_4$ spinel ferrite using DNN-WSO for intelligent analysis of sensor applications[217]

Narkhede and his team utilised multimodal AI fusion approach to detect hazardous chemical leaks through gas and sensors array and thermal images of 6400 gas samples. The work used Convolutional Neural Network (CNN) for extracting thermal images and long short-term memory (LSTM) for gas sensing sequences individually and then used multimodal data fusion in case of any system failures, so that, one modal works while another repairs. Late fusion also applied individual fusion result adding with max or average fusion modal results to achieve maximum accuracy. Their work achieved 96% accuracy through this innovative approach and exemplifies the AI prediction integration in real-world problem of detecting gas leaks in factories, home or labs and other leak-prone environments.[218] In dynamic settings, hybrid AI-electrochemical devicing setups show promises real-time feedback to maintain accuracy when conditions change unexpectedly. Such advances help move these electrochemical materials beyond experimental stages implying deployment of AI outside laboratories now seems within reach.

### 6.6 Commercialization

In order to turn laboratory-scale innovations into practical technology for real-world applications across several sectors, a deliberate focus on the commercialisation of spinel ferrite-based sensors is crucial. Despite significant advancements in research, successful market integration requires focused efforts in the areas of scalability, reproducibility, cost optimisation, and regulatory compliance.

## 7. Conclusion

This review highlights the significant potential of spinel ferrite-based materials in the field of electrochemical sensing specifically, detection and quantification of a wide range of analytes in metal oxide gas sensors. Spinel ferrites exhibit distinctive physicochemical and electrochemical properties that make them highly suitable for pseudocapacitance and prominent active sites in supercapacitors. In particular, the integration of spinel ferrites with highly conductive matrices and various nanostructured materials significantly improves electrochemical performance. Recent studies have shown that ferrites containing transition metal ions such as Ni, Mn, and Zn demonstrate enhanced catalytic activity, favourable redox behaviour, and improved electron transport characteristics. The formation of such hybrid and composite systems promotes synergistic effects, leading to enhanced sensitivity, faster electron transfer kinetics, and capacitance. Therefore, spinel ferrite-based nanocomposites have emerged as promising candidates for the development of high-performance electrochemical devices.

Moreover, the incorporation of spinel ferrites into practical electrochemical reactions-based platforms, including screen-printed electrodes and other advanced substrates, offers considerable advantages in terms of portability, cost-effectiveness, mechanical stability, and scalability. These developments contribute to the broader applicability and commercialization potential of electrochemical sensing and storage devices. Spinel ferrite based electrochemical devices- such as sensors and supercapacitors represent a rapidly evolving and promising research domain. Ongoing interdisciplinary efforts are anticipated to facilitate their practical implementation in real-world applications, enabling rapid, accurate, and reliable analytical solutions across diverse fields.

**Conflict of interest**

There is no conflict to declare.

**Acknowledgement:**

Authors did not have any financial funding for this work.